# Single-Source Nets
# of Algebraically-Quantized Reflective Liouville Potentials on the Line
# I. Almost-Everywhere Holomorphic Solutions of Rational Canonical Sturm-Liouville Equations with Second-Order Poles

G. Natanson
ai-solutions Inc.
2232 Blue Valley Dr.
Silver Spring MD 20904 U.S.A.
greg_natanson@yahoo.com

The paper presents the unified technique for constructing SUSY ladders of rational Liouville potentials (RLPs) starting from the so-called ‘Gauss-reference’ (GRef) potentials exactly quantized on the line via classical Jacobi, classical (generalized) Laguerre, or Romanovski-Routh polynomials with energy-dependent indexes. Each RLP is obtained by means of the Liouville transformation (LT) of the appropriate rational canonical Sturm-Liouville equation (RCSLE) with second-order poles. The presented analysis takes advantage of the generic factorization of canonical Sturm-Liouville equations (CSLEs) in terms of intertwining ‘generalized’ Darboux operators. We refer to the latter operators as the canonical Liouville-Darboux transformations (CLDTs) to stress that they are equivalent to three-step operations: i) the LT from the CSLE to the Schrödinger equation; ii) the Darboux transformation (DT) of the appropriate LP; and iii) the inverse LT from the Schrödinger equation to the new CSLE. It is proven that the CLDT preserves the rational form of the RCSLE if its factorization function (FF) is an almost-everywhere holomorphic (AEH) solution of the RCSLE (or, in other words, a solution with a rational logarithmic derivative). As explained in the paper there are up to four gauge transformations which convert each RCSLE of our interest into the second-order differential equations with energy-dependent polynomial coefficients. The most important result of the paper is that polynomial solutions of these equations belong to sequences of Heine polynomials obtained by varying free terms at fixed values of singular points and the appropriate characteristic exponents. This allows us to construct networks of polynomial solutions -- the so-called ‘Gauss-seed (GS) Heine’ or ‘*c*-Heine’ polynomials -- starting from Jacobi, (generalized) Laguerre, or Routh polynomials.

## Introduction

About thirty years ago Cooper, Ginoccio, and Khare [1] opened a new direction in the theory of quantized-by-polynomial rational potentials by applying the state-erasing Darboux transformation (DT) to the generic potential solvable by hypergeometric functions [2]. Unfortunately this fundamentally significant consequence of their work has not attracted proper attention, perhaps due to the fact that their study was mainly focused on Gendenshtein's controversial statements [3] concerning the interrelationship between shape-invariance and exact solvability of rational potentials. As a result of this very specific focus the mentioned authors [1] overlooked the most striking feature of the eigenfunctions describing the discrete energy spectrum in the new exactly solvable potential, namely, these eigenfunctions are expressible in terms of polynomials of a completely new type never analyzed before in either mathematical or quantum-mechanical literature.

A renewed interest in this problem was only recently stimulated by Quesne's pioneering work [4] who constructed rational potentials quantized by $X_1$-Laguerre and $X_1$-Jacobi orthogonal polynomials discovered by Gomez-Ullate, Kamran and Milson [5, 6] a year earlier. In a separate publication [7] Quesne presented a detailed analysis showing that the exactly quantized potentials discovered by her are nothing but rational SUSY partners of the isotonic oscillator and Darboux/Pöschl-Teller (D/PT) potential [8, 9]. Her discovery was later broadened by Odake and Sasaki [10, 11] who constructed families of rational potentials quantized by $X_m$-Laguerre and $X_m$-Jacobi orthogonal polynomials [12]. In this connection it seems also important to point to Grandati's [13] mathematically scrupulous analysis of this problem for the isotonic oscillator.

Since all the mentioned potentials have singularities on the real axis we skip further discussion of any issues related to exceptional orthogonal polynomials focusing solely on reflective rational potentials on the line. The epithet 'reflective' is used here to stress that we only consider potentials vanishing either at $+\infty$ or alternatively at $-\infty$. In particular this assumption allows us to avoid any reference to rational SUSY partners of the harmonic oscillator having parabolic barriers at both ends of the infinite quantization interval.

Before outlining main results of this paper let us mention three works by Quesne [14-16] who developed a similar technique for all three shape-invariant potentials on the line: Rosen-Morse (RM) [17], Morse [18], and Gendenshtein [3] (conventionally referred to as 'Scarf II') potentials. As explained in next Section the latter represent three different types of boundary conditions for rational Sturm-Liouville problems associated with three families of 'Gauss-reference' (*r*-GRef, *c*-GRef, and *i*-GRef) potentials in our terms [19, 20].

For each of the aforementioned exactly solvable GRef potentials we, in following Bose [21], start from the rational 'canonical'[x)] Sturm-Liouville equation (RCSLE) either with three regular singular points (including infinity) or -- in the confluent case -- with the second-order pole at 0 and an irregular singular point at infinity. As pointed to in [22] the technique adopted by us [2] from Bose's paper [21] is nothing but the conventional [25, 24, 26] Liouville transformation (LT) applied to the given RCSLE. The potential obtained in such a way is thus referred to in our publications as the rational Liouville potential (RLP). In next Section we outline most important elements of our theory of RCSLEs under the following very strict constraints:

i) the given density function has second-order poles at the finite end-points of the quantization interval and is holomorphic everywhere else;
ii) the order of the so-called 'tangent polynomial' (TP) in the numerator of the given density function may not be larger than 2.

As illuminated in Section 5 sign of TP discriminant plays a fundamentally important role in the concept of polynomial-fraction beams (PFr beams), $_{\iota}\mathbf{B}$, which are obtained by varying parameters of the appropriate Bose invariants (including the energy ε) at fixed values of TP coefficients. (Here the index ι specifies the appropriate family of the RLPs, namely, ι =1, 0, and *i* for SUSY partners of the *r*-GRef, *c*-GRef, and *i*-GRef potentials, respectively.)

After the RLP is constructed we convert the RCSLE into a solved-by-polynomials differential equation using the appropriate gauge transformation. (For PFr beams $_{1}\mathbf{B}$ and $_{i}\mathbf{B}$

---

x) In following [22] we use the epithet 'canonical' to indicate that the given SLE is written in the 'normal' [23, 24] form, with no first-derivative term.

the resultant solved-by-polynomials differential equations have only regular singular points and therefore belong to the class of Fuschian equations with rational coefficients referred to below as Heine-type equations.) In other words we move in the direction exactly opposite to the conventional one [23, 27] utilized by Levai [28, 29] and more recently by Quesne [4].

While analyzing the cited work by Cooper, Ginoccio, and Khare [1] the author made several important observations briefly summarized in [19]. Possibly the most important of them is the fact that eigenfunctions for the Cooper-Ginoccio-Khare (CGK) potential have the so-called 'quasi-algebraic' [30] form in the sense that they turn into algebraic functions if all (generally irrational) power exponents happened to be rational numbers. In following [31], Gomez-Ullate, Kamran and Milson [12] (as well as Grandati in [13]) recently referred to these solutions as 'quasi-rational' to stress that their logarithmic derivatives are rational functions. However, for the purpose of this paper it seems more essential to emphasize another important feature of these solutions, namely, as pointed to in [32] with regard to the Heun equation they are 'valid in the entire plane, except of course, at the singularities and, in most cases, with various cuts made to ensure single-valueness'. For this reason we prefer to refer to them as 'almost-everywhere holomorphic' (AEH) solutions (especially keeping in mind that the original term 'quasi-algebraic' introduced in [30] is inapplicable to confluent rational potentials since the appropriate AEH solutions contain exponential factors). We explicitly distinguish the AEH solutions of CRSLEs with GRef Bose invariants by referring to them as Gauss-seed (GS) solutions.

Another important inference based on our analysis of specific structure of the CGK potential [1] and already illustrated by a few examples in [19, 33, 34] is that canonical Liouville-Darboux transformations (CLDTs) with AEH factorization functions (FFs) preserve the rational form of the CRSLE. We use the term 'CLDT' to stress that the transformation in question is induced by a Darboux deformation of the appropriate LP followed by the inverse LT from the Schrödinger equation with the new potential to the CSLE.

As a matter of fact transformations of such kind were originally introduced by Rudyak and Zakhariev [35] in the scattering theory and then studied more cautiously by Leib and Schnizer [36-38] and independently by Suzko [39-41] in the nineties. In this connection it seems also useful to point to a more recent paper by Suzko and Giorgadze [42] who explicitly re-formulated this problem in a way very close to our approach though using a slightly different terminology. In Appendix A we briefly summarize this universal approach dealing with the so-called 'generalized Darboux transformations' of the 'generalized' [42] Schrödinger equation – CLDTs of the CSLE in our terms. However, anywhere else in this paper we only discuss its reduction to RCSLEs.

We found it necessary to change the term 'GDT' for the 'CLDT' because the former is used by different authors with different meanings even in the theory of linear ordinary differential equations (see, i.g., [43, 44]) -- let alone the theory of the linear [45] and nonlinear [46-48] non-stationary Schrödinger equation and the theory of coupled linear ordinary differential equations [49].) In addition, since the RCSLE is nothing but a specific (no first derivative) case of the generic second-order differential equation with rational coefficients its factorization in terms of 'generalized Darboux' operators represents a particular example of a more general factorization scheme recently suggested by Gomez-Ullate, Kamran, and Milson [50].

The paper is organized as follows. In Section 2 we specify three major types of quantization intervals for the RCSLEs of our interest:

i) finite interval (0, 1) or (−1, +1) for SUSY partners of the *r*-GRef potential;
ii) semi-axis (0, ∞) for SUSY partners of the *c*-GRef potential;
iii) infinite interval (–∞, +∞) for SUSY partners of the *i*-GRef potential.

The TP used to generate each family of RLPs is chosen in such a way that the LT converts each RCSLEs into the Schrödinger equation with a rational potential V(x) on the line: $-\infty < x < +\infty$. SUSY partners of radial *r*- and *c*-GRef potentials will be analyzed in a separate publication [51].

In Section 3 we eliminate second-order poles from the free term of the RCSLE by means of a gauge transformation which transforms the given RCSLE into either the Fuschian differential equation with rational coefficients ($|\iota| =1$) or its confluent counter-part ($\iota=0$). We then show that the polynomial component of any AEH solution is a solution of the derived equation. Since the latter is a second-order differential equation of Heine's type [52, 53] for $\iota = 1, i$ we refer to these components as Gauss-seed (GS) Heine and *c*-Heine polynomials. The epithet 'GS' is added to emphasize that we deal with a very special type of Heine and *c*-Heine polynomials which are generated via multi-step CLDTs using seed solutions of the CSLEs with GRef Bose invariants.

In Section 4 we explicitly take advantage of the fact that rational SUSY partners of both *r*- and *i*-GRef potentials are all linked to CRSLEs with regular singular points (including infinity). As a result we can make use of the theory of Fuschian differential equations [23] to demonstrate that there are generally two classes of AEH solutions ($\alpha$- and $\beta$-Classes in our terms) depending on which of two energy-dependent parameters $\alpha(\varepsilon|_{\iota}\mathbf{B})$ or $\beta(\varepsilon|_{\iota}\mathbf{B})$ is equal to the non-positive integer –m. In particular commonly used eigenfunctions for the *r*- and *i*-GRef potentials belong to the $\alpha$-Class. (Some examples of AEH solutions from the $\beta$-Class are presented in subsection 5.1 of Section 5 dealing exclusively with *r*-GRef potentials.)

In Section 5 we derive quartic equations to compute energies of co-existent AEH solutions formed by Jacobi, generalized Laguerre, and Routh [54, 20] polynomials for the *r*-GRef, *c*-GRef, and Milson [22] potentials covered by subsection 5.1, 5.2, and 5.3, respectively. For bound energy states in *r*-GRef and *c*-GRef potentials the quartic equations of this type were originally obtained by Grosche [55]. In our recent paper [20] we have proven that one can write a similar quartic equation for bound energy states in the Milson's [22] reduction of the *i*-GRef potential.

In Subsection 5.1 we show that the m$^{\text{th}}$ eigenfunction of the RCSLE with the *r*-GRef Bose invariant (BI) generated using the second-order TP with positive discriminant $\Delta_T > 0$ is always accompanied by three GS solutions formed by polynomials of the same order m. It is crucial that this quartet of AEH functions is composed of solutions of four different types, namely, two solutions **a**m and **b**m regular at the opposite ends, the already mentioned normalizable solution **c**m and a solution **d**m irregular at both ends. Factorization functions (FFs) of these four types **a**, **b**, **c**, and **d** ($T_3$, $T_4$, $T_1$, and $T_2$ in Sukumar's terms [56]) were originally introduced in our studies [57, 58] on Darboux transformations (DTs) of centrifugal-barrier potentials. We found this notation especially convenient for labeling different AEH solutions of the RCSLE with *r*-GRef BI taking into account that the aforementioned double indexes **t**m, with **t** = **a**, **b**, **c**, or **d**, unambiguously determine these solutions in the region where the RLP generated by the TP with $\Delta_T > 0$ has at least m bound energy levels. Each solution is associated with one of four simple real roots of the quartic polynomial introduced in Section 5.1 instead of the radical equation initially derived by us in [2] for eigenfunctions.

As shown in Appendix B the upper bound energy levels in the *r*-GRef potential disappear one-by-one along the sequence of the straight lines and as a result the eigenfunction turns into an GS solution of type **a** on the outer part of each line (referred to as "**c**/**a**′ separatrix", where we mark **a** by prime to distinguish the latter solution, **a**′m, from the one in the principal sequence **a**m starting from the basis solution **a**0).

Since the generic *c*-GRef potential analyzed in subsection 5.2 has a Coulomb tail (CT) at infinity it has infinitely many bound energy levels [2] provided that the potential asymptotically approaches 0 at the given end of the quantization interval. We refer to this *c*-GRef potential of as the CT0-branch. It is proven that the RCSLE associated with the CT0-branch has infinitely many GS solutions of each of four types and that all GS solutions of type **a** are nodeless which implies that multi-step rational SUSY partners of this potential form an infinite single-source net of isospectral potentials. The CT+ branch with a CT approaching a positive barrier height at $-\infty$ represents a more challenging example which will be addressed separately.

In subsection 5.3 we briefly summarize our results [20] for the *i*-GRef potential, with emphasis on its algebraically quantized reduction discovered by Milson [22]. The important feature of the Milson potential is that each eigenfunction **c**m is accompanied by an GS solution **d**m. In particular, since the basic solution **d**0 is necessarily nodeless it can be used as the FF for constructing the exactly quantized rational SUSY partner with an inserted energy level.

In Section 6 we prove that any CLDT with an AEH FF converts the given RCSLE into another one while keeping the density function unchanged. Some specific properties of RefPFrs generated by single-step CLDTs of GRef potentials are further exploited in Appendix C.

The crucial common feature of multi-step rational SUSY partners of all GRef potentials on the line, *excluding the Gendenshtein potential*, is that any CLDT with an AEH FF preserves exponent differences (ExpDiffs) of finite singular points. As for the Gendenshtein potential it behaves to a large extent similarly to radial GRef potentials [19, 51], namely, the exponent differences of the singular points $-i$ and $+i$ change by 1 with each step. In Appendix D we illustrate this anomaly using single-step SUSY partners of a nearly-symmetric reduction of the Gendenshtein potential.

Since any AEH solution preserves its 'weighted-polynomial' form under action of the CLDT in question its monomial-product component satisfies Heine's or *c*-Heine's equation. (In this series of publications we only consider the general case when the polynomial solution of our interest has only simple zeros.)

Some important features of polynomials solutions of these equations – Gauss seed (GS) Heine and *c*-Heine polynomials as we refer to them -- are outlined in Section 7.

In Section 8 we again restrict our analysis to Fuschian differential equations with energy-dependent rational coefficients and extend the results of Section 4 to SUSY pairs of Heine's equations constructed using CLDTs with AEH FFs. It is proven that the latter transformations change the $\alpha$- and $\beta$-parameters evaluated at energies of AEH solutions by exactly the same integer so that the partner AEH solutions both belong to the same $\alpha$- or $\beta$-Class.

Finally, in Section 9 we summarize the most important results of this paper and also outline its further developments which will be covered in shortly coming publications.

## 2. Three types of the quantization intervals for RCSLEs with algebraic energy spectra

Let us start our discussion from the general expression for the 'canonical' (no first-derivative term) SLE with a rational free term (RCSLE):

$$\left\{ \frac{d^2}{d\xi^2} + I^o[\xi \,|\, \bar{\lambda}_o, {}_\iota O^o_{N-|\iota|-1}] + {}_\iota\wp[\xi; {}_\iota T_{K_*}]\, \varepsilon \right\} \Phi[\xi; \varepsilon \,|\, {}_\iota\mathbf{B}] = 0\,. \tag{2.1}$$

where the energy-dependent superposition

$$I^o[\xi; \varepsilon \,|\, {}_\iota\mathbf{B}] \equiv I^o[\xi \,|\, {}_\iota\bar{\lambda}_o, {}_\iota O^o_{N-|\iota|-1}] + {}_\iota\wp[\xi; {}_\iota T_{K_*}]\, \varepsilon \tag{2.2}$$

of the reference PFr (RefPFr)

$$I^{o}[\xi\,|\,{}_{\iota}\bar{\lambda}_{o},{}_{\iota}O^{o}_{N-|\iota|-1}] = \sum_{r=0}^{N-1}\frac{1-{}_{\iota}\lambda^{2}_{o;r}}{4(\xi-{}_{\iota}e_{r})^{2}} + \frac{{}_{\iota}O^{o}_{N-|\iota|-1}[\xi]}{4\prod_{r=0}^{N-1}(\xi-{}_{\iota}e_{r})} - {}^{1}\!/_{4}\,\delta_{\iota,0}\nu_{o}^{2} \qquad (2.3)$$

and the rational density function

$${}_{\iota}\wp[\xi;\,{}_{\iota}T_{K_*}] = \frac{{}_{\iota}T_{K_*}[\xi]}{4\prod_{r=0}^{N-1}(\xi-{}_{\iota}e_{r})^{2}} \qquad (K_* \le 2N-2\,|\,\iota\,|) \qquad (2.4)$$

is referred to as 'Bose invariant' to emphasize Bose's impact [21] on our initial studies on *r*- and *c*-GRef potentials ($N = \iota + 1$) exactly solvable by hypergeometric ($\iota = 1$) and *c*-hypergeometric ($\iota = 0$) functions [2], respectively. The family of exactly quantized *i*-GRef potentials with two complex-conjugated regular singular points ($\iota = i$) was more recently discovered by Milson [22] (see also [20] for further developments).

We thus consider three types of quantization intervals:

i) finite interval :

$\iota=1,\ {}_{1}e_{0}=0,\ {}_{1}e_{1}=1\ (0 < \xi \equiv z < 1)$ (2.5)

or

$\iota=\pm1,\ {}_{\pm1}e_{0} = -1,\ {}_{\pm1}e_{1} = +1\ (-1 < \xi \equiv \eta_{-} < 1);$ (2.5*)

ii) positive semi-axis:

$\iota=0,\ {}_{0}e_{0} = 0\ (0 < \xi \equiv \zeta < \infty);$ (2.5')

iii) real axis:

$\iota = i,\ {}_{i}e_{1} = -e_{0} = i\ (-\infty < \xi \equiv \eta_{+} \equiv \eta < +\infty).$ (2.5")

The RCSLE of our interest thus has N second-order poles at singular points ${}_{\iota}e_{0}, {}_{\iota}e_{1}, \ldots, {}_{\iota}e_{N-1}$ and either regular ($|\iota| = 1$) or irregular ($\iota = 0$) singularity at infinity. In all the cases singular points ${}_{\iota}e_{r}$ and ${}_{\iota}e_{r+1}$ may form a complex conjugated pair, which is then also true for the pair of the exponent differences ${}_\iota\lambda_{0;r}$ and ${}_\iota\lambda_{0;r+1}$. (If $\iota = i$ this is necessarily true for the first pair of singular points ${}_i e_0 = -i$ and ${}_i e_1 = i$.) For all real singular points $e_r$ the parameter $\lambda_{0;r}$ can be any positive number other than 1.

It has been pointed by Roychoudhury et al [59] in the particular case of $r$-GRef potential that one needs to differentiate between parameters ${}_\iota Q^o \equiv {}_\iota\bar{\lambda}_o, {}_\iota O^o_{N-|\iota|-1}$ defining the RefPFr (2.3) and the coefficients of the polynomial numerator of the PFr in the right-hand side of (2.4) which specifies the change of variable $\xi(x)$ used to convert the RCSLE with BI (2.2) into the 1D Schrödinger equation. For this reason we introduce a concept of PFr beams (PFr beams), ${}_\iota\mathbf{B}$, which are obtained from Bose invariants (2.2) by varying the parameters ${}_\iota Q^o$ at fixed values of the polynomial coefficients. A set of parameters ${}_\iota Q^o$ defines individual PFr rays in the given beam and for this reason are referred to below as 'ray identifiers' (RIs). The term 'PFr ray' is thus an alternative name for BI (2.2) with fixed values of all the parameters except the energy ε.

As initially drawn to our attention by Milson [22], the LT [25, 24, 26] converts RCSLE (2.1) into the Schrödinger equation with the RLP:

$$V_L[{}_\iota\xi(x;T_{K_*}) \,|\, {}_\iota\mathbf{B}] = -{}_\iota\wp^{-1}[{}_\iota\xi(x;T_{K_*});T_{K_*}]\, I^o[{}_\iota\xi(x) \,|\, \bar{\lambda}_o, {}_\iota O^o_{N-|\iota|-1}] - \tfrac{1}{2}\{{}_\iota\xi, x\}, \tag{2.6}$$

where $\{{}_\iota\xi, x\}$ stands for the so-called 'Schwartz derivative' (see, i.g., [60]) and the variable ${}_\iota\xi(x; {}_\iota T_{K*})$ is determined via the first-order differential equation

$${}_\iota\xi'(x; {}_\iota T_{K_*}) = {}_\iota\wp^{-1/2}[{}_\iota\xi; {}_\iota T_{K_*}], \tag{2.7}$$

with prime standing for the derivative with respect to x. Each PFr ray just corresponds to a single potential curve drawn on RLP (2.6).

In this series of publications we are only interested in Bose invariants with the density function

$$
{}_{\iota}\wp[\xi;\,{}_{\iota}T_{K}] = \begin{cases} \dfrac{{}_{\iota}T_{K}[\xi]}{4\,\xi^{2}(1-\iota\xi)^{2}} & \text{for } \iota = 0,1, \\[2ex] \dfrac{{}_{\iota}T_{K}[\xi]}{4(\xi-\iota^{2})^{2}} & \text{for } \iota = \pm1, i, \end{cases} \tag{2.8}
$$

where the order K of the tangent polynomial (TP), ${}_{\iota}T_{K}[\xi]$, does not exceed 2. The PFr beams ${}_{1}\mathbf{B}$ and ${}_{\pm1}\mathbf{B}$ correspond to the same RLP expressed in terms of two different variables ${}_{1}\xi(x) = z(x)$ and ${}_{\pm1}\xi(x) = \eta_{-}(x) = 2z(x) - 1$ utilized in our [2, 61, 62] and Levai's [28, 29] papers, accordingly. It is assumed that the singular points in RefPFr (2.3) differ from any TP zero ${}_{\iota}\xi_{T;k'}$.

While we still use the conventional parameterization of TP

$$
{}_{0}T_{K}[\varsigma] = {}_{0}a_{2}\zeta^{2} + {}_{0}b\,\zeta + {}_{0}c_{0} \tag{2.9′}
$$

for ι=0, in other cases (ι=1, ±1, or *i*) it is more convenient to parameterize the TP ${}_{\iota}T_{K}[\xi]$ as

$$
{}_{\iota}T_{K}[\xi] = {}_{\iota}c_{0}\,(\xi - {}_{\iota}e_{1})^{2} + {}_{\iota}c_{1}\,(\xi - {}_{\iota}e_{0})^{2} + {}_{\iota}d\,(\xi - {}_{\iota}e_{0})(\xi - {}_{\iota}e_{1})\,. \tag{2.9}
$$

The leading coefficient is thus given by the linear relation

$$
{}_{\iota}a_{2} = {}_{\iota}c_{0} + {}_{\iota}c_{1} + {}_{\iota}d \quad (|\iota|=1). \tag{2.9*}
$$

[For the reasons explained below we will also use an alternative notation ${}_{0}d$ for the linear coefficient ${}_{0}b$ in (2.9′).] As a direct consequence of (2.9) one finds

$$
{}_{\iota}c_{r} = {}_{\iota}T_{K}[{}_{\iota}e_{r}]/({}_{\iota}e_{1} - {}_{\iota}e_{0})^{2} \qquad \text{for } r = 0,\ |\iota|=1 \tag{2.10}
$$

so that

$$
{}_{1}c_{r} = {}_{1}T_{K}[r] \quad \text{for } r = 0,1. \tag{2.10†}
$$

In particular such a parameterization assures that

$$ {}_{\pm1}c_r = {}_1c_r \;\; (r = 0,1), \;\; {}_{\pm1}d = {}_1d \tag{2.11} $$

and

$$ {}_\iota\xi'(\infty; {}_\iota T_K) = {}_\iota\wp^{-1/2}[1; {}_\iota T_{K_*}] = 2/\sqrt{{}_1c_1} \text{ for } \iota = \pm1 \text{ or } 1. \tag{2.12} $$

We also require that ${}_i c_0^* = {}_i c_1 \equiv c_R + i\, c_I$ and ${}_\iota d$ is real if $\iota = i$.

It will be demonstrated in Section 6 (and then elaborated in more detail in Part II) that rational SUSY partners of GRef potentials (N=|ι|+1, n=0) are described by Ref PFrs of the following form:

$$ \begin{aligned} I^o[\xi \,|\, {}^\ell_\iota\mathbf{B}] = &-\frac{{}_\iota h_{o;0}}{4(\xi - {}_\iota e_0)^2} - |\iota| \frac{{}_\iota h_{o;1}}{4(\xi - {}_\iota e_1)^2} - \tfrac{1}{4}\delta_{\iota,0}\nu_o^2 \\ &- \sum_{k'=1}^{\ell\mathfrak{I}} \frac{{}_\iota\rho_T({}_\iota\rho_T + 1)}{(\xi - {}_\iota\xi_{T;k'})^2} - \sum_{k=1}^{n} \frac{2}{(\xi - \xi_{o;k}^{(n)})^2} + \Delta I^o[\xi \,|\, {}^\ell_\iota\mathbf{B}], \end{aligned} \tag{2.13} $$

where

$$ {}_\iota e_0 = \begin{cases} 0 & \text{for } \iota = 0 \text{ or } 1, \\ -1 & \text{for } \iota = \pm1, \\ -i & \text{for } \iota = i, \end{cases} \qquad {}_\iota e_1 = \begin{cases} 1 & \text{for } \iota = 1 \text{ or } \pm1, \\ \\ +i & \text{for } \iota = i. \end{cases} \tag{2.14} $$

It is essential that the PFr

$$ \Delta I^o[\xi \,|\, {}_\iota\mathbf{B}] \equiv \frac{O^o_{N-|\iota|-1}[\xi \,|\, {}_\iota\mathbf{B}]}{4\Pi_N[\xi; \overline{e}^{(N)}]} \tag{2.15} $$

has only first order poles. The parameters appearing in the right-hand side of (2.13) are defined as follows:

$\mathfrak{I}$ is the number of distinct TP zeros (other than the singular points ${}_\iota e_0$ and ${}_\iota e_1$);

${}_\iota\rho_T$ is equal to one half of the order of the TP zero ${}_\iota e_0$, i. e.,

$$ {}_\iota\rho_T \equiv \tfrac{1}{2}\,\delta_T, \tag{2.16} $$

where

$$\delta_T \equiv \Im - K + 1; \tag{2.16′}$$

$$e_r^{(N)} \equiv \begin{cases} {}_\iota e_r & \text{for } r = 0, |\iota|, \\ \xi_{r-|\iota|}^{(n)} & \text{for } r = |\iota|+1, ..., N - \ell\Im - 1, \\ {}_\iota \xi_{T;k'} & \text{for } r = N - \ell(\Im + 1 - k') \text{ and } k' = 1 \text{ or } \Im; \end{cases} \tag{2.17}$$

$$\ell = 0 \text{ or } 1; \tag{2.18}$$

$$N = |\iota| + 1 + n + \ell\Im . \tag{2.19}$$

We also use the generic notation for monomial products:

$$\Pi_n[\xi; \bar{e}] \equiv \prod_{k=1}^{n} (\xi - e_k) \tag{2.20}$$

so that

$$\Pi_N[\xi; \bar{e}^{(N)}] \equiv \prod_{r=0}^{|\iota|} (\xi - {}_\iota e_r) \Pi_n[\xi; \bar{\xi}_o^{(n)}] \Pi_\Im^\ell[\xi; {}_\iota \bar{\xi}_T], \tag{2.21}$$

where the monomial product

$$\Pi_\Im[\xi; {}_\iota \bar{\xi}_T] \equiv \prod_{k'=1}^{\Im} (\xi - {}_\iota \xi_{T;k'}). \tag{2.22}$$

is proportional to the TP unless the latter has a double root. In the latter case (K=$\Im$+1=2)

$$_\iota T_2[\xi] = {}_\iota a_2 \, \Pi_1^2[\xi; {}_\iota \xi_T]. \tag{2.23}$$

We use symbol ${}_\iota \boldsymbol{G}^{K\Im\aleph}$, with $\aleph$ standing for the degeneracy of the root ${}_\iota e_0$, to classify PFr beams associated with the GRef potentials (n = 0, $\ell = 0$, N = |ι|+1). In particular any of shape-invariant GRef potentials belongs to the class $V[\xi|{}_\iota \boldsymbol{G}^{K0\aleph}]$. Three shape-invariant GRef potentials on the line are represented by the RM potential $V[z|{}_1 \boldsymbol{G}^{000}]$, the Morse potential $V[\zeta|{}_0 \boldsymbol{G}^{000}]$, and the Gendenshtein potential $V[\eta|{}_i \boldsymbol{G}^{201}]$ for ι=1, 0, and *i*, respectively.

It has been pointed to by Groshe [55] that bound-state energies for *r*- and *c*-GRef potentials are given by one of the roots of quartic equations. A similar quartic equation was derived by the author [20] for the Milson potential. Explicit representations for all three quartic polynomials will be presented in Section 5.

The Bose invariants of our interest can be thus represented as:

$$I[\xi;\varepsilon\,|\,{}^{\ell}_{\iota}\mathbf{B}] \equiv -\sum_{r=0}^{|\iota|} \frac{h_r({}_{\iota}c_r\,\varepsilon;\,{}_{\iota}h_{o;r})}{4(\xi-{}_{\iota}e_r)^2} - \tfrac{1}{4}\,\delta_{\iota,0}\,(\nu_o^2 - {}_0a_2\,\varepsilon)$$
$$-\sum_{k=1}^{n}\frac{2}{(\xi-\xi_{o;k}^{(n)})^2} - \sum_{k'=1}^{\ell\Im}\frac{{}_{\iota}\rho_T({}_{\iota}\rho_T+1)}{(\xi-{}_{\iota}\xi_{T;k'})^2} + \Delta I^{o}[\xi\,|\,{}^{\ell}_{\iota}\mathbf{B}]$$
$$+\frac{{}_{\iota}d\,\varepsilon}{4\prod_{r=0}^{|\iota|}(\xi-{}_{\iota}e_r)} \tag{2.24}$$

where

$$h_r(\varepsilon;\,{}_{\iota}h_{o;r}) \equiv {}_{\iota}h_{o;r} - \varepsilon \quad \text{for } r = 0, |\iota|. \tag{2.25}$$

We require that the potential $V[z|{}_1\mathcal{G}^{K\Im 0}]$ and therefore each of its rational SUSY partner vanish at $+\infty$ while having an exponential reflection barrier at large negative x which gives

$${}_1h_{o;0} \geq {}_1h_{o;1} = -1 \tag{2.26}$$

(making the transformation $z \to 1-z$, if necessary). For simplicity we also set ${}_1c_1$=1 so that result the coefficient of the second-order pole $(1-z)^{-2}$ in the BI $I[z;\varepsilon|{}_1\mathbf{B}]$ takes the form:

$$-\tfrac{1}{4}\,{}_1h_1(\varepsilon;-1) = \tfrac{1}{4}(\varepsilon+1). \tag{2.27}$$

An analysis of the indicial equation

$${}_1\rho_1({}_1\rho_1 - 1) + \tfrac{1}{4}(\varepsilon+1) = 0 \tag{2.28}$$

for the singular point z=1 reveals that that the appropriate RCSLE has real or complex-conjugated characteristic exponents at negative energies and in the scattering region, respectively. We can thus choose the necessary asymptotic behavior of the BI without making the LT to the Schrödinger equation. Taking advantage of the condition

$$ {}_1h_{o;0} + 1 \geq 0 \tag{2.29} $$

it is convenient to introduce a new real RI

$$ \lambda_o \equiv \sqrt{{}_1h_{o;0} + 1} \geq 0 \tag{2.30} $$

which is nothing but the exponent difference for the singular point z=0 at zero energy.

An analysis of the BI [2]

$$ I[z;\varepsilon\,|\,{}_1\boldsymbol{G}] = \frac{1-\mu^2({}_1c_0\,\varepsilon;\lambda_o)}{4z^2} + \frac{1}{4(1-z)^2} - \frac{\lambda_o^2 - 1 - \mu^2({}_1a_2\,\varepsilon;\mu_o)}{4z(1-z)}, \tag{2.31} $$

where

$$ \mu(a\varepsilon;\nu) = \sqrt{\nu^2 - a\,\varepsilon} \tag{2.32} $$

and

$$ \mu_o \equiv \sqrt{f_o + 1}. \tag{2.32′} $$

shows that

$$ I[z;\varepsilon\,|\,{}_1\boldsymbol{G}] \sim \tfrac{1}{4}[1-\mu^2({}_1a_2\varepsilon;\mu_o)]z^{-2} \text{ at large z,} \tag{2.33} $$

(In following our recent works [19, 20] we use symbol $f_o$ instead of f in [2].) Thereby the BI for the RCSLE expressed in terms of the reciprocal variable $\tilde{z} = 1/z$ can be approximated at small $\tilde{z}$ as

$$ \tilde{I}[\tilde{z};\varepsilon\,|\,{}_1\boldsymbol{G}] = \tilde{z}^{-4} I[\tilde{z}^{-1};\varepsilon\,|\,{}_1\boldsymbol{G}] \sim \tfrac{1}{4}[1-\mu^2({}_1a_2\varepsilon;\mu_o)]\tilde{z}^{-2} \tag{2.33*} $$

at small $\tilde{z}$.

We thus conclude that energy-dependent parameter (2.32) and RI (2.32′) are nothing but the exponent differences for the singular point $\tilde{z} = 0$ (z = ∞) at the energies ε and 0, respectively.

Similarly, for ι = *i* we require that

$$ {}_{i}O^{o}_{n+2\ell;n+2\ell} \mid {}^{\ell}_{i}\mathbf{B}) = 2h_{o;R} + 1 + 8n + 4\ell\rho_T(\rho_T - 1). \qquad (2.34)$$

It has been already proven in [20] that the condition

$$ {}_{i}O^{o}_{0} = 2h_{o;R} + 1 \qquad (2.35)$$

assures that the *i*-Gref potential vanishes at infinity as required. It will proven in Section 6 that CLDTs of our interest automatically preserve condition (2.34) for any of its multi-step SUSY partners. Since we only consider SUSY partners of Milson's reduction of the *i*-GRef potential (${}_{i}c_1 = {}_{i}c_0$) we choose ${}_{i}c_0 = {}_{i}c_1 = 1$.

Below we also restrict our analysis solely to *c*-RLPs $V[\zeta(x)|_0\mathbf{B}]$ with a Coulomb tail at +∞ which is only true if the leading coefficient of TP (2.9′) is positive. For simplicity we choose ${}_0a_2 = 1$. As explained in subsection 5.2 below, rational potentials with an exponential reflection barrier at large negative x (and the Coulomb tail approaching 0 as x→+∞) represent only one branch CT0 of *c*-RLPs $V[\zeta(x)|_0\mathbf{B}]$. There is the second branch CT+ formed by *c*-RLPs with a Coulomb reflection barrier at large positive x which are all exponentially approach 0 as x→–∞.

In next section we discuss in detail GS solutions which allow one to construct multi-step SUSY partners of GRef potentials either exactly (n =0) or conditionally exactly (n > 0) quantized by the 'GS Heine' and 'GS *c*-Heine' polynomials mentioned in Introduction. When applicable we will also explicitly distinguish between GS Heine polynomials $\mathrm{Hi}[z \mid {}_1\boldsymbol{G}]$ and $\mathrm{Hi}[z \mid {}_i\boldsymbol{G}]$ by referring to them as Jacobi-seed ($\mathscr{J}$S) and Routh-seed ($\mathfrak{R}$S) Heine polynomials, respectively.

## 3. Almost-everywhere holomorphic solutions and related SUSY pairs of solved-by-polynomials differential equations

Suppose that the RCSLE with BI (2.24) has a solution of the form

$$\phi[\xi\,|\,{}^{\ell}_{\iota}\mathbf{B}_{\downarrow\dagger m};\dagger m] = {}_{\iota}\Theta[\xi;2\,{}_{\iota}\rho_{0;\dagger m}-1,2\,{}_{\iota}\rho_{1;\dagger m}-|\,\iota\,|] \times \frac{\Pi_{n_{\dagger m}}[\xi;\,{}^{*}\bar{\xi}^{(n_{\dagger m})}_{\dagger m}]}{\Pi^{\ell}_{\mathfrak{I}}{}^{\iota}\rho_{T}[\xi;\,{}_{\iota}\bar{\xi}_{T}]\,\Pi_{n}[\xi;\bar{\xi}^{(n)}_{o}]} \tag{3.1}$$

at the energy ${}_{\iota}\varepsilon_{\dagger m} < 0$. It is essential that the non-rational component of this solution,

$${}_{\iota}\Theta[\xi;2\,{}_{\iota}\rho_{0}-1,2\,{}_{\iota}\rho_{1}-|\,\iota\,|] \equiv \begin{cases} (1-i\xi)^{i\rho_0}(1+i\xi)^{i\rho_1} & \text{for } \iota = i \ \ ({}_{i}\rho_{1} = {}_{i}\rho_{0}^{*}), \\ \prod\limits_{r=0}^{|\iota|} |\,\xi - {}_{\iota}e_{r}\,|^{\iota\rho_r}\ \ e^{-\delta_{\iota;0}\,{}_{\iota}\rho_1\xi} & \text{otherwise}, \end{cases} \tag{3.2}$$

is holomorphic at any finite point except $\xi$=0 for $\iota$=0 or both singular points $\xi = {}_{\iota}e_0$ and $\xi = {}_{\iota}e_1$ for $|\iota| = 1$. By definition the sub-beam ${}_{\iota}\mathbf{B}_{\downarrow\dagger m}$ is formed by PFr rays described by a subset of the RI in question. In this series of publications we only discuss AEH solutions which are obtained from GS solutions

$$\phi[\xi\,|\,{}_{\iota}\boldsymbol{G}_{\downarrow\dagger m};\dagger m] = {}_{\iota}\Theta[\xi;2\,{}_{\iota}\rho_{0;\dagger m}-1,2\,{}_{\iota}\rho_{1;\dagger m}-|\,\iota\,|]\Pi_{m}[\xi;\bar{\xi}_{\dagger m}] \tag{3.3}$$

via (generally multi-step) DTs. As a result the energy ${}_{\iota}\varepsilon_{\dagger m}$ mentioned above always coincides with the energy of the appropriate seed solution. As explained below the index **†** specifies behavior of AEH solution (3.1) at the ends of the quantization interval and is referred to as the solution type. It will be proven in Section 7 that any regular AEH solution retains its type under CLDTs with AEH FFs for $\iota$ = 0 or 1. As for the *i*-GRef potential we are only consider either eigenfunctions or AEH solutions irregular at ±∞ and, according to the conventional rules of SUSY quantum mechanics, these two types (respectively, **c** or **d** in our terms [57, 58]) cannot be converted into each other.

In this series of publications we only consider the typical case when the polynomials in the right-hand side of (3.1) have only simple zeros and therefore can be represented as monomial products. It will be proven in the end of this Section that the polynomial solutions of our interest, including Jacobi, Laguerre, and Routh polynomials appearing in the right-hand side of (3.3), have only simple zero at any regular point of the given RCSLE. We also assume that AEH solution (3.1) is irregular at any of the outer singular points which assures that the numerator of the PFr in the right-hand side does not have zeros at these points. It is however possible that AEH solution (3.1) becomes regular at one of the outer singular points at some specific values of the RIs. In those 'exotic cases' the polynomial numerator of the PFr in the right-hand side must have triple zero which assures that the ChExp of the solution at the aforementioned singular point is equal to 2. We postpone an analysis of those anomalous cases unless we directly run into them in one of the currently studied examples.

Absolute values of the exponent differences,

$$ {}_{\iota}\lambda_{r;\dagger m} = 2\,{}_{\iota}\rho_{r;\dagger m} - 1, \tag{3.4}$$

and the exponent $\nu_{\dagger m}$ are determined by the relations

$$ {}_{\iota}\lambda^{2}_{r;\dagger m} = h_r({}_{\iota}\varepsilon_{\dagger m};\, {}_{\iota}h_{o;r}) + 1 \tag{3.5}$$

and

$$ \nu^{2}_{\dagger m} = \lambda^{2}({}_{0}\varepsilon_{\dagger,m};\, \nu_{o}), \tag{3.5$'$}$$

accordingly.

Since we are currently interested only in RLPs on line the coefficient ${}_{\iota}c_0$ for $\iota = 0$, 1, or $\pm 1$ is required to be positive so that density function (2.8) has the second order pole at the origin for $\iota = 0$ or 1 and at 1 for $\iota = \pm 1$. This implies that AEH solution (3.1) is squarely integrable with the given density function near the lower end point iff ${}_{\iota}\lambda_{0;\dagger m} > 0$. Similarly AEH solution (3.1) is squarely integrable with density function (2.8) near the upper end point iff ${}_{\iota}\lambda_{1;\dagger m} > 0$ for $\iota = 1, \pm 1$ or iff $\nu_{\dagger m} > 0$ for $\iota = 0$. In following our original studies [57, 58] on the Darboux transformations of centrifugal-barrier potentials (years before the birth of the SUSY quantum mechanics [62, 63]), we label the four possible types of AEH solutions as

$$\mathbf{t}=\mathbf{a} \text{ for } {}_\iota\lambda_{0;\mathbf{t}\mathrm{m}} > 0,\ {}_\iota\lambda_{1;\mathbf{t}\mathrm{m}} < 0\ (\iota = 1, \text{ or } \pm 1) \tag{3.6a}$$

$$\text{or } {}_\iota\lambda_{0;\mathbf{t}\mathrm{m}} > 0,\ \nu_{\mathbf{t}\mathrm{m}} < 0; \tag{3.6′a}$$

$$\mathbf{t}=\mathbf{b} \text{ for } {}_\iota\lambda_{0;\mathbf{t}\mathrm{m}} < 0,\ {}_\iota\lambda_{1;\mathbf{t}\mathrm{m}} > 0\ (\iota = 1, \text{ or } \pm 1) \tag{3.6b}$$

$$\text{or } {}_\iota\lambda_{0;\mathbf{t}\mathrm{m}} < 0,\ \nu_{\mathbf{t}\mathrm{m}} > 0; \tag{3.6′b}$$

$$\mathbf{t}=\mathbf{c} \text{ for } {}_\iota\lambda_{0;\mathbf{t}\mathrm{m}} > 0,\ {}_\iota\lambda_{1;\mathbf{t}\mathrm{m}} > 0\ (\iota = 1, \text{ or } \pm 1) \tag{3.6c}$$

$$\text{or } {}_\iota\lambda_{0;\mathbf{t}\mathrm{m}} > 0,\ \nu_{\mathbf{t}\mathrm{m}} > 0; \tag{3.6′c}$$

$$\mathbf{t}=\mathbf{d} \text{ for } {}_\iota\lambda_{0;\mathbf{t}\mathrm{m}} < 0,\ {}_\iota\lambda_{1;\mathbf{t}\mathrm{m}} < 0\ (\iota = 1, \text{ or } \pm 1) \tag{3.6d}$$

$$\text{or } {}_\iota\lambda_{0;\mathbf{t}\mathrm{m}} < 0,\ \nu_{\mathbf{t}\mathrm{m}} < 0. \tag{3.6′d}$$

Since this paper deals solely with the RLPs on the line the TP leading coefficient ${}_i a_2$ is requested to be positive so that the density function ${}_i\wp[\eta; {}_iT_{\mathrm{K}}]$ for $\iota = i$ behaves as $\eta^{-2}$ at large $|\eta|$. For *i*-GRef potentials we found only GS solutions of the following types [20]:

$$\mathbf{t}=\mathbf{c} \text{ for } 0 \le \mathrm{m} < -{}_i\rho_{\mathbf{t}\mathrm{m};\mathrm{R}} - 1/2, \tag{3.7c}$$

$$\mathbf{t}=\mathbf{d} \text{ for } {}_i\rho_{\mathbf{t}\mathrm{m};\mathrm{R}} > 1/2, \tag{3.7d}$$

and

$$\mathbf{t}=\mathbf{d}' \text{ for } -\mathrm{m} - 1/2 < {}_i\rho_{\mathbf{t},\mathrm{m};\mathrm{R}} < 1/2, \tag{3.7d′}$$

where ${}_i\rho_{\mathbf{t}\mathrm{m};\mathrm{R}}$ is the common real part of the characteristic exponents ${}_i\rho_{r;\mathbf{t}\mathrm{m}}$ with $r = 0$ and 1. Note that the common real part ${}_i\lambda_{\mathbf{c}\mathrm{m};\mathrm{R}}$ of the complex conjugated parameters ${}_i\lambda_{0;\mathbf{c}\mathrm{m}}$ and ${}_i\lambda_{1;\mathbf{c}\mathrm{m}}$ is chosen to be negative. In the particular case of the symmetric potential both characteristic exponents ${}_i\rho_{0;\mathbf{t}\mathrm{m}}$ become real and the absolute value of the parameter ${}_i\lambda_{\mathbf{c}\mathrm{m};\mathrm{R}}$ coincides with the conventionally defined exponent difference.

The monomial products $\Pi_{\mathrm{m}}[\xi; {}_\iota\bar{\xi}^{(\mathrm{m})}_{\mathbf{t}\mathrm{m}}]$ forming GS solutions of types **a**, **b**, and **d** are referred to by Quesne [14-16] as Cases I, II, or III, respectively. It should be however stressed that the type of any AEH solution for the *r*-GRef potential and its SUSY partners depends on the choice of the quantization interval so that the same polynomial may form an AEH solution of a different type after the quantization interval was changed. For example, the pair of the shape-invariant RM [17] and Eckart/Manning-Rosen (E/MR) [65, 66] potentials can be described by the same BI (see Appendix C in [33] for details) and as a result share the same sets of AEH solutions. However the type of the solution depends on the choice of the quantization interval which distinguishes one rational potential from another. The direct consequence of this ambiguity is that the 'Case II' polynomial for one problem, for example, can describe a bound state for another and so on.

Before proceeding with our analysis let us make some comments on nodeless of the lowest-energy eigenfunction $\mathbf{c}0$ and regular solutions (types $\mathbf{a}$ and $\mathbf{b}$) below the ground energy level for a generic 1D potential on the line. This is the well-known feature [67] of regular solutions of Sturm-Liouville equations (SLEs) with a nonsingular behavior within a closed infinite interval [*a*, *b*]. However the standard treatment fails if either the quantization interval becomes infinite or if the differential operator is singular at *a* or *b* [68]. The latter cases are all regarded as singular and we refer the reader to Ch. 9 in [68] for a detailed analysis of the underlying mathematical problems.

An extension the Sturm Oscillation Theorem to the half-line in case the Direchlet boundary condition $\mathrm{u}(a;\varepsilon) = 0$ has been discussed, for example, by Simon [69]. However, as clarified below the transition $b \to \infty$ deserves a more cautious analysis.

Let us consider the Schrödinger equation with a nonsingular potential on $L^2(-\infty,+\infty)$. It is assumed that the solution regular at $-\infty$ or $+\infty$ is unambiguously determined by the Direchlet boundary condition

$$\lim_{x\to-\infty} \Psi_{\mathbf{a}}(x;\varepsilon\,|\,{}_\iota\mathbf{B}) = 0 \tag{3.8a}$$

or

$$\lim_{x\to+\infty} \Psi_{\mathbf{b}}(x;\varepsilon\,|\,{}_\iota\mathbf{B}) = 0, \tag{3.8b}$$

respectively. This assumption assures that the regular solutions of this type can be nequivocally obtained via the limiting transitions:

$$\Psi_{\mathbf{a}}(x;\varepsilon\,|\,{}_{\iota}\mathbf{B}) = \lim_{a\to-\infty} \Psi_{\mathbf{a}}(a \le x;\varepsilon\,|\,{}_{\iota}\mathbf{B}) \tag{3.9a}$$

and

$$\Psi_{\mathbf{b}}(x;\varepsilon\,|\,{}_{\iota}\mathbf{B}) = \lim_{b\to+\infty} \Psi_{\mathbf{b}}(x \le b;\varepsilon\,|\,{}_{\iota}\mathbf{B}), \tag{3.9b}$$

where $\Psi_{\mathbf{a}}(a \le x;\varepsilon\,|\,{}_{\iota}\mathbf{B})$ and $\Psi_{\mathbf{b}}(x \le b;\varepsilon\,|\,{}_{\iota}\mathbf{B})$ are regular solutions of the given SLE with the boundary conditions defined on the finite quantization interval [*a*, *b*]. It is essential that both solutions are defined independently of the choice of the upper or lower end-point. Another remarkable feature of the solutions regular at the end-points *a* and *b* is that their nodes move to the opposite end as the energy ε decreases [67]. In other words, the ground state energy must monotonically decrease as $a\to-\infty$ and $b\to+\infty$. Therefore the eigenfunction

$$\psi_{\mathbf{c},0}[x\,|\,{}_{\iota}\mathbf{B}] = \lim_{\substack{a\to-\infty,\\ b\to+\infty}} \psi_{\mathbf{c},0}(a \le x \le b\,|\,{}_{\iota}\mathbf{B}) \tag{3.9c}$$

corresponding to the lowest eigenvalue $\varepsilon_{\mathbf{c},0}$ does not have any node $x_0$ on the real axis -- otherwise the Sturm-Liouville problem in question would have a solution vanishing at the ends of the quantization interval $(-\infty, x_0]$ or $[x_0, +\infty)$, respectively, at $\varepsilon < \varepsilon_{\mathbf{c},0}$. Similarly we conclude that solutions (3.9a*) and (3.9b*) regular at $-\infty$ and $+\infty$, respectively, are nodeless at any energy $\varepsilon < \varepsilon_{\mathbf{c},0}$. An analysis presented above thus confirm that either ground energy eigenfunction or any regular solution below the ground energy level do not have nodes on the real axis.

For $|\iota|=1$ the RLPs $V[\xi(x)|_{\iota}\mathbf{B}]$ have exponential tails at both ends so that the general solution at $\varepsilon < 0$ exponentially grows at infinity and only the exponentially decreasing solutions satisfy Direchlet conditions (3.9a) and (3.9b). In case of the *c*-RLPs $V[\zeta(x)|_{0}\mathbf{B}]$ this is also true for $x \to -\infty$. To prove that the latter potentials also have a limit point (LP) singularity at $+\infty$ it is easier to study behavior of solutions of the appropriate RCSLE at large values of $\zeta$. Again one finds that the general solution at $\varepsilon < 0$ exponentially grows for $\zeta \gg 1$ whereas only the exponentially decreasing solution $\Phi_{\mathbf{b}}[\zeta;\varepsilon\,|\,{}_{0}\mathbf{B}]$ satisfy the Direchlet condition in the limit $\zeta \to \infty$. This should be also true for the solution $\Psi_{\mathbf{b}}[\zeta(x);\varepsilon\,|\,{}_0\mathbf{B}]$ of the appropriate the Schrödinger equation in the limit $x \to \infty$.

Let us now come back to an analysis of the monomial products $\Pi_m[\xi;\,{}_\iota\bar{\xi}_{\dagger m}]$ and prove that they satisfy the second-order differential equation with energy-dependent polynomial-in-ξ coefficients. For ι=0 or 1 (±1) these equations can be obtained by applying one of the four ($\sigma_r$ = + or − for $r$ = 0, 1) energy-dependent gauge transformations

$$\Phi[\xi;\varepsilon\,|\,{}^{\ell}_{\iota}\mathbf{B}^{K\Im 0};\bar{\sigma}] = {}_\iota\Theta[\xi;\sigma_0\,\lambda({}_\iota c_0\,\varepsilon;\lambda_o),\sigma_1\sqrt{\delta_{\iota,0}\nu_o^2-\varepsilon}\,;\bar{\xi}_o^{(n)};\ell\,{}_\iota\bar{\xi}_T] \times F[\xi;\varepsilon\,|\,{}^{\ell}_{\iota}\mathbf{B}^{K\Im 0};\bar{\sigma}] \qquad (3.10)$$

to the appropriate RCSLE, where the gauge weight is constructed using only smaller characteristic exponents at any of the outer singular points including (if any) the singularities at the TP zeros:

$${}_\iota\Theta[\xi;\lambda,\nu;\bar{\xi}_o^{(n)};{}_\iota\bar{\xi}_T] \equiv \frac{{}_\iota\Theta[\xi;\lambda,\nu]}{\Pi_{\Im}^{\ell}{}^{\iota\rho_T}[\xi;{}_\iota\bar{\xi}_T]\,\Pi_n[\xi;\bar{\xi}_o^{(n)}]} \qquad (3.11)$$

The 'basic GRef' weight function in the numerator of the fraction in the right-hand side of (3.11) is defined via (3.2).

It will be proven in Section 5 that the resultant four equations are exactly solvable by polynomials (P-ES) within a certain range of RIs of the *r*- and *c*-GRef PFr beams (n = 0, $\ell = 0$) provided that the appropriate PFr beams are generated by means of TPs with positive determinants. One of the main results of this paper is that each of these polynomial solutions can be used as a Gauss-seed (GS) function to construct a SUSY ladder of rational potentials conditionally exactly solvable by polynomials (P-CES). If none of the used GS functions has nodes within the quantization intervals then the appropriate SUSY partner of the GRef potential turns out to be conditionally exactly quantized by polynomials (P-CEQ). In other words we use the epithet 'CEQ', instead of the term 'CES' in Levai and Roy's pioneering paper [70] presenting the first example of this extremely rich family of rational SUSY partners with positions of outer singular points dependent on two RIs of the original GRef potential (two parameters of the isotonic oscillator in the particular case of Levai-Ray potentials [70]).

For SUSY partners $V[\eta \mid {}^{\ell}_{i}\mathbf{B}^{220}]$ of the *i*-GRef potential $V[\eta \mid {}_{i}\mathcal{G}^{220}]$ we consider only two energy-dependent gauge transformations

$$\Phi[\eta;\varepsilon \mid {}^{\ell}_{i}\mathbf{B}^{220};\sigma] = {}_{i}\Theta[\eta;\lambda_{\sigma}(\varepsilon;h_{o;0}),\lambda^{*}_{\sigma}(\varepsilon;h_{o;0});\bar{\xi}^{(n)}_{o};{}_{\ell\,i}\bar{\xi}_{T}]$$

$$\times F[\eta;\varepsilon \mid {}^{\ell}_{i}\mathbf{B}^{220};\sigma], \qquad (3.12)$$

where index $\sigma = \pm$ specifies the generally complex square root

$$\lambda_{\sigma}(\varepsilon;h_{o}) \equiv \sigma\sqrt{h_{o}+\varepsilon+1} \qquad (3.13)$$

via the condition

$$\sigma\, Re\lambda_{\sigma}(\varepsilon;h_{o}) > 0. \qquad (3.13^{*})$$

We require the characteristic exponents of the BI at the singular points $-i$ and $+i$ to be complex conjugated so that the basic GRef weight function in the numerator of the fraction in the right-hand side of (3.1) for $\iota = i$ is real:

$${}_{i}\Theta[\eta;2\rho-1,2\rho^{*}-1] \equiv (\eta+i)^{\rho}\ (\eta-i)^{\rho^{*}}. \qquad (3.14)$$

One can easily verify that the function $F[\xi;\varepsilon \mid {}^{\ell}_{\iota}\mathbf{B}^{K\Im 0};\bar{\sigma}]$ satisfies the equation

$$\left\{{}^{\ell}_{\iota}\hat{D}\{\bar{\rho}(\varepsilon \mid {}_{\iota}\mathcal{G}^{K\Im 0};\bar{\sigma});\bar{\xi}^{(n)}_{o};{}_{\iota}\bar{\xi}_{T}\} + C_{n+\ell\Im}[\xi;\varepsilon \mid {}^{\ell}_{\iota}\mathbf{B}^{K\Im 0};\bar{\sigma}]\right\}F[\xi;\varepsilon \mid {}^{\ell}_{\iota}\mathbf{B}^{K\Im 0};\bar{\sigma}] = 0, \qquad (3.15)$$

where the energy-dependent characteristic exponents ${}_{\iota}\bar{\rho}(\varepsilon)$ in the second-order differential operator

$$ {}^{\ell}_{\iota}\hat{D}\{{}_{\iota}\bar{\rho}(\varepsilon);\bar{\xi}_{o}^{(n)};{}_{\iota}\bar{\xi}_{T}\} \equiv \prod_{r=0}^{|\iota|}(\xi-{}_{\iota}e_{r})\,\Pi_{n}[\xi;\bar{\xi}_{o}^{(n)}]\Pi_{\mathfrak{J}}^{\ell}[\xi;{}_{\iota}\bar{\xi}_{T}]\frac{d^{2}}{d\xi^{2}} + 2\,{}^{\ell}_{\iota}B_{n+\ell\mathfrak{J}+|\iota|}[\xi;{}_{\iota}\bar{\rho}(\varepsilon);\bar{\xi}_{o}^{(n)};{}_{\iota}\bar{\xi}_{T}]\frac{d}{d\xi} \tag{3.16} $$

are defined as radical functions of the RIs of the appropriate GRef PFr beam:

$$ 2\bar{\rho}(\varepsilon\,|\,{}_{\iota}\boldsymbol{G}^{K\mathfrak{J}0};\bar{\sigma}) \equiv \begin{cases} \sigma_{0}\sqrt{\lambda_{o}^{2}-{}_{1}c_{0}\,\varepsilon}+1,\ \ \sigma_{1}\sqrt{-\varepsilon}+1 & \text{for } \iota=1 \text{ or } \pm 1, \\ \sigma_{0}\sqrt{\lambda_{o}^{2}-{}_{0}c_{0}\,\varepsilon}+1,\ \ \sigma_{1}\sqrt{\nu_{o}^{2}-{}_{0}c_{0}\,\varepsilon} & \text{for } \iota=0, \\ 1+\lambda_{\sigma}(\varepsilon;h_{o}),\ \ 1+\lambda_{\sigma}^{*}(\varepsilon;h_{o}) & \text{for } \iota=i. \end{cases} \tag{3.17} $$

It is essential that the polynomial coefficient of the first derivative in the right-hand side of (3.16),

$$ {}^{\ell}_{\iota}B_{n+\ell\mathfrak{J}+|\iota|}[\xi;{}_{\iota}\bar{\rho}(\varepsilon);\bar{\xi}_{o}^{(n)};{}_{\iota}\bar{\xi}_{T}] \equiv \Pi_{n}[\xi;\bar{\xi}_{o}^{(n)}]\Pi_{\mathfrak{J}}^{\ell}[\xi;{}_{\iota}\bar{\xi}_{T}]\prod_{r=0}^{|\iota|}(\xi-{}_{\iota}e_{r}) \times\left(\sum_{r=0}^{|\iota|}\frac{{}_{\iota}\rho_{r}(\varepsilon)}{\xi-{}_{\iota}e_{r}}-\sum_{k=1}^{n}\frac{1}{\xi-\xi_{o;k}^{(n)}}-\frac{1}{\mathfrak{J}}\sum_{k'=1}^{\ell\mathfrak{J}}\frac{1}{\xi-{}_{\iota}\xi_{T;k'}}-{}^{1}\!/_{2}\,\delta_{\iota,0}\,{}_{0}\rho_{1}(\varepsilon)\right) \tag{3.18} $$

except the aforementioned limiting case of the Gendenshtein potential ($\iota = i$, $\xi_{T;1} = -\xi_{T;2} = i$) and its rational SUSY partners [16].

To obtain an explicit expression for the free term of differential equation (3.15) it is convenient to re-arrange RefPFr (2.13) in an alternative form referred to in [19] as 'gauge partial decomposition' (GPD). [We use the term '*partial* decomposition' to stress that the PFr (3.20) below contains both second- and first-order poles.] Namely, we represent last three terms in the right-hand side of (2.13) as

$$-\sum_{k=1}^{n}\frac{2}{(\xi-\xi_{o;k}^{(n)})^2}-\sum_{k'=1}^{\ell\mathfrak{I}}\frac{{}_\iota\rho_T({}_\iota\rho_T+1)}{(\xi-{}_\iota\xi_{T;k'})^2}+\Delta I^o[\xi\,|\,{}_\iota^\ell\mathbf{B}]$$

$$=2\hat{Q}^{K\mathfrak{I}0}[\xi;\bar{\xi}_o^{(n)},\ell\,{}_\iota\bar{\xi}_T]+\frac{\hat{O}^{\downarrow}_{n+\ell\mathfrak{I}}[\xi\,|\,{}_\iota^\ell\mathbf{B}]}{4\prod_{r=0}^{|\iota|}(\xi-{}_\iota e_r)\Pi_n[\xi;\bar{\xi}_o^{(n)}]\Pi_{\mathfrak{I}}^{\ell}[\xi;{}_\iota\bar{\xi}_T]},\tag{3.19}$$

where

$$\hat{Q}^{K\mathfrak{I}0}[\xi;\bar{\xi}_o^{(n)};\ell\,{}_\iota\bar{\xi}_T)]$$
$$\equiv-\tfrac{1}{2}\Pi_n[\xi;\bar{\xi}_o^{(n)}]\Pi_{\mathfrak{I}}^{\ell\,{}_\iota\rho_T}[\xi;{}_\iota\bar{\xi}_T]\frac{d^2}{d^2\xi}\frac{1}{\Pi_n[\xi;\bar{\xi}_o^{(n)}]\Pi_{\mathfrak{I}}^{\ell\,{}_\iota\rho_T}[\xi;{}_\iota\bar{\xi}_T]}\tag{3.20}$$

so that

$$\hat{Q}^{K\mathfrak{I}0}[\xi;\bar{\xi}_o^{(n)};\bar{0}_{\mathfrak{I}}]\equiv\hat{Q}[\xi;\bar{\xi}_o^{(n)};1]$$
$$\equiv\frac{\ddot{\Pi}_n[\xi;\bar{\xi}_o^{(n)}]}{2\Pi_n[\xi;\bar{\xi}_o^{(n)}]}-\frac{\dot{\Pi}_n^2[\xi;\bar{\xi}_o^{(n)}]}{\Pi_n^2[\xi;\bar{\xi}_o^{(n)}]},\tag{3.21}$$

and

$$\hat{Q}^{K\mathfrak{I}0}[\xi;\bar{\xi}_o^{(n)};{}_\iota\bar{\xi}_T]\equiv\hat{Q}[\xi;\bar{\xi}_o^{(n)};1]+\hat{Q}[\xi;{}_\iota\bar{\xi}_T;{}_\iota\rho_T]+\Delta\hat{Q}^{K\mathfrak{I}0}[\xi;\bar{\xi}_o^{(n)};{}_\iota\bar{\xi}_T],\tag{3.22}$$

where

$$\hat{Q}[\xi;{}_\iota\bar{\xi}_T;{}_\iota\rho_T]\equiv\tfrac{1}{2}\,{}_\iota\rho_T\frac{\ddot{\Pi}_{\mathfrak{I}}[\xi;{}_\iota\bar{\xi}_T]}{\Pi_{\mathfrak{I}}[\xi;{}_\iota\bar{\xi}_T]}-\tfrac{1}{2}\,{}_\iota\rho_T({}_\iota\rho_T+1)\frac{\dot{\Pi}_{\mathfrak{I}}^2[\xi;{}_\iota\bar{\xi}_T]}{\Pi_{\mathfrak{I}}^2[\xi;{}_\iota\bar{\xi}_T]}\tag{3.23}$$

and

$$\Delta\hat{Q}^{K\mathfrak{I}0}[\xi;\bar{\xi}_o^{(n)};{}_\iota\bar{\xi}_T]\equiv-\,{}_\iota\rho_T\frac{\dot{\Pi}_n[\xi;\bar{\xi}_o^{(n)}]\dot{\Pi}_{\mathfrak{I}}[\xi;{}_\iota\bar{\xi}_T]}{\Pi_n[\xi;\bar{\xi}_o^{(n)}]\,\Pi_{\mathfrak{I}}[\xi;{}_\iota\bar{\xi}_T]}\tag{3.23$^\dagger$}$$

with

$${}_\iota\xi_T\equiv\frac{1}{\mathfrak{I}}\sum_{k'=1}^{\mathfrak{I}}{}_\iota\xi_{T;k'}\quad(\mathfrak{I}=1\text{ or }2).\tag{3.24}$$

We thus come to the following alternative representation for the RefPFs of our interest:

$$I^{o}[\xi \mid {}^{0}_{\iota}\mathbf{B}] = I^{o}[\xi \mid {}_{\iota}\boldsymbol{G}] + 2\hat{Q}[\xi; \bar{\xi}_{o}^{(n)}] + \frac{\hat{O}_{n}^{\downarrow}[\xi \mid {}^{0}_{\iota}\mathbf{B}]}{4\prod_{r=0}^{|\iota|}(\xi - {}_{\iota}e_{r})\Pi_{n}[\xi; \bar{\xi}_{o}^{(n)}]} \tag{3.25a}$$

and

$$I^{o}[\xi \mid {}^{1}_{\iota}\mathbf{B}] = I^{o}[\xi \mid {}_{\iota}\boldsymbol{G}] + 2\hat{Q}^{K\Im 0}[\xi; \bar{\xi}_{o}^{(n)}, {}_{\iota}\bar{\xi}_{T}] + \frac{\hat{O}_{n+\Im}^{\downarrow}[\xi \mid {}^{1}_{\iota}\mathbf{B}]}{4\prod_{r=0}^{|\iota|}(\xi - {}_{\iota}e_{r})\Pi_{n}[\xi; \bar{\xi}_{o}^{(n)}]\Pi_{\Im}[\xi; {}_{\iota}\bar{\xi}_{T}]}. \tag{3.25b}$$

Alternatively RefPFr (3.25a) can be represented as

$$I^{o}[\xi \mid {}^{0}_{\iota}\mathbf{B}] = I^{o}[\xi \mid {}_{\iota}\boldsymbol{G}] + 2Q[\xi; \bar{\xi}_{o}^{(n)}] + \frac{O_{n}^{\downarrow}[\xi \mid {}^{0}_{\iota}\mathbf{B}]}{4\prod_{r=0}^{|\iota|}(\xi - {}_{\iota}e_{r})\Pi_{n}[\xi; \bar{\xi}_{o}^{(n)}]}, \tag{3.25a$'$}$$

where the PFr

$$Q[\xi; \bar{\xi}_{o}^{(n)}] \equiv l\dot{d} \mid \Pi_{n}[\xi; \bar{\xi}_{o}^{(n)}] \mid = \frac{\ddot{\Pi}_{n}[\xi; \bar{\xi}_{o}^{(n)}]}{\Pi_{n}[\xi; \bar{\xi}_{o}^{(n)}]} - \frac{\dot{\Pi}_{n}^{2}[\xi; \bar{\xi}_{o}^{(n)}]}{\Pi_{n}^{2}[\xi; \bar{\xi}_{o}^{(n)}]} \tag{3.26}$$

$$= \hat{Q}[\xi; \bar{\xi}_{o}^{(n)}] + \frac{\ddot{\Pi}_{n}[\xi; \bar{\xi}_{o}^{(n)}]}{2\Pi_{n}[\xi; \bar{\xi}_{o}^{(n)}]} \tag{3.26*}$$

was adopted by us from Quesne's works [14, 15, 71-73] (see also her joint works with Grandati [74] and Marquette [75]) and for this reason we refer to it as QPFr. Similarly we use the term 'Quesne partial decomposition' (QPD) of the RefPFr to indicate that second-order poles at the outer singular points ${}_{\iota}\bar{\xi}_{\{\dagger m\}_{p}}$ in the right-hand side of (3.25a′) are combined with the first-order poles to form the QPFr.

Taking into account that

$$\dot{\Pi}_{\Im}[\xi; {}_{\iota}\bar{\xi}_{T}] = 2\delta_{\Im,2}(\xi - {}_{\iota}\xi_{T}) + \delta_{\Im,1} \tag{3.27}$$

one can directly verify that

$$\lim_{\xi_1,\xi_2\to\xi_T} \hat{Q}^{220}[\xi;\xi_1,\xi_2] = \hat{Q}^{210}[\xi;\xi_T] \tag{3.28}$$

and

$$\lim_{\xi_1,\xi_2\to\xi_T} \Delta\hat{Q}^{220}[\xi;\bar{\xi}_o^{(n)};\xi_1,\xi_2] = \Delta\hat{Q}^{210}[\xi;\bar{\xi}_o^{(n)};\xi_T] \tag{3.28$^\dagger$}$$

so that the PFr $\hat{Q}^{220}[\xi;\bar{\xi}_o^{(n)};{}_\iota\bar{\xi}_T]$ turns into $\hat{Q}^{210}[\xi;\bar{\xi}_o^{(n)};{}_\iota\xi_T]$ as $|{}_\iota\xi_{T;1} - {}_\iota\xi_{T;2}| \mapsto 0$.

Coming back to the generic PFr beam ${}^{\ell}_{\iota}\mathbf{B}^{K\Im 0}$ we thus represent the free term of differential equation (3.15) as sum of three energy-dependent terms

$$C_{n+\ell\Im}[\xi;\varepsilon\,|\,{}^{\ell}_{\iota}\mathbf{B}^{K\Im 0};\bar{\sigma}] = {}^{1}\!/_{4}\,\hat{O}^{\downarrow}_{n+\ell\Im}[\xi\,|\,{}^{\ell}_{\iota}\mathbf{B}^{K\Im 0}] \tag{3.29}$$

$$+\{{}^{1}\!/_{4}\,{}_\iota d\,\varepsilon + 2\rho_0(\varepsilon\,|\,{}_\iota\boldsymbol{G}^{K\Im 0};\bar{\sigma})[|\iota|\rho_1(\varepsilon\,|\,{}_\iota\boldsymbol{G}^{K\Im 0};\bar{\sigma}) - {}^{1}\!/_{2}\,\delta_{\iota,0}\,\sigma_1\,\lambda(\varepsilon;\nu_o)]\}$$

$$\times\Pi_n[\xi;\bar{\xi}_o^{(n)}]\Pi^{\ell}_{\Im}[\xi;{}_\iota\bar{\xi}_T]$$

$$-2\,B_1[\xi;\bar{\iota};\bar{\rho}(\varepsilon\,|\,{}^{\ell}_{\iota}\mathbf{B};\bar{\sigma})]$$

$$\times\{\dot{\Pi}_n[\xi;\bar{\xi}_o^{(n)}]\Pi^{\ell}_{\Im}[\xi;{}_\iota\bar{\xi}_T] + \ell\,{}_\iota\rho_T\,\Pi_n[\xi;\bar{\xi}_o^{(n)}]\dot{\Pi}^{\ell}_{\Im}[\xi;{}_\iota\bar{\xi}_T]\}$$

where, by analogy with (3.18), we put

$$B_1[\xi;\bar{\iota};{}_\iota\bar{\rho}] \equiv \prod_{r=0}^{|\iota|}(\xi - {}_\iota e_r)\left(\sum_{r'=0}^{|\iota|}\frac{{}_\iota\rho_{r'}}{\xi - {}_\iota e_{r'}} - \delta_{\iota,0}\,{}^{1}\!/_{2}\,{}_0\rho_1\right) \tag{3.30}$$

$$= \delta_{|\iota|,1}[{}_\iota\rho_0(\xi - {}_\iota e_1) + {}_\iota\rho_1(\xi - {}_\iota e_0)] + \delta_{\iota,0}({}_0\rho_0 - {}^{1}\!/_{2}\,{}_0\rho_1\xi). \tag{3.30$'$}$$

[Note also that the energy-dependent characteristic exponents at singular points ${}_\iota e_r$ are given by (3.17) for any SUSY partner ${}^{\ell}_{\iota}\mathbf{B}^{K\Im 0}$ of the GRef PFr beam ${}^{\ell}_{\iota}\boldsymbol{G}^{K\Im 0}$.

We thus conclude that the monomial product $\Pi_{n_{\dagger m}}[\xi;{}_\iota\overset{*}{\bar{\xi}}{}^{(n_{\dagger m})}_{\dagger m}]$ in the right-hand side of (3.1) satisfies the following second-order differential equation with polynomial coefficients:

$$\left\{ {}^{\ell}_{\iota}\hat{D}\{ {}_{\iota}\bar{\rho}_{\dagger m}; \bar{\xi}_{o}^{(n)}; {}_{\iota}\bar{\xi}_{T}\} + C_{n+\ell\Im}[\xi; {}_{\iota}\varepsilon_{\dagger m} \mid {}^{\ell}_{\iota}\mathbf{B}_{\downarrow\dagger m}; \bar{\sigma}_{\dagger}]\right\}\Pi_{n_{\dagger m}}[\xi; {}^{*}_{\iota}\bar{\xi}^{(n_{\dagger m})}_{\dagger m}] = 0, \tag{3.31}$$

where

$${}_{\iota}\bar{\rho}_{\dagger m} \equiv \bar{\rho}({}_{\iota}\varepsilon_{\dagger m}; {}^{\ell}_{\iota}\mathbf{B}_{\downarrow\dagger m}; \bar{\sigma}_{\dagger}). \tag{3.32}$$

Note that polynomial solutions of the latter equation may have only a simple root in any regular point unless it is a zero solution. In this series of publications we restrict our analysis solely to AEH solutions irregular in any outer singular point so that polynomial solutions of our interest are also required to have simple roots at all the outer singular points. With this restrictions, we assure that polynomial components of AEH solutions have only simple roots and therefore can be represented as monomial products in agreement with (3.1).

In the particular case of GRef potentials (N=|ι|+1) polynomial (3.27) turns into the constant

$$C_{0}({}_{\iota}\varepsilon_{\dagger m} \mid {}_{\iota}\boldsymbol{G}_{\downarrow\dagger m}; \bar{\sigma}_{\dagger}) = {}_{\iota}C^{0}_{0}({}_{\iota}\lambda_{0;\dagger m}, {}_{\iota}\lambda_{1;\dagger m}) + \tfrac{1}{4}({}_{\iota}O^{o}_{0} + {}_{\iota}d\, {}_{\iota}\varepsilon_{\dagger m}), \tag{3.33}$$

where

$${}_{\iota}C^{0}_{0}(\lambda_{0}, \lambda_{1}) \equiv \tfrac{1}{2}(\lambda_{0} + 1)[|\iota| - (-1)^{|\iota|}\lambda_{1}]. \tag{3.34}$$

Polynomial solutions of the resultant differential equation

$$\prod_{r=0}^{|\iota|}(\xi - {}_{\iota}e_{r})\ddot{\Pi}_{m}[\xi; {}_{\iota}\bar{\xi}_{\dagger m}] + 2\hat{B}_{1}[\xi; \bar{\iota}; {}_{\iota}\bar{\lambda}_{\dagger m}]\dot{\Pi}_{m}[\xi; {}_{\iota}\bar{\xi}_{\dagger m}] + C_{0}({}_{\iota}\varepsilon_{\dagger m} \mid {}_{\iota}\boldsymbol{G}_{\downarrow\dagger m}; \bar{\sigma}_{\dagger})\Pi_{m}[\xi; {}_{\iota}\bar{\xi}_{\dagger m}] = 0 \tag{3.35}$$

are investigated in detail in Section 5.

## 4. Two classes of polynomial solutions for energy-dependent Fuschian equations

It directly follows from the conventional theory of Fuschian second-order differential equations [23] that the leading coefficient of polynomial (3.28) can be represented as

$$C_{n+\ell\Im;n+\ell\Im}(\varepsilon\,|\,{}^{\ell}_{\iota}\mathbf{B};\bar{\sigma}) = \alpha(\varepsilon\,|\,{}^{\ell}_{\iota}\mathbf{B};\bar{\sigma})\,\beta(\varepsilon\,|\,{}^{\ell}_{\iota}\mathbf{B};\bar{\sigma}) \quad (|\iota|=1), \tag{4.1}$$

where $\alpha(\varepsilon|{}^{\ell}_{\iota}\mathbf{B};\bar{\sigma})$ and $\beta(\varepsilon|{}^{\ell}_{\iota}\mathbf{B};\bar{\sigma})$ are two (presumed to be real) roots of the quadratic equation:

$$X^2(\varepsilon\,|\,{}^{\ell}_{\iota}\mathbf{B};\bar{\sigma}) - \left[2\,{}^{\ell}_{\iota}B_{n+\ell\Im+1;n+\ell\Im+1}\left({}_{\iota}\bar{\rho}(\varepsilon\,|\,{}_{\iota}\boldsymbol{G};\bar{\sigma});\bar{\xi}_{o}^{(n)};{}_{\iota}\bar{\xi}_{T}\right)-1\right]X(\varepsilon\,|\,{}_{\iota}\mathbf{B}^{(\ell)};\bar{\sigma})$$
$$+\,C_{n+\ell\Im;n+\ell\Im}(\varepsilon\,|\,{}^{\ell}_{\iota}\mathbf{B};\bar{\sigma}) = 0 \tag{4.2}$$

so that

$$\alpha(\varepsilon\,|\,{}^{\ell}_{\iota}\mathbf{B};\bar{\sigma}) + \beta(\varepsilon\,|\,{}^{\ell}_{\iota}\mathbf{B};\bar{\sigma})$$
$$= 2\,{}^{\ell}_{\iota}B_{n+\ell\Im+1;n+\ell\Im+1}\left(\bar{\rho}(\varepsilon\,|\,{}_{\iota}\boldsymbol{G};\bar{\sigma});\bar{\xi}_{o}^{(n)};{}_{\iota}\bar{\xi}_{T}\right)-1 \tag{4.3}$$
$$\equiv 2\Xi(\varepsilon\,|\,{}_{\iota}\boldsymbol{G};\bar{\sigma}) - 2\ell\Im\,{}_{\iota}\rho_{T} - 2n - 1, \tag{4.3*}$$

where

$$\Xi(\varepsilon\,|\,{}_{\iota}\boldsymbol{G};\bar{\sigma}) \equiv \rho_0(\varepsilon;{}_{\iota}\boldsymbol{G};\bar{\sigma}) + \rho_1(\varepsilon;{}_{\iota}\boldsymbol{G};\bar{\sigma}) \text{ for } |\iota| = 1. \tag{4.4}$$

We select the roots via the requirement:

$$\beta(\varepsilon_{\iota}|{}^{\ell}_{\iota}\mathbf{B};\bar{\sigma}) > \alpha(\varepsilon\,|\,{}^{\ell}_{\iota}\mathbf{B};\bar{\sigma}). \tag{4.5}$$

[In this series of publications we are only interested in the range of energies where quadratic equation (4.2) has a positive discriminant.] A polynomial of order m satisfies differential equation (3.31) if either

$$\alpha(\varepsilon_{\mathbf{t}m}\,|\,{}^{\ell}_{\iota}\mathbf{B};\bar{\sigma}) = -m \tag{4.6a}$$

or

$$\beta(\varepsilon_{\mathbf{t}m}\,|\,{}^{\ell}_{\iota}\mathbf{B};\bar{\sigma}) = -m. \tag{4.6b}$$

We say that the polynomial in question and the appropriate AEH solution belongs to α- or β-Class depending on which condition (4.6a) or (4.6b) is fulfilled. In will be proven in Section 8 that the CLDTs of our interest keep each AEH solution within the same Class.

## 5. Quartic-equation algorithms to obtain energies of GS solutions

It was Cooper, Ginocchio, and Khare [1] who first made use of normalizable nodeless (m=0) *r*-GS solutions to construct exactly-quantized rational SUSY partners of the generic r-GRef potential. As mentioned in Introduction this fundamentally significant work stimulated our initial interest in this problem. Next important development drastically affecting the direction of our studies was the paper by Gomez-Ullate, Kamran, and Milson [76] who demonstrated existence of non-normalizable *r*- and *c*-GS solutions formed by Jacobi and generalized Laguerre polynomials with no zeros inside the quantization intervals [−1,+1] and on the positive semi-axis, respectively. A certain setback of their approach comes from the fact that the latter is based solely on the known theorems for distributions of polynomial zeros: the Klein formula [77, 53] and the so-called 'Kienast-Lawton-Hahn (KLH) theorem' [78, 79, 80] in Grandati's terms [13, 81].

A more promising technique for generating multi-step CEQ potentials is to take advantage of regular AEH solutions which are necessarily nodeless iff they lie below the ground energy level. It was Levai and Roy [70] who first used regular AEH solutions **a**m and **b**m below the ground energy level to construct two construct two sequences of CEQ SUSYpartners of the isotonic oscillator. It was more recently discovered by Quesne [4, 7] that the very first potential in each sequence is quantized by $X_1$-Laguerre polynomials [5, 6]. Quesne's discovery was later extended by Odake and Sasaki [10, 11] to other potentials in both sequences under assumption that the AEH solutions in question are nodeless. Their results were promptly rectified by Grandati [13]) using the KLH theorem to select nodeless AEH solutions. (In [19] we explicitly confirmed that the regular AEH solutions selected by Grandati are precisely the ones used by Levai and Roy to construct the aforementioned sequences of CEQ potentials.)

Under influence of Grandati's works [13, 81] Quesne [15] use the KLH theorem to select nodeless GS solutions **b**m using them as FFs for single-step CLDTs of the Morse potential $V[\zeta; {}_0\boldsymbol{G}^{000}]$. In a parallel work [14] she also used the Klein formula [77, 53] to select nodeless regular GS solutions for the RM potential. A more detailed comparison between two approaches will be made in Part III where both shape-invariant potentials $V[\xi; {}_{\iota}\mathcal{G}^{000}]$, with $\iota = 0$ and 1, are treated as the limiting cases of the linear TP (LTP) *r*- and *c*-GRef potentials $V[\xi; {}_{\iota}\mathcal{G}^{110}]$, respectively.

In this series of publications we will only consider multi-step CLDTs of rational potentials using AEH solutions $\mathbf{t}_k m_k$ of the RCSLEs with GRef Bose invariants as seed solutions. As proven in Section 7 these CLDTs do not change characteristic exponents (ChExps) of regular AEH solutions at finite singular end points of the quantization interval, when applied to GRef potentials $V[\xi; {}_{\iota}\mathcal{G}^{K\Im 0}]$ *on the line* for $\iota = 0$ or 1. The CLDTs under discussion also keep unchanged the asymptotics of the AEH solutions regular at infinity for rational SUSY partners of *c*-GRef potential $V[\zeta; {}_{0}\mathcal{G}^{K\Im 0}]$. As a direct consequence of this proof we conclude that each GS solutions $\mathbf{a}m_k$ and $\mathbf{b}m_k$ are converted by the given multi-step CLDT into regular AEH solutions of the same type $\mathbf{a}$ and $\mathbf{b}$, respectively. Since the transformation does not affect the position of the ground energy level and both initial and resultant AEH solutions have the same energy the latter solution must be necessarily nodeless if this is true for the GS solutions in question.

In Part II we will take advantage of this remarkable feature of regular AEH solutions to develop the general theory of nets of CEQ potentials on the line starting from *r*- and *c*-GRef potentials $V[\xi; {}_{\iota}\mathcal{G}^{K\Im 0}]$, with $\iota = 0$ or 1. Some elements of this theory have been recently utilized by Odake and Sasaki [82] for the shape-invariant limiting cases – the RM and Morse potentials -- making use of the fact that both potentials are explicitly expressible in terms of the variable x used in the corresponding Schrödinger equations. Obviously we deal here solely with the theory of CEQ Sturm-Liouville problems and all the results can be obtained without converting the given RCSLE into the Schrödinger equation. A more comprehensive analysis of their work will be performed in Part III specifically dealing with these two potentials.

In subsections 5.1 and 5.2 below we present a preliminary analysis of the equations determining energies of GS solutions mainly to demonstrate that RCSLE with *r*- and *c*-GRef

Bose invariants do have nodeless regular AEH solutions which can be used as seed solutions to generate ladders of CEQ potentials.   A more comprehensive study will be presented in Part II.

For completeness subsection 5.3 briefly outlines the results of our analysis of AEH solutions of the RCSLE associated with the Milson potential – the reduction of the *i*-GRef potential constructed by Milson [22] using the symmetric TP $({}_i c_0 = {}_i c_0^* = {}_i c_1)$.    As proven in [20] the generic *i*-GRef potential  is exactly quantized by an orthogonal subset of Routh polynomials [54] referred to by us as 'Romanovski-Routh polynomials', by analogy  with the terms 'Romanovski-Bessel' and 'Romanovski-Jacobi' polynomials in Lesky's classification scheme [83]  of three families of orthogonal polynomials discovered by Romanovsky [84].  To a certain extent an extension of Milson's work [22] presented in [20] is done along the lines of our original study [2] on the *r*- and *c*-GRef potentials.  However the energy spectrum for the generic *i*-GRef potential is described by a set of two radical equations unless we focus solely on its reduction generated by means of the symmetric TP.   By analogy with the *r*- and *c*-GRef potentials, this reduction (referred to by us as the Milson potential) does have the energy spectrum unambiguously determined by one of real roots of a quartic polynomial and, what is even more important in the current framework, bound energy states in the Milson potential (if exist) are always accompanied by infinitely many GS solutions of type **d** [20].   At least one of these solutions, **d**0, is nodeless and therefore can be used as the FF to construct a rational potential exactly quantized by ℜS Heine polynomials.  As demonstrated in [20] the Milson potential has two branches referred to as 'inside' and 'outside' depending on positions of TP zeros relative to the unit circle.  The two intersect along the shape-invariant Gendenshtein potential.  The remarkable feature of the RCSLE associated with the inner branch of the Milson potential (as well as its shape-invariant limit) is that it has two sequences of nodeless AEH solutions which can be used as FFs for constructing new Hi-CEQ potentials. One can then extend the ladder using pairs of sequential eigenfunctions **c**,v;  **c**,v+1 [85-90].

To be able to treat all three GRef PFr beams ${}_\iota\boldsymbol{G}^{K\Im 0}$ ($\iota$= 1, 0, or *i*) in a uniform fashion we represent AEH solutions (3.1) as

$$\phi[\xi\,|\,{}_{\iota}\boldsymbol{G}^{\mathrm{K}\Im 0}_{\downarrow\dagger\mathrm{m}};\dagger\mathrm{m}] = {}_{\iota}\Theta[\xi;\ {}_{\iota}\lambda_{0;\dagger\mathrm{m}},\ {}_{\iota}\lambda_{1;\dagger\mathrm{m}}]\,\Pi_{\mathrm{m}}[\xi;\ {}_{\iota}\bar{\xi}_{\dagger\mathrm{m}}], \tag{5.1}$$

where ${}_{0}\lambda_{1;\dagger\mathrm{m}} \equiv \nu_{\dagger\mathrm{m}}$, ${}_{i}\lambda_{1;\dagger\mathrm{m}} = {}_{i}\lambda^{*}_{0;\dagger\mathrm{m}}$, and the basic GRef weight function is defined via (3.2). As mentioned above we refer to (5.1) as GS solutions which fall into the three general classes: Jacobi-seed ($\mathscr{J}$S) for ι=1 or ±1, Laguerre-seed ($\mathscr{L}$S) for ι=0, and Routh-seed ($\mathscr{R}$S) for ι= *i*. We use the term 'primary' for the sequence of polynomial solutions starting from a constant, i.e. each primary sequence of GS solutions starts from the necessarily nodeless 'basic' GS solution †0.

The main purpose of this Section is to outline the general algorithms to determine energies of all possible AEH solutions of the RCSLE with the GRef PFrs

$$\begin{aligned} {}_{\iota}I^{o}[\xi;\,{}_{\iota}G^{o}] = &-\frac{{}_{\iota}h_{o;0}}{4(\xi-{}_{\iota}e_{0})^{2}} - |\iota|\frac{{}_{\iota}h_{o;1}}{4(\xi-{}_{\iota}e_{1})^{2}} - \tfrac{1}{4}\delta_{\iota,0}\nu_{o}^{2} \\ &+ \frac{{}_{\iota}O^{o}_{0}}{4(\xi-{}_{\iota}e_{0})(\xi-{}_{\iota}e_{1})} \end{aligned} \tag{5.2}$$

associated with for three families of the GRef potentials defined in Section 2. (The symbol ${}_{\iota}G^{o}$ here is used for two independent RIs which will be explicitly defined below for each of three cases ι=0, 1 and *i*.) After making the appropriate gauge transformation we thus come to second-order differential equation (3.33) exactly solvable by hypergeometric, generalized Laguerre, and Routh polynomials for ι= 1, 0, and *i*, respectively. Use of hypergeometric equation [2], compared with Levai's [28, 29] usage of the Jacobi equation (ι=±1) for similar purposes, makes it easier to formulate the necessary condition for existence of polynomial solutions so that below we only consider the realization of the generic *r*-GRef potential as a PFr in the variable z varying between 0 and 1.

## 5.1 Zero-factorization-energy separatrices for eigenfunctions of the Schrödinger equation with the generic *r*-GRef potential

In the particular case of the r-GRef potential $V[z \mid {}_1\boldsymbol{G}]$ differential equation (3.15) takes the form

$$\left\{ z(z-1)\frac{d^2}{dz^2} + 2\,{}_1B_1[z;\bar\rho(\varepsilon \mid {}_\iota\boldsymbol{G};\bar\sigma)]\frac{d}{dz} + C_0(\varepsilon \mid {}_1\boldsymbol{G};\bar\sigma) \right\} F[z;\varepsilon \mid {}_1\boldsymbol{G};\bar\sigma] = 0,$$

(5.1.1)

where the first-order polynomial is defined via (3.30′), with ι = 1:

$$ {}_1B_1[z;\bar\rho(\varepsilon \mid {}_1\boldsymbol{G};\bar\sigma)] = \Xi(\varepsilon \mid {}_1\boldsymbol{G};\bar\sigma)\, z - \rho_0(\varepsilon;{}_1\boldsymbol{G};\bar\sigma) \qquad (5.1.2)$$

and the free term is represented by the energy-dependent parameter

$$C_0[z;\varepsilon \mid {}_1\boldsymbol{G};\bar\sigma] = \tfrac{1}{4}({}_1O_0^0 + {}_1d\varepsilon) + 2\rho_0(\varepsilon \mid {}_1\boldsymbol{G};\bar\sigma)\rho_1(\varepsilon \mid {}_1\boldsymbol{G};\bar\sigma), \qquad (5.1.3)$$

as a direct consequence of (3.29) for n = $\ell$ = 0. Comparing (5.1.1) with the conventional expression for the hypergeometric equation we find

$$F[z;\varepsilon \mid {}_1\boldsymbol{G};\bar\sigma] = F[\alpha(\varepsilon \mid {}_1\boldsymbol{G};\bar\sigma), \beta(\varepsilon \mid {}_1\boldsymbol{G};\bar\sigma); \sigma_0\sqrt{\lambda_o^2 - {}_1c_0\,\varepsilon} + 1; z], \qquad (5.1.4)$$

where

$$\alpha(\varepsilon \mid {}_1\boldsymbol{G};\bar\sigma) + \beta(\varepsilon \mid {}_1\boldsymbol{G};\bar\sigma) = 2\Xi(\varepsilon \mid {}_1\boldsymbol{G};\bar\sigma) - 1 \qquad (5.1.5a)$$

$$= \sigma_0\sqrt{\lambda_o^2 - {}_1c_0\,\varepsilon} + \sigma_1\sqrt{-\varepsilon} + 1 \qquad (5.1.5a^*)$$

and

$$\alpha(\varepsilon \mid {}_1\boldsymbol{G};\bar\sigma)\beta(\varepsilon \mid {}_1\boldsymbol{G};\bar\sigma) = C_0[z;\varepsilon \mid {}_1\boldsymbol{G};\bar\sigma] \qquad (5.1.5b)$$

under assumption that the α- and β-roots of quadratic equation (4.2) are selected by condition (4.5).

Function (5.1.4) turns into a polynomial of order m at the energy ${}_1\varepsilon_{\dagger m}$ if one of the parameters

$$ {}_1\alpha^{G}_{\dagger m} \equiv \alpha({}_1\varepsilon_{\dagger m} \mid {}_1\boldsymbol{G};\bar{\sigma}_{\dagger}) \quad \text{or} \quad {}_1\beta^{G}_{\dagger m} \equiv \beta({}_1\varepsilon_{\dagger m} \mid {}_1\boldsymbol{G};\bar{\sigma}_{\dagger}) \tag{5.1.6} $$

is equal to –m. It is convenient to represent positive difference between ${}_1\beta_{\dagger m}$ and ${}_1\alpha_{\dagger m}$ as

$$ {}_1\beta_{\dagger m} - {}_1\alpha_{\dagger m} = \sqrt{\mu_o^2 - {}_1a_2\,{}_1\varepsilon_{\dagger m}} \tag{5.1.7} $$

where we use RI (2.32′),

$$ \mu_o = \sqrt{\lambda_o^2 - {}_1O_0^o - 1}, \tag{5.1.8} $$

instead of the parameter

$$ f \equiv f_o = \mu_o^2 - 1 \tag{5.1.8$^\dagger$} $$

in [2]. We thus come to the following sufficient condition for existence of the *r*-GS solution $\dagger$m at the energy ${}_1\varepsilon_{\dagger m}$:

$$ {}_1\lambda_{0;\dagger m} + {}_1\lambda_{1;\dagger m} + 2m + 1 = o_{\dagger m}\sqrt{\mu_o^2 - {}_1a_2\,{}_1\varepsilon_{\dagger m}} \tag{5.1.9} $$

$$ = \mu_{\dagger m}, \tag{5.1.9$^\dagger$} $$

where the signed exponent differences ${}_1\lambda_{r;\dagger m}$ are defined via (3.5) or, to be more precise,

$$ {}_1\lambda_{r;\dagger m} = \sigma_{r;\dagger}\sqrt{h_{o;r} + 1 - {}_1c_r\,{}_1\varepsilon_{\dagger m}}. \tag{5.1.10} $$

The *r*-GS solution $\dagger$m thus belongs to either α- or β-Class depending on sign $o_{\dagger m}$ of the parameter $\mu_{\dagger m}$ defined by the condition

$$\mu_{\dagger m} = \begin{cases} \sqrt{\mu_o^2 - {}_1a_2\, {}_1\varepsilon_{\dagger m}} & \text{if } {}_1\alpha_{\dagger m} = -m, \\ -\sqrt{\mu_o^2 - {}_1a_2\, {}_1\varepsilon_{\dagger m}} & \text{if } {}_1\beta_{\dagger m} = -m. \end{cases} \qquad (5.1.10^{\dagger})$$

The indicial equation

$$\rho_r(\varepsilon \mid {}_1\mathbf{B})[\rho_r(\varepsilon \mid {}_1\mathbf{B}) - 1] - \tfrac{1}{4} h_r({}_1c_r\varepsilon; {}_1h_{o;r}) = 0 \qquad (5.1.11)$$

for the singular point z = r of RCSLE (2.1) with ι=1 has real roots

$$\rho_r(\varepsilon \mid {}_1\mathbf{B},\pm) = \tfrac{1}{2}[1 \pm \sqrt{h_r({}_1c_r\varepsilon; {}_1h_{o;r}) + 1}] \qquad (5.1.12)$$

iff

$$h_r({}_1c_r\varepsilon; {}_1h_{o;r}) \ge -1. \qquad (5.1.13)$$

As mentioned in the end of Section 2 we can choose without loss of generality:

$${}_1h_{o;0} \ge {}_1h_{o;1} \qquad (5.1.14)$$

(making the transformation $z \to 1 - z$, if necessary). By setting

$${}_1h_{o;1} = -1 \ , \qquad (5.1.15)$$

we thus assure that indicial equations (5.1.11) for the *r*-PFr beam has two different real roots at both end points for any negative value of the energy parameter ε.

Taking into account that the Schwartz derivative is invariant with respect to scaling of the variable z, one can easily verify that

$$\lim_{z \to r} \{z, x\} = -2 / {}_1c_r \qquad (5.1.16)$$

so that condition (5.1.15) is equivalent to the requirement that the *r*-GRef potential vanishes as x→+∞, whereas inequality (5.1.14) assures that the potential has a reflection barrier at large negative x, excluding the limiting case of the asymptotically-leveled (*r*-AL) potential curves

$$h_0^o = h_1^o = -1 \ \ (\lambda_o = 0). \qquad (5.1.17)$$

Keeping in mind that

$$_1\varepsilon_{\dagger m} = -\,_1\lambda^2_{1;\dagger m} \tag{5.1.18}$$

and making use of (2.9*) we can represent transcendental equation (5.1.9) as the following algebraic equation

$$\Lambda_2^{(m)}(_1\lambda_{1;\dagger m};\lambda_o,\mu_o;\,_1d) = 2\,_1\lambda_{0;\dagger m}(_1\lambda_{1;\dagger m} + 2m + 1) \tag{5.1.19}$$

linear in $_1\lambda_{0;\dagger m}$ and quadratic in $_1\lambda_{1;\dagger m}$, where the quadratic polynomial in $_1\lambda_{1;\dagger m}$, is defined via the identity

$$\Lambda_2^{(m)}(\lambda;\lambda_o,\mu_o;\,_1d) \equiv (1+\,_1d)\lambda^2 - \left(\lambda + 2m + 1\right)^2 + \mu_o^2 - \lambda_o^2 \tag{5.1.19*}$$

Equation (5.1.19) has to be solved together with the second equation

$$_1\lambda^2_{0;\dagger m} = \lambda_o^2 + \,_1c_0\,_1\lambda^2_{1;\dagger m} \tag{5.1.19′}$$

quadratic in both $_1\lambda_{0;\dagger m}$ and $_1\lambda_{1;\dagger m}$. Taking square of both sides of the first equation and substituting (5.1.19*) into the left-hand side of the resultant expression, one can then verify that the signed exponent $_1\lambda_{1;\dagger m}$ at the upper end z = 1 must coincide with a real root of the quartic polynomial:

$$G_4^{(m)}[\lambda\,|\,_1\boldsymbol{G}] = [\,_1d\lambda^2 - 2(2m+1)\lambda + \mu_o^2 - \lambda_o^2 - (2m+1)^2]^2 - 4(\lambda + 2m+1)^2(\,_1c_0\,\lambda^2 + \lambda_o^2). \tag{5.1.20a}$$

As seen from (5.1.20a) the quartic equation of our interest is reduced to the two quadratic equations in the limiting case of the radial [61, 62, 33, 51] potentials ($_1c_0 = 0$) as well as for *r*-AL potential curves defined via (5.1.15). As explained below the *r*-AL potential curves $V[z\,|\,_1\breve{\boldsymbol{G}}]$ play a crucial role in our analysis since it allows us to make possible some quantitative predictions concerning *r*-GS solutions $\dagger$m coexistent with the m$^{th}$ bound energy level in the generic *r*-GRef potential on line (provided that the TP has nonnegative discriminant $_1\Delta_T$ ). The most important result proven below is existence of $v_o$+1 quartets of *r*-GS solutions **t**m (**t** = **a**, **b**, **c**, and **d**) in the region where the *r*-GRef potential generated using a TP with ${}_1\Delta_T$ >0 has at least $v_o$+1 bound energy levels.

Quartic polynomial (5.1.20a) can be also represented in the following alternative form:

$$G_4^{(m)}[\lambda \mid {}_1\boldsymbol{G}] = [{}_1c_0\lambda^2 + \lambda_o^2 - {}_1a_2\lambda^2 - \mu_o^2 - (\lambda + 2m + 1)^2]^2 - 4(\lambda + 2m + 1)^2({}_1a_2\lambda^2 + \mu_o^2), \tag{5.1.20b}$$

which can be analytically decomposed into the product of second-order polynomials at a=0. Some remarkable features of PFr beams generated using the LTP will be explored in Part III.

One can also convert radical equation (5.1.9) into the quartic equation with respect to $|{}_1\varepsilon_{\mathbf{t}m}|$:

$$(G_{4;4}^{(m)} |{}_1\varepsilon_{\mathbf{t}m}|^2 + G_{4;2}^{(m)} |{}_1\varepsilon_{\mathbf{t}m}| + G_{4;0}^{(m)})^2 = |{}_1\varepsilon_{\mathbf{t}m}| (G_{4;3}^{(m)} |{}_1\varepsilon_{\mathbf{t}m}| + G_{4;1}^{(m)})^2. \tag{5.1.20†}$$

In a slightly different form the latter has been already suggested by Grosche [55] as the quantization condition for bound energy levels (**t** = **c**) in the generic *r*-GRef potential. It should be however stressed that use of Grosche's quartic polynomial to determine energies of bound states the discrete energy spectrum is complicated by the fact that quartic polynomial (5.1.20†) has generally 4 positive roots and selection of the one associated with a bound energy level is by no means a trivial problem.

Note that the leading coefficient of the quartic polynomial has the same sign as the TP discriminant. In particular it vanishes if the TP has a double root.

Let us now prove the $m^{th}$ eigenfunction **c**m of the RCSLE with BI (2.31) is accompanied by three *r*-GS solutions **t**m of distinct types **t** = **a**, **b**, and **d** if the BI in question is generated using a TP with ${}_1\Delta_T$ > 0. As a starting point let us first prove this assertion for the *r*-AL potential curves. Setting $\lambda_o$ to 0 in the right-and side of (5.1.20a) allows one to analytically decompose this quartic polynomial into the product of two quadratic polynomials

$$^{\pm}\breve{G}_2^{(m)}[\lambda\,|\,_1\breve{\mathcal{G}}] = {}^{\pm}_{1}\breve{g}_2({}_1a_2, {}_1c_0)\lambda^2 + {}^{\pm}_{1}\breve{g}_1^{(m)}({}_1c_0)\lambda + {}_1\breve{g}_0^{(m)}(\mu_o) \qquad (5.1.21)$$

with the coefficients

$$^{\pm}_{1}\breve{g}_2({}_1a_2, {}_1c_0) \equiv \pm 2\sqrt{{}_1c_0} - {}_1d = (1 \pm \sqrt{{}_1c_0})^2 - {}_1a_2, \qquad (5.1.21')$$

$$^{\pm}_{1}\breve{g}_1^{(m)}({}_1c_0) \equiv 2(\pm\sqrt{{}_1c_0} + 1)(2m+1), \qquad (5.1.21'')$$

$${}_1\breve{g}_0^{(m)}(\mu_o) \equiv (2m+1)^2 - \mu_o^2\,. \qquad (5.1.21''')$$

Note that both leading coefficients are positive if the TP has a positive discriminant and that the leading coefficient $^{+}_{1}\breve{g}_2({}_1a_2, {}_1c_0)$ remains positive as far as the TP does not have zeros inside the quantization interval [0, 1]. Each quadratic equation has roots of opposite sign as far as common free term (5.1.21‴) remains negative ($2m < \mu_o - 1$).

Comparing quadratic polynomials (5.1.19*), with $\lambda_o$ set to 0, and (5.1.21), one finds

$$\Lambda_2^{(m)}(\lambda; 0, \mu_o; {}_1d) = \pm 2\sqrt{{}_1c_0}(\lambda + 2m + 1)\lambda - {}^{\pm}\breve{G}_2^{(m)}[\lambda\,|\,_1\breve{\mathcal{G}}] \qquad (5.1.22)$$

so that

$$\Lambda_2^{(m)}({}_1\breve{\lambda}_{1;\breve{\mathbf{t}}_{\pm}m}; 0, \mu_o; {}_1d) = \pm 2\sqrt{{}_1c_0}({}_1\breve{\lambda}_{1;\breve{\mathbf{t}}_{\pm}m} + 2m + 1)\,{}_1\breve{\lambda}_{1;\breve{\mathbf{t}}_{\pm}m}, \qquad (5.1.23)$$

where ${}_1\breve{\lambda}_{1;\breve{\mathbf{t}}_{\pm}m}$ is one of the roots of the quadratic equations

$$^{\pm}\breve{G}_2^{(m)}[{}_1\breve{\lambda}_{1;\breve{\mathbf{t}}_{\pm}m}\,|\,{}_1\breve{\mathcal{G}}_{\downarrow\breve{\mathbf{t}}_{\pm}m}] = 0\,. \qquad (5.1.24)$$

Substituting (5.1.23) into the left-hand side of (5.1.19), with $\lambda_o$ set to 0, gives

$$ {}_1\breve{\lambda}_{0;\breve{t}_\pm m} = \pm \sqrt{{}_1c_0}\, {}_1\breve{\lambda}_{1;\breve{t}_\pm m} \qquad (5.1.25) $$

which implies that

$$ \breve{t}_- = \mathbf{a} \text{ or } \mathbf{b}, \quad \breve{t}_+ = \mathbf{c} \text{ or } \mathbf{d}. \qquad (5.1.25^\dagger) $$

We thus conclude that the RCSLE with BI (2.31) generated using a TP with ${}_1\Delta_T > 0$ has the quartet of *r*-GS solutions **t**m of four distinct types: **t** = **a**, **b**, **c**, **d** for

$$ 2m < \mu_o - \lambda_o - 1 \qquad (5.1.26) $$

in the limit $\lambda_o \to 0$, excluding the *sym*-GRef potential, i.e., iff ${}_1c_0 \neq 1$.

Our next step is to prove that this assertion holds for positive values of the RI $\lambda_o$ as far as the *r*-GRef potential in question has at least m bound energy levels. The proof is based on the observation that the signed exponent difference at the origin, ${}_1\lambda_{0;\mathbf{t}m}$ is unambiguously determined by the relation

$$ {}_1\lambda_{0;\mathbf{t}m} = \frac{\mu^2_{\mathbf{t}m} - {}_1\lambda^2_{0;\mathbf{t}m}}{2({}_1\lambda_{1;\mathbf{t}m} + 2m + 1)} + \tfrac{1}{2}[{}_1d\,{}_1\lambda_{1;\mathbf{t}m} - (2 + {}_1d)(2m+1)] \qquad (5.1.27) $$

for each real zero ${}_1\lambda_{1;\mathbf{t}m}$ of quartic polynomial (5.1.20a) unless two m-dependent parameters

$$ \mu_m \equiv \sqrt{\mu_o^2 + {}_1a_2(2m+1)^2} \qquad (5.1.28a) $$

and

$$ \lambda_m \equiv \sqrt{\lambda_o^2 + {}_1c_0\,(2m+1)^2}\,. \qquad (5.1.28b) $$

coincide. In the latter case

$$ {}_1\lambda_{1;\mathbf{t}m} = -2m - 1 \qquad (5.1.29) $$

and the denominator of the fraction in the right-hand side of (5.1.27) vanishes. It then immediately follows from the structure of quartic polynomial (5.1.20a) that the latter has a double root at each point of the hyperbola

$$\mu_o^2 - \lambda_o^2 + (1 + {}_1d)(2m+1)^2 = 0 \qquad (5.1.29^*)$$

in the 2D space of the RIs $\mu_o$ and $\lambda_o$. In particular this implies that the number of *r*-GS solutions $\mathbf{t}$m for the given order m of the seed polynomial $\Pi_m[z; \bar{z}^{(m)}_{\mathbf{t}m}]$ may not exceed 4 despite the mentioned ambiguity.

By approximating two vanishing roots $y_{m;+}$ and $y_{m;-}$ of the quartic polynomial

$$G_4^{(m)}[y - 2m - 1 \mid {}_1\boldsymbol{G}] \approx \{({}_1c_0 - {}_1a_2)[y - 2(2m+1)]y - y^2 + \lambda_m^2 - \mu_m^2\}^2 - 4y^2\mu_o^2$$

in y by the following linear formulas:

$$y_{m;\pm} \approx \frac{\mu_m^2 - \lambda_m^2}{2\left[(1+{}_1d)(2m+1) \pm \mu_m\right]} \text{ for } \mu_m \approx \lambda_m \qquad (5.1.30)$$

we confirm that the pair of the merging roots must remain real on both sides of the hyperbola. Substituting ${}_1\lambda_{1;\mathbf{t}_\pm m} = y_{m;\pm} - 2m - 1$ into the right-hand side of (5.1.27) and then approximating $y_{m;\pm}$ via (5.1.30) one finds

$${}_1\lambda_{0;\mathbf{t}_\pm m} \approx \pm\lambda_m \text{ for } \mu_m \approx \lambda_m \qquad (5.1.30^\dagger)$$

so that the AEH solutions co-existent at the same energy ${}_1\varepsilon_{\tilde{\mathbf{t}}_\pm m} = -(2m+1)^2$ have two distinct types: $\mathbf{t}_+ = \mathbf{a}$ and $\mathbf{t}_- = \mathbf{d}$ and for this reason we refer to curve (5.1.29) as the m^th^ **a**/**d**-DRt hyperbola.

Since ${}_1c_0 > 0$ for any *r*-GRef potential on the line the exponent difference at the singular points 0 and 1 obey the inequality $|\,{}_1\lambda_{0;\mathbf{t}m}\,| > |\,{}_1\lambda_{1;\mathbf{t}m}\,|$. This implies that two real roots ${}_1\lambda_{1;\mathbf{t}m}$ and ${}_1\lambda_{1;\mathbf{t}'m}$ of the quartic polynomial $G_4^{(m)}[\lambda \mid {}_1\boldsymbol{G}]$ may not merge with each other giving rise to pair of complex-conjugated roots so that a pair of AEH solutions of different types may not simply disappear. Excluding the curves ${}_1\lambda_{0;\mathbf{t}m}(\lambda_o, \mu_o) = -\tilde{m}$ ($\mathbf{t} = \mathbf{b}$ or $\mathbf{d}$) and ${}_1\lambda_{1;\mathbf{t}m}(\lambda_o, \mu_o) = -\tilde{m}$ ($\mathbf{t} = \mathbf{a}$ or $\mathbf{d}$) discussed below any AEH solution from the m^th^

**abcd**-quartet may thus change its type iff the exponent difference $|{}_1\lambda_{1;\mathbf{t}m}|$ vanishes. The latter is possible only along the straight-lines

$$\mu_o = |2m+1\pm\lambda_o| \tag{5.1.31}$$

where the free term of quartic polynomial (5.1.20a) vanishes:

$$G_4^{(m)}[0\,|\,{}_1\boldsymbol{G}] \equiv [\mu_o^2-(\lambda_o+2m+1)^2]\times[\mu_o^2-(\lambda_o-2m-1)^2]=0. \tag{5.1.31*}$$

As proven in Appendix B the eigenfunction associated with the highest bound energy level is the first one to change its type so that there are two *r*-GS solutions of type **a** on the upper side of the straight-line

$$\mu_o = 2m+1+\lambda_o \tag{5.1.31A}$$

in the $\mu_o\lambda_o$ plane. We refer to these straight-line as the **c**/**a**′ zero-factorization-energy (ZFE) separatrix where prime is used to indicate that we deal with the supplementary sequence of R@O *r*-GS solutions which does not start from the basic solution **a**0. Excluding the curves mentioned above, the $m^{th}$ **abcd**-quartet thus always exists in the so-called 'Area $A_m$' cut by the **c**/**a**′ ZFE separatrix from the quadrant $\mu_o > 0,\ \lambda_o > 0$ according to (5.1.26).

If the curve ${}_1\lambda_{r;\mathbf{t}m}(\lambda_o,\mu_o) = -\tilde{m}$ for $\tilde{m}$ = 1, …, or m crosses Area $A_m$ then the *r*-GS solution still labelled by us as **t**m has a different type $\mathbf{t}_\Diamond$ at each point of the curve, namely,

$$\mathbf{t}_\Diamond = \begin{cases} \mathbf{c} \text{ if } \mathbf{t} = \mathbf{a} \text{ or } \mathbf{b}, \\ \mathbf{a} \text{ if } \mathbf{t} = \mathbf{d} \text{ and } r = 0, \\ \mathbf{b} \text{ if } \mathbf{t} = \mathbf{d} \text{ and } r = 1, \end{cases} \tag{5.1.32}$$

Because, as a direct consequence of (22.3.3) in [91], the Jacobi polynomial $P_m^{(-\tilde{m},\mu)}(\eta)$ has zero of order $\tilde{m}$ at η = 1. Along the curve (and only there) the *r*-GS solution **t**m turns into the *r*-GS solution $\mathbf{t}_\Diamond m_\Diamond$ with the ChExp

$$ {}_1\rho_{r;\mathbf{t}_\Diamond \mathrm{m}_\Diamond}(\lambda_o,\mu_o) = 1 - 2\mathrm{m} + \tilde{\mathrm{m}} \qquad (5.1.32^*) $$

at the endpoint z = $r$ so that $\mathrm{m}_\Diamond = \mathrm{m} - \tilde{\mathrm{m}}$.

There is no anomalous points for the quartets formed by basic solutions (m=0) so that the generic *r*-GRef potential $V[z \mid {}_1\mathcal{G}^{K\Im 0}_{\downarrow \mathbf{c}0}]$ generated using the TP with $\Delta_T > 0$ the ground energy eigenfunction **c**0 is always accompanied by three basis solutions **a**0, **b**0, and **d**0 nodeless by definition. This implies that the exactly-quantized-by-polynomials (P-EQ) CGK $V[z \mid {}^1_1\mathcal{G}^{220}_{\mathbf{c}0}]$ potential has three P-EQ 'siblings' $V[z \mid {}^1_1\mathcal{G}^{220}_{\mathbf{t}0}]$, with **t** = **a**, **b**, and **d**: two of them $V[z \mid {}^1_1\mathcal{G}^{220}_{\mathbf{a}0}]$ and $V[z \mid {}^1_1\mathcal{G}^{220}_{\mathbf{b}0}]$ has the same discrete energy spectrum as $V[z \mid {}_1\mathcal{G}^{220}_{\downarrow\mathbf{c}0}]$ whereas the third has an extra bound energy state inserted below the energy level ${}_1\varepsilon_{\mathbf{c}0}$. Note that we refer to the potentials generated using CLDTs with basic FFs as 'P-EQ' (rather than as P-CEQ) to stress that positions of the singularities in the appropriate RCSLEs are unaffected by allowed variations of the RIs.

An analysis of discriminants

$$ {}_1\breve{\Delta}^{(m)}_{\pm} = 4[{}_1a_2(2m+1)^2 + {}^{\pm}_1\breve{g}_2({}_1a_2, {}_1c_0)\mu_o^2] \qquad (5.1.33) $$

of quadratic polynomials (5.1.21) shows that they monotonically increase with m for positive values of the TP leading coefficient ${}_1a_2$. Since the linear coefficient ${}^{-}_1g^{(m)}_1({}_1c_0)$ is positive (negative) if ${}_1c_0 < 1$ (${}_1c_0 > 1$) the eigenfunction **c**m if exists is always accompanied by m+1 nodeless solutions of type **a** or **b** for ${}_1c_0 < 1$ or ${}_1c_0 > 1$, respectively, if the TP leading coefficient ${}_1a_2$ lies within the range

$$ 0 < {}_1a_2 < |1 - \sqrt{{}_1c_0}\,|. \qquad (5.1.34) $$

As $\lambda_o$ increases these solutions must stay below the ground energy level ${}_1\varepsilon_{\mathbf{c}0}$ and therefore remain nodeless until the eigenfunction **c**m disappears on the **c**/**a**′ ZFE separatrix. All these solutions can be used *c*-GS functions for multi-step CSLTs to construct a single-source net of P-CEQ potentials.

Since quadratic polynomial ${}^{+}\breve{G}_2^{(m)}[\lambda\,|\,{}_1\breve{G}]$ has a positive linear coefficient the AL potential curves have the discrete energy spectrum only in the basic domain $A_0$ ($\mu_o > 1$). In theory it may happen that the bound energy states potentials re-appear somewhere beyond the basic domain $A_0$ in the generic case. However, at least in this series of publications we will restrict our analysis solely to the range $\mu_o > \lambda_o + 1$. Based on this restriction it is convenient to introduce another RI

$$ {}_1v_o \equiv \tfrac{1}{2}(\mu_o - \lambda_o - 1) \tag{5.1.35} $$

which directly specifies the number of bound energy levels

$$ {}_1n_o = [{}_1v_o] + 1 \tag{5.1.35*} $$

so that we refer to this new RI as the bound energy measure (BEM). As mentioned above we are only interested in its positive range.

Let us now demonstrate that *r*-GS solutions $\dagger m$ of Class β introduced in previous Section do exist in the subdomain $A_0$ at least if ${}_1c_0 > 1$ and the TP leading coefficient ${}_1a_2$ is chosen to be sufficiently small depending on the value of m. While keeping a more detailed analysis of the LTP PFr beams for Part III let us simply notice that condition (5.1.9) defining *r*-GS solutions $\dagger m$ of Class β ($o_{\dagger m} = -$) in the limit ${}_1a_2$ takes the form

$$ {}_1\tilde{\lambda}_{0;\dagger m} = -\mu_o - {}_1\tilde{\lambda}_{1;\dagger m} - 2m - 1 \tag{5.1.36} $$

where ${}_1\tilde{\lambda}_{1;\dagger m}$ is one of the roots of the quadratic equation

$$ \lambda_o^2 + {}_1c_0\,{}_1\tilde{\lambda}_{1;\dagger m}^2 - (\mu_o + {}_1\tilde{\lambda}_{1;\dagger m} + 2m + 1)^2 = 0. \tag{5.1.37} $$

For ${}_1c_0 > 1$ and $\mu_o > \lambda_o$ the latter has roots of opposite sign. It then directly follows from (5.1.36) that the positive root corresponds to the *r*-GS solution **b**m whereas the type of the second solution **d** or **a** may depend on m. We thus conclude that the RCSLE with the BI $I[z; \varepsilon \mid {}_1\boldsymbol{G}^{110}]$ has infinitely many *r*-GS solutions **b**m of Class β. Only a finite number of these solutions may retain at nonzero values of ${}_1a_2$.

If the TP has a DRt $z_T$ (${}_1\Delta_T$=0) then the leading coefficient of quartic polynomial (5.1.20a) vanishes and the maximum number of co-existent *r*-GS solutions reduces to 3. The cubic equation determining the energies of these solutions has been already discussed in [34]. It is remarkable that the first-order differential equation describing the DRtTP PFr beam, with $z_T = 2$:

$$(z')^2 = \frac{4z^2(1-z)^2}{(2-z)^2} \tag{5.1.38}$$

allows integration in an explicit form:

$$\hat{z}(x) = \frac{2}{\hat{\eta}(x)+1} = \frac{2}{1+\sqrt{1+e^{-2x}}}, \tag{5.1.39}$$

where the variable $\hat{\eta}(x)$ satisfies first-order differential equation (3.9) in [59] with C=1:

$$\hat{\eta}' = \hat{\eta}(x) - \hat{\eta}^{-1}(x). \tag{5.1.40}$$

This brings us to the Dutt-Khare-Varsni (DKV) potential [92, 59] which turns out [34] to be nothing but the particular case of the DRtTP potential with $z_T$ = 2. We thus have a very rare example of a non-shape-invariant P-EQ potential which can be represented as an explicit function of its argument x.

Taking into account that quadratic equation (5.1.24) for ${}_1\lambda_{1;\breve{\mathbf{t}}_m}$ turns into the linear equation

$${}_1\lambda_{1;\breve{\mathbf{t}}_m} = \frac{\mu_0^2 - (2m+1)^2}{2(1-\sqrt{{}_1c_0})(2m+1)} \tag{5.1.41}$$

in the limiting case ${}_1\Delta_T = 0$ we conclude that the m[th] eigenfunction, if exists, is always accompanied by m+1 *r*-GS solutions of type **a** or **b** for ${}_1c_0 > 1$ or ${}_1c_0 < 1$, respectively. In the particular case of the TP ${}_1T_2[z] = (z-2)^2$ the coefficient ${}_1c_0$ is equal to 4 so that the appropriate RCSLE has a primary sequence of *r*-GS solutions of type **a**. One can use any of three basic solutions **a**0, **c**0, and **d**0 to construct a P-EQ SUSY partner which can be also written as an explicit function of x. Furthermore, following the prescriptions outlined in in Part II one can construct multi-step P-CEQ SUSY partners which can be again explicitly expressed in terms of x.

## 5.2 Two intersecting branches of the c-GRef potential

For ι=0 RefPFr (5.2) takes the form:

$$ {}_0I^o[\zeta;\lambda_o,g_o,\nu_o] = \frac{1-\lambda_o^2}{4\zeta^2} - \frac{g_o}{4\zeta} - \tfrac{1}{4}\nu_o^2, \tag{5.2.1}$$

with $g_o \equiv -{}_0O_0^o$. In terms of [93, 94]

$$ g_1 = g_o,\ \nu_o^2 = g_2,\ \eta = \lambda_o^2. \tag{5.2.2}$$

(While adopting his notation from [93], Grosche changed the meaning of the third parameter η, namely, η in his works [55, 95] stands for $h_{o;0} = \lambda_o^2 - 1$.)

The distinguished feature of the *c*-GRef potential on the line,

$$ V[\zeta\,|\,{}_0\boldsymbol{G}^{K\Im 0}] = \frac{\lambda_o^2 + g_o\,\zeta + \nu_o^2\,\zeta^2}{{}_0T_K[\zeta;\Im]} + \frac{({}_0d - {}_0a\zeta)\zeta}{{}_0T_K^2[\zeta;\Im]} - \frac{5\Delta_T\zeta^2}{4\,{}_0T_K^3[\zeta;\Im]}, \tag{5.2.3}$$

compared with its regular-at-infinity (R@∞) counterparts (ι=1 or *i*), is that it has a quantitatively different asymptotic behavior near the end points (with the Morse potential as the only exception). As a result, one has to choose the zero point energy depending on the relative position $V_{+\infty} - V_{-\infty}$ of potential tails in the limits $x \to \pm\infty$, where

$$V_{\pm\infty} \equiv \lim_{x \to \pm\infty} V[\zeta(x) \,|\, {}_0\boldsymbol{G}^{K\Im 0}]. \tag{5.2.4}$$

In the general case of the second-order TP (${}_0a_2 > 0$, K=2) the potential has the Coulomb tail (CT) at $+\infty$:

$$V[\zeta(x) \,|\, {}_0\boldsymbol{G}^{2\Im 0}] = V_{+\infty} + q_1 x^{-1} + 0(x^{-2}) \quad \text{for } x >> 1 \tag{5.2.5}$$

where we put

$$2q_1 \equiv \lim_{\zeta \to \infty} \left\{ \zeta \left( V[\zeta(x) \,|\, {}_0\boldsymbol{G}^{K\Im 0}] - V_{+\infty} \right) \right\} = g_o \tag{5.2.5$^\dagger$}$$

and the potential has an infinite discrete energy spectrum for $g_o < 0$ if

$$\text{CT0:} \quad V_{-\infty} > V_{+\infty} = 0 \; (\nu_o = 0). \tag{5.2.6a}$$

On other hand, the potential may have only a finite number of bound energy states if

$$\text{CT+:} \quad V_{+\infty} > V_{-\infty} = 0 \; (\lambda_o = 0), \tag{5.2.6b}$$

by analogy with its R@∞ counterparts. The CT0 and CT+ branches of the *c*-GPFr potential intersect along the *c*-AL potential curves

$$c\text{-AL:} \quad V_{-\infty} = V_{+\infty} = 0 \; (\lambda_o = \nu_o = 0). \tag{5.2.6}$$

In the LTP limit (${}_0a_2 = 0$, ${}_0b \equiv {}_0d > 0$, K = $\Im$ = 1) the *c*-GPFr potential $V[\zeta(x) \,|\, {}_0\boldsymbol{G}^{110}]$ has a parabolic barrier at $+\infty$ so that we refer to this potential as the parabolic-barrier (PB) *r*-GRef potential. We shall come back to an analysis of this anomalous case in Part III. In this paper (as already mentioned in Section 2) we assume that the TP order, K, is equal to 2 and set the

leading coefficient ${}_0a_2$ to 1. Since the TP may not have positive roots we need to consider only two possibilities:

a) ${}_0b \equiv {}_0d > 0;$ (5.2.7a)

b) ${}_0b \equiv {}_0d < 0, \ \Delta_T < 0.$ (5.2.7b)

By setting ι=0 in (3.33) we come to the following second-order differential equation with an irregular singular point at infinity:

$$\left\{\zeta\frac{d^2}{d\zeta^2}+(2\,{}_0\rho_{0;\mathbf{t}m}\,\zeta-\nu_{\mathbf{t}m})\frac{d}{d\zeta}+{}_0C_0(\mathbf{t},m)\right\}\Pi_m[\zeta;\zeta_{\mathbf{t}m}]=0, \tag{5.2.8}$$

where

$${}_0C_0(\mathbf{t},m)=\tfrac{1}{4}({}_0d\,{}_0\varepsilon_{\mathbf{t}m}-g_o)-{}_0\rho_{0;\mathbf{t}m}\,\nu_{\mathbf{t},m}. \tag{5.2.8$^\dagger$}$$

Since $\nu_{\mathbf{t}m} > 0$ for bound energy levels ($\mathbf{t} = \mathbf{c}$) scaling the variable $\zeta$ by a positive factor:

$$\varsigma_+ \equiv \nu_{\hat{\mathbf{t}}_+m}\zeta \qquad (\hat{\mathbf{t}}_+ = \mathbf{b} \text{ or } \mathbf{c}) \tag{5.2.9}$$

turns differential equation (5.2.8) into the *c*-hypergeometric equation on the positive semi-axis. This is also true for *c*-GS solutions **b**m. However, to be able to express solutions of types **a** and **d** in terms of c-hypergeometric functions on the positive semi-axis one has to reflect the argument as it was first pointed to by Junker and Roy [96, 97] in their analysis of DTs of the Morse and isotonic oscillators using nodeless FFs of type **d**. More recently nodeless generalized Laguerre polynomials in the reflected variable

$$\varsigma_- \equiv \nu_{\hat{\mathbf{t}}_m}\zeta \tag{5.2.9*}$$

were used in [76, 81] and in [13] to construct rational SUSY partners of the Morse potential ($\hat{\mathbf{t}}_- = \mathbf{d}$) and the isotonic oscillator ($\hat{\mathbf{t}}_- = \mathbf{a}$ and $\mathbf{d}$), respectively.

By expressing (5.2.6) in terms of variables (5.2.9) and (5.2.9*) one comes to the following *c*-hypergeometric equation

$$\{\varsigma_{\pm}\frac{d}{d\varsigma_{\pm}^{2}}+(\gamma_{\hat{\mathbf{t}}_{\pm}m}+1-\varsigma_{\pm})\frac{d}{d\varsigma_{\pm}}-\alpha_{\hat{\mathbf{t}}_{\pm}m}\}F[\alpha_{\hat{\mathbf{t}}_{\pm}m};\gamma_{\hat{\mathbf{t}}_{\pm}m};\varsigma_{\pm}], \tag{5.2.10}$$

where

$$\alpha_{\mathbf{t}m}\equiv -{}_{0}C_{0}(\mathbf{t},m)/\nu_{\mathbf{t}m}=-m \tag{5.2.10a}$$

and

$$\gamma_{\mathbf{t}m}\equiv {}_{0}\lambda_{0;\mathbf{t}m}+1. \tag{5.2.10′}$$

We thus come to the following set of three algebraic equations with respect to the unknown quantities ${}_{0}\varepsilon_{\mathbf{t}m}$, ${}_{0}\lambda_{0;\mathbf{t}m}$, and $\nu_{\mathbf{t}m}$:

$$g_{o}-{}_{0}d\,{}_{0}\varepsilon_{\mathbf{t}m}=-2\nu_{\mathbf{t}m}({}_{0}\lambda_{0;\mathbf{t}m}+2m+1), \tag{5.2.11a}$$

$${}_{0}\lambda_{0;\mathbf{t}m}^{2}=\lambda^{2}({}_{0}c_{0}\,{}_{0}\varepsilon_{\mathbf{t}m};\lambda_{o}), \tag{5.2.11b}$$

$$\nu_{\mathbf{t}m}^{2}=\nu^{2}({}_{0}\varepsilon_{\mathbf{t}m};\nu_{o}), \tag{5.2.11c}$$

where the energy-dependent quantitates in the right-hand-sides of (5.2.11b) and (5.2.11c) are defined via (5.1.16) and (3.5′), respectively.

At this point it is convenient to analyze two branches CT0 and CT+ of the c-GRef potential separately. Namely, by setting $\nu_{o}=0$ in (5.2.11b) we can eliminate the factorization energy ${}_{0}\varepsilon_{\mathbf{t}m}$ via the relation

$${}_{0}\varepsilon_{\mathbf{t}m}=-\nu_{\mathbf{t}m}^{2} \tag{5.2.12*}$$

which leads to the quartic equation

i) CT0 ($\nu_{o}=0$): $$[g_{o}+{}_{0}d\,\nu_{\mathbf{t}m}^{2}+2(2m+1)\nu_{\mathbf{t}m}]^{2}-4\nu_{\mathbf{t}m}^{2}(\lambda_{o}^{2}+{}_{0}c_{0}\nu_{\mathbf{t}m}^{2})=0 \tag{5.2.12}$$

with the leading coefficient equal to $\Delta_{T}$ unless $\Delta_{T}=0$. Since the equation has a positive free term it may have only an even (odd) number of negative roots if $\Delta_{T}>0$ ($\Delta_{T}<0$). If the TP has a double root ($\Delta_T = 0$) then quartic equation (5.2.12) turns into the cubic equation, by analogy with the regular case analyzed in [34].

For each real nonzero root of this equation the characteristic exponent of the given *c*-GS solution at the origin is computed via the relation

$$
{}_0\lambda_{0;\mathbf{t}m} = -\frac{g_o + {}_0d\,\nu_{\mathbf{t}m}^2 + 2(2m+1)\nu_{\mathbf{t}m}}{2\nu_{\mathbf{t}m}}. \tag{5.2.12′}
$$

Note that quartic equation (5.2.12) does not have a zero root unless $g_o = 0$ and therefore the numerator of fraction (5.2.12′) may not vanish. The important corollary directly followed from this observation is that *c*-GS solutions $\mathbf{t}m$ for the CT0-branch of the *c*-GRef potential may not change their type as the RI $\lambda_o$ increases starting from 0. Again, similarly to the regular case $\iota = 1$, the curves ${}_0\lambda_{0;\mathbf{t}m}(\lambda_o, g_o) = -\tilde{m}$ for $\tilde{m} = 1, \ldots, m$ and $\mathbf{t} = \mathbf{b}$ or $\mathbf{d}$ represent an exception from this general rule because the Laguerre polynomial $L_m^{(-\tilde{m})}(\varsigma_-)$ has zero of order $\tilde{m}$ at $\varsigma_- = 0$ (see (22.3.9) in [91]). As a result, along each curve the *c*-GS solution $\mathbf{t}m$ turns into another *c*-GS solution $\mathbf{t}_\Diamond\, m_\Diamond$ of type

$$
\mathbf{t}_\Diamond = \begin{cases} \mathbf{c} \ \text{ if } \ \mathbf{t} = \mathbf{b}, \\ \mathbf{a} \ \text{ if } \ \mathbf{t} = \mathbf{d}, \end{cases} \tag{5.2.13}
$$

with $m_\Diamond < m$. Keeping in mind that the ChExp of the solution $\mathbf{t}_\Diamond\, m_\Diamond$ at $\varsigma = 0$ is given by the relation

$$
{}_0\rho_{0;\mathbf{t}_\Diamond m_\Diamond}(\lambda_o, g_o) = 1 - 2m + \tilde{m} \tag{5.2.13*}
$$

we again find that $m_\Diamond = m - \tilde{m}$.

Since the TP with zero discriminant is allowed to have only a negative double root its linear coefficient ${}_0d$ and as a result the leading coefficient of the appropriate cubic equation must be positive. Therefore the latter equation has an odd number of negative roots. In particular, if the CT0 branch of the potential $V[\zeta|_0 G^{210}]$ has at least m+1 bound energy levels then the $m^{th}$ eigenfunction is always accompanied by the *c*-GS solutions **b**m and **d**m.

Likewise, setting $\lambda_o$ to 0 turns (5.2.11a) into the quartic equation

ii) CT+ ($\lambda_o$= 0): $$[g_o + ({}_0d/{}_0c_0)\, {}_0\lambda^2_{0;\mathbf{t}m}]^2 - 4({}_0\lambda_{0;\mathbf{t}m} + 2m + 1)^2 (\nu_o^2 + {}_0\lambda^2_{0;\mathbf{t}m}/{}_0c_0) = 0. \qquad (5.2.14)$$

For each real nonzero root of this equation the energy of the *c*-GS solution is unambiguously determined by the formula

$$ {}_0\varepsilon_{\mathbf{t}m} = -{}_0\lambda^2_{0;\mathbf{t}m} / {}_0c_0. \qquad (5.2.14^*)$$

After the energy is computed the exponent $\nu_{\mathbf{t}m}$ is obtained via the fractional relation and

$$\nu_{\mathbf{t}m} = \frac{{}_0d\; {}_0\varepsilon_{\mathbf{t}m} - g_o}{2({}_0\lambda_{0;\mathbf{t}m} + 2m + 1)} \qquad (5.2.14')$$

assuming that

$$g_o \neq -({}_0d/{}_0c_0)(2m+1)^2 \qquad (5.2.14'')$$

so that the right-hand side of (5.2.14) does not vanish and as result the denominator of the fraction in the right-hand side (5.2.14′) differs from 0.

If

$$\frac{{}_0c_0\, g_o}{{}_0d} = -(2m+1)^2 \qquad (5.2.15)$$

then two *c*-GS solutions **b**m and **d**m with the same signed exponent difference at the origin,

$${}_0\lambda_{0;\mathbf{t}m} = -2m - 1, \qquad (5.2.15')$$

co-exist at the energy

$$ {}_0\varepsilon_{\mathbf{t}m} = -(2m+1)^2 / {}_0c_0 \quad (\mathbf{t} = \mathbf{b}, \mathbf{d}). \tag{5.2.15*} $$

Each of quartic equations (5.2.12) and (5.2.14) are analytically decomposed into a pair of quadratic equations in the limiting case of the *c*-AL potential curve $V[\zeta|{}_0\breve{G}^{210}]$. In particular, setting $\nu_o$ to 0 in (5.2.14) leads to the quadratic equation

iii) *c*-AL ($\lambda_o = 0$, $\nu_o = 0$):

$$ {}^{\pm}_{0}\breve{g}_2({}_0d, {}_0c_0)\,{}_0\breve{\lambda}^{2}_{0;\breve{\mathbf{t}}_{\pm}m} \pm 2\sqrt{{}_0c_0}\,(2m+1)\;{}_0\breve{\lambda}_{0;\breve{\mathbf{t}}_{\pm}m} + {}_0c_0\,g_o = 0, \tag{5.2.16} $$

with the leading coefficients

$$ {}^{\pm}_{0}\breve{g}_2({}_0d, {}_0c_0) \equiv {}_0d \pm 2\sqrt{{}_0c_0} \tag{5.2.16′} $$

and discriminants

$$ {}_0\Delta^{(m)}_{\pm} = 4[{}_0c_0(2m+1)^2 - {}_0c_0\,g_o\;{}^{\pm}_{0}\breve{g}_2({}_0d, {}_0c_0)]. \tag{5.2.16″} $$

If the TP with two distinct real roots (${}_0\Delta_T > 0$) has a positive linear coefficient ${}_0d$ then the leading coefficients of quadratic equations (5.2.16) are both positive. Since the common free term is negative for $g_o < 0$ we conclude that each quadratic equation has a pair of real roots of opposite sign. Combining (5.2.16) with (5.2.14*) one can represent the numerator of fraction (5.2.14′) as

$$ {}_0d\;{}_0\breve{\varepsilon}_{\breve{\mathbf{t}}_{\pm}m} - g_o = \pm 2({}_0\breve{\lambda}_{0;\breve{\mathbf{t}}_{\pm}m} + 2m+1)\,{}_0\breve{\lambda}_{0;\breve{\mathbf{t}}_{\pm}m} / \sqrt{{}_0c_0} \tag{5.2.16*} $$

which gives

$$ \sqrt{{}_0c_0}\,\breve{\nu}_{\breve{\mathbf{t}}_{\pm}m} = \pm\,{}_0\breve{\lambda}_{0;\breve{\mathbf{t}}_{\pm}m} \tag{5.2.16†} $$

so that the types $\breve{\mathbf{t}}_+$ and $\breve{\mathbf{t}}_-$ are defined via (5.1.25†), similarly to the regular case.

On other hand, if the TP has a pair of complex conjugated roots ($_0\Delta_T < 0$) while the linear coefficient is still positive, then the leading coefficient of the quadratic equation for $_0\breve{\lambda}_{0;\breve{\mathbf{t}}_-m}$ changes its sign so that two real roots of the latter equation (if exist) must be both negative.

Finally, if condition (5.2.7b) holds then the quadratic equation for $_0\breve{\lambda}_{0;\breve{\mathbf{t}}_+m}$ has two negative roots so that the potential does not have the discrete energy spectrum.

We thus proved that the RCSLE in the limit $\lambda_o = 0$, $\nu_o = 0$ has infinite number of *c*-GS solutions of each of four types **a**, **b**, **c**, and **d** iff the TP has positive discriminant. As stressed above *c*-GS solutions for the CT0-branch of the *c*-GRef potential may not change their type as $\lambda_o$ increases. This implies that the quartet **abcd** of *c*-GS solutions **t**m must exist for any non-negative value of $\lambda_o$ provided that the reflective barrier of the potential $V[\zeta|_0\mathcal{G}^{210}]$ is located at small $\zeta$. Since each sequence of *c*-GS solutions starts from the necessarily nodeless basic solution the CT0-branch has four P-EQ SUSY partners. It will be shown in a separate publication that all *c*-GS solutions of type $\breve{\mathbf{t}}_- = \mathbf{a}$ lie below the ground energy level of the AL *c*-GRef potential curve and therefore this must be also true for the CT0-extension of this curve. This implies that there is an infinite ladder of P-CEQ potentials $V[\zeta|\,^p_0\mathcal{G}^{210}_{\mathbf{a}m_{k=1,\ldots,p}}]$ with a Coulomb tail approaching zero as $\zeta \to \infty$. One can also add a finite number of steps using nodeless *c*-GS solutions **b**m starting from the basic solution **b**0.

Selection of nodeless regular *c*-GS solutions for the CT+ branch represents a more challenging problem and will be addressed in a separate study.

Since discriminant $_0\Delta_-^{(m)}$ is a monotonically increasing function m and the linear coefficient of the given quadratic polynomial is negative all *c*-GS solutions of type $\breve{\mathbf{t}}_- = \mathbf{a}$ are necessarily lie below the ground energy level of the AL *c*-GRef potential curve. Again, as $\lambda_o$ increases these solutions must stay below the ground energy level $_1\varepsilon_{\mathbf{c}0}$ in the CT0 GRef potential on the line and therefore remain nodeless. We thus conclude that the latter potential has infinitely many rational

SUSY partners $V[\zeta|{}^{p}_{0}\breve{G}^{210}_{\{\mathbf{a}m_k\}_p}]$. In particular this infinite family of Hi-CEQ potentials contains the potential $V[\zeta|{}^{1}_{0}\breve{G}^{210}_{\mathbf{a}0}]$ exactly quantized by *c*-Heun polynomials.

As pointed to by Grandati [13, 81] the KLH theorem assures that any Laguerre polynomial $L_{2j}^{({}_0\breve{\lambda}_{0;\mathbf{d}2j})}(\varsigma_-)$ of even order 2j is nodeless on the negative semi-axis if

$$|{}_0\breve{\lambda}_{0;\mathbf{d}2j}| = -\frac{\sqrt{{}_0c_0}\,(4j+1) + \tfrac{1}{2}\sqrt{{}_0\Delta_+^{(2j)}}}{{}^{+}_{0}\breve{g}_2({}_0d,\,{}_0c_0)} > 2j. \tag{5.2.17}$$

Making use of (5.2.16′) and (5.2.16′′) we can re-write (5.2.17) as

$$({}_0d - 2\sqrt{{}_0c_0}\,)\,j^2 - \sqrt{{}_0c_0}\,(j + \tfrac{1}{4}\sqrt{{}_0c_0}\,|g_o|) < 0. \tag{5.2.17′}$$

An analysis of this inequality shows that there is a finite number of nodeless GS solutions $\mathbf{d}2j$ within the range

$$0 \le 2j < \frac{\sqrt{{}_0c_0} + \sqrt{\Delta_{\mathbf{d}}(g_o;{}_0d,{}_0c_0)}}{{}_0d - 2\sqrt{{}_0c_0}}, \tag{5.2.17*}$$

where $\Delta_{\mathbf{d}}(g_o;{}_0d,{}_0c_0)$ is discriminant of the quadratic polynomial in j in the left-hand side of (5.2.17′). Note that the discriminant $\Delta_{\mathbf{d}}(g_o;\,{}_0d,{}_0c_0)$ and therefore the upper bound for j increase as $|g_o|$ grows. Each of these nodeless solutions can be used as the FF to construct CEQ SUSY partner $V[\zeta|{}^{1}_{0}\breve{G}^{210}_{\mathbf{d}2j}]$ of the *c*-AL potential curve $V[\zeta|{}_0\breve{G}^{210}]$.

To be able to extend the constructed family of CEQ potentials beyond the limiting case $\lambda_o = 0$ one needs to study more carefully behavior of the curves ${}_0\lambda_{0;\mathbf{d}2j}(\lambda_o, g_o) = -2j$ in the plane $\lambda_o g_o$ as $\lambda_o$ grows. In fact, each curve determines the upper bound for the range of $\lambda_o$ (at a fixed value of the RI $g_o$) where the potential $V[\zeta|{}^{1}_{0}G^{210}_{\mathbf{d}2j}]$ is definitely quantized by GS *c*-Heine polynomials. There is no simple answer to this problem so that we just postpone its discussion for future (more scrupulous) studies of this very special family of P-CEQ potentials on the line.

Coming back to quartic equation (5.2.13), note that its leading coefficient is equal to $\Delta_T / {}_0c_0$ so that the quartic equation again turns into a cubic equation if the TP has a double root. One directly verify that the latter equation has a negative leading coefficient whereas its free term remains positive for

$$0 < \nu_o < \frac{|g_o|}{2(2m+1)}. \qquad (5.2.18)$$

which implies that the equation must have an even number of negative roots and therefore the m[th] eigenfunction is accompanied by the *c*-GS solutions **b**m and **d**m as far as the parameter $\nu_o$ lies within the specified range. According to (5.2.13′) the exponent $\nu_{\dagger m}$ is necessarily positive at zero energy (${}_0\lambda_{0;\dagger m} = 0$) so that the *c*-GS solution changing its type at the upper bound of this range must be regular at infinity. Since the number of nodes of this solution does not exceed the number of nodes m of the eigenfunction **c**m this cannot be the breaking point where a new bound state arises. We thus conclude that bound energy levels disappear one-by-one as $\nu_o$ increases from 0 to $\frac{1}{6}|g_o|$. In theory the discrete energy spectrum can re-appear again beyond the range for $0 < \nu_o < \frac{1}{6}|g_o|$. To unequivocally verify that the CT+ potential $V[\zeta\,|\,{}_0G^{210}]$ does not have the discrete energy spectrum for $\nu_o > \frac{1}{6}|g_o|$ one need to re-examine more cautiously the roots of the appropriate cubic equation for large values of $\nu_o$ which will be done in a separate publication.

### 5.3 Milson potential

In case of the generic *i*-GRef potential $V[\eta\,|\,{}_i\boldsymbol{G}]$ RefPFr (5.2) takes the form [20]

$$ {}_iI^o[\eta;h_o] = -\tfrac{1}{4}[h_o/(\eta+i)^2 + h_o^*/(\eta-i)^2 - {}_iO_0^o/(\eta^2+1)], \tag{5.3.1}$$

where $h_o \equiv h_{o;R} + i\,h_{o;I}$. Differential equation (3.15) turns into the equation with three regular singular points – *i*, + *i*, and ∞:

$$\left\{(\eta^2+1)\frac{d^2}{d\eta^2} + 2\,{}_iB_1[\eta;\bar\rho(\varepsilon\,|\,{}_i\boldsymbol{G};\sigma)]\frac{d}{d\eta} + C_0(\varepsilon\,|\,{}_i\boldsymbol{G};\sigma)\right\}F[\eta;\varepsilon\,|\,{}_i\boldsymbol{G};\sigma] = 0, \tag{5.3.2}$$

where, according to (3.17), (3.30′), and (3.29) with ι = *i*,

$$2\,{}_iB_1[\eta;\bar\rho(\varepsilon\,|\,{}_i\boldsymbol{G};\sigma)] = [\,{}_i\lambda_\sigma(\varepsilon;h_o)+1](\eta-i) + [\,{}_i\lambda_\sigma^*(\varepsilon;h_o)+1](\eta+i) \tag{5.3.3}$$

and

$$C_0[\eta;\varepsilon\,|\,{}_i\boldsymbol{G};\bar\sigma] = \tfrac{1}{4}(\,{}_iO_0^o + {}_id\varepsilon) + 8\,|\,{}_i\lambda_\sigma(\varepsilon;h_o)+1|^2\,. \tag{5.3.4}$$

As initially pointed to by Routh [54] (and extended in a more general context in previous Section) differential equation (5.3.2) has a polynomial solution of order m at the energy ${}_i\varepsilon_{\dagger m}$ only if one of the parameters

$${}_i\alpha^G_{\dagger m} \equiv \alpha(\,{}_i\varepsilon_{\dagger m}\,|\,{}_i\boldsymbol{G}_{\downarrow\dagger m};\sigma_{\dagger m}) \quad \text{or} \quad {}_i\beta^G_{\dagger m} \equiv \beta(\,{}_i\varepsilon_{\dagger,m}\,|\,{}_i\boldsymbol{G}_{\downarrow\dagger m};\sigma_{\dagger m}) \tag{5.3.5}$$

is equal to –m, where

$$\alpha(\varepsilon\,|\,{}_i\boldsymbol{G};\sigma) = Re\,\lambda_\sigma(\varepsilon;h_o) - \tfrac{1}{2}\sqrt{Re^2\,\lambda_\sigma(\varepsilon;h_o) - C_0(\varepsilon\,|\,{}_i\boldsymbol{G};\sigma)} \tag{5.3.6a}$$

and

$$\beta(\varepsilon\,|\,{}_i\boldsymbol{G};\sigma) = Re\,\lambda_\sigma(\varepsilon;h_o) + \tfrac{1}{2}\sqrt{Re^2\,\lambda_\sigma(\varepsilon;h_o) - C_0(\varepsilon\,|\,{}_\iota\boldsymbol{G};\sigma)} \tag{5.3.6b}$$

are two roots of the quadratic equation

$$X^2(\varepsilon\,|\,{}_i\boldsymbol{G};\sigma) - [2Re\,\lambda_\sigma(\varepsilon;h_o) + 1]X(\varepsilon\,|\,{}_i\boldsymbol{G};\sigma) + C_0(\varepsilon\,|\,{}_\iota\boldsymbol{G};\sigma) = 0. \qquad (5.3.7)$$

As for the m-dependent symbol $\sigma_{\mathbf{t}\mathrm{m}}$ it is defined as follows

$$\sigma_{\mathbf{t}\mathrm{m}} \equiv \begin{cases} - \quad \text{for } \mathbf{t} = \mathbf{c} \text{ or } \mathbf{d}', \\ + \quad \text{for } \mathbf{t} = \mathbf{d}, \end{cases} \qquad (5.3.8)$$

where the types $\mathbf{c}$, $\mathbf{d}$, and $\mathbf{d}'$ are given by (3.7c), (3.7d), and (3.7d′), respectively. As mentioned in previous Section we are only interested in the range of energies where this quadratic equation has a positive discriminant:

$$C_0(\varepsilon\,|\,{}_\iota\boldsymbol{G};\sigma) < Re^2\,\lambda_\sigma(\varepsilon;h_o) \qquad (5.3.9)$$

so that both roots (5.3.6a) and (5.3.6b) are real. It has been proven by Routh [54] that the quadratic equation in m:

$$m^2 + m(4\rho - 1) + C_0 = 0 \qquad (5.3.10)$$

is also a sufficient condition for the Fuschian equation with real polynomial coefficients:

$$\left\{(\eta^2+1)\frac{d^2}{d\eta^2} + 2[\rho(\eta - i) + \rho^*(\eta + i)]\frac{d}{d\eta} + C_0\right\}\Re_m^{(\rho)}(\eta) = 0, \qquad (5.3.11)$$

to have a polynomial solution $\Re_m^{(\rho)}(\eta)$ of order m. To give credit to this fundamentally significant result of Routh' paper we [20] refer to $\Re_m^{(\rho)}(\eta)$ as 'Routh polynomial'. Later Romanovsky [84] discovered a set of orthogonal polynomials which turned out to satisfy Routh condition (5.3.10) and therefore form an orthogonal subset of Routh polynomials referred to by us [20] as Romanovski-Routh polynomials.

Setting the energy ε in (5.3.2) and the complex number ρ in (5.3.7) respectively to ${}_i\varepsilon_{\mathbf{t}\mathrm{m}}$ and to the characteristic exponent

$${}_i\rho_{\mathbf{t}\mathrm{m}} \equiv \rho_{\mathbf{t}\mathrm{m};R} + i\,\rho_{\mathbf{t}\mathrm{m};I} \equiv {}_i\rho_{0;\mathbf{t}\mathrm{m}} \qquad (5.3.12)$$

of the *i*-GS solution

$$\phi_{\dagger m}[\eta \,|\, {}_i\boldsymbol{G}^{220}_{\downarrow\dagger m}] = (1 - i\,\eta)^{\;{}_i\rho_{\dagger m}}(1 + i\,\eta)^{{}_i\rho^*_{\dagger m}}\,\Pi_m\left[\eta;\bar{\eta}_{\dagger m}\right] \qquad (5.3.13)$$

at the singular point $-i$ we conclude that the energies of all possible *i*-GS solutions are unambiguously determined by a coupled set of generally transcendental and algebraic equations

$$\tfrac{1}{2}\,{}_i d\;{}_i\varepsilon_{\dagger m} = -h_{o;R} - \tfrac{1}{2} - |\,{}_i\lambda_{\dagger m} + 1|^2 - 2m(m + 1 + 2\,{}_i\lambda_{\dagger m;R}) \qquad (5.3.14a)$$

and

$${}_i c_0\;{}_i\varepsilon_{\dagger m} = \lambda^2_{\dagger m} - h_{o;0} - 1 \qquad (5.3.14b)$$

with respect to ${}_i\varepsilon_{\dagger m}$ and the complex parameter

$${}_i\lambda_{\dagger m} \equiv \lambda_{\sigma\dagger}({}_i\varepsilon_{\dagger m}; h_o) \equiv \lambda_{\dagger m;R} + i\,\lambda_{\dagger m;I} \equiv 2\,{}_i\rho_{\dagger m} - 1 \qquad (5.3.15)$$

related to the characteristic exponent (5.3.12) in a simple fashion

$${}_i\lambda_{\dagger m} = 2\,{}_i\rho_{\dagger m} - 1. \qquad (5.3.16)$$

Note that the parameter ${}_i O^0_0$ was excluded from the right-hand side of (5.3.14a) via (2.35).

As mentioned in Introduction we are only interested in Milson's [22] reduction $V[\eta|\boldsymbol{M}]$ of the *i*-GRef potential $V[\eta|\,{}_i\boldsymbol{G}^{220}]$ generated by means of the symmetric TP

$${}_iT_{sym}[\eta] = {}_i a_2\,(\eta^2 + \kappa_+), \qquad (5.3.17)$$

with ${}_i c_0 = {}_i c^*_0 = {}_i a_2(\kappa_+ - 1)$. The aforementioned between 'inside' and 'outside' branches of the Milson potential correspond to the following ranges of the parameter $\kappa_+$: $0 < \kappa_+ < 1$ and $\kappa_+ > 1$, respectively, with the border case $\kappa_+ = 1$ represented by the Gendenshtein potential. As explained below, the remarkable feature of the RCSLE associated with the inside branch is that it has two infinite sequences of nodeless $\Re$S solutions. Each of these solutions can be thus used for constructing a new Hi-CEQ potential.

It has been proven in [20] that a number of excited bound energy states in the given potential is equal to

$$n_{max} = [\lambda_{o;R} - 1/2], \tag{5.3.18}$$

where

$$_i\lambda_o \equiv \lambda_{o;R} + i\ \lambda_{o;I} \equiv \sqrt{h_o + 1} \quad (\lambda_{o;R} > 0,\ -\infty < \lambda_{o;I} < \infty). \tag{5.3.19}$$

To derive the cited quartic equation one first needs to split complex algebraic equation (5.3.14b) into two real algebraic equations:

$$\lambda^2_{\dagger m;R} - \lambda^2_{\dagger m;I} = h_{o;R} + 1 - {_i c_{0;R}}\ {_i\varepsilon_{\dagger m}} \tag{5.3.20a}$$

and

$$2\lambda_{\dagger m;R}\lambda_{\dagger m;I} = h_{o;I} - {_i c_{0;I}}\ {_i\varepsilon_{\dagger m}}, \tag{5.3.20b}$$

with respect to $\lambda_{\dagger m;R}$, $\lambda_{\dagger m;I}$, and ${_i\varepsilon_{\dagger m}}$, where ${_i c_{0;R}}$ and ${_i c_{0;I}}$ are the real and imaginary parts of the coefficient ${_i c_0}$. Summing up (5.3.14a) with (5.3.20a) and making use of (2.9*), one comes to the following quadratic formula [20]

$${_i a_2}\ {_i\varepsilon_{\dagger m}} = -(2\lambda_{\dagger m;R} + 2m + 1)^2. \tag{5.3.21}$$

for factorization energies of *i*-GS solutions (5.3.4) which is applicable to any *i*-GRef potential Keeping in mind that the coefficient ${_i c_0}$ of TP (5.3.17) is real one can eliminate $\lambda_{\dagger m;I}$ from (5.3.21a) via the relation

$$\lambda_{\dagger m;I} = \frac{h_{o;I}}{2\lambda_{\dagger m;R}} \tag{5.3.22}$$

and then use (5.3.21) to exclude the energy which leads us to the quartic equation sought for [20]

$$\lambda^4_{\dagger m;R} - [h_{o;R} + 1 + (1 - \kappa_+)\ (\lambda_{\dagger m;R} + m + 1/2)^2]\lambda^2_{\dagger m;R} - h^2_{o;I} = 0 \tag{5.3.23}$$

with respect to $\lambda_{\dagger m;R}$.

It is essential that the derived equation has a positive leading coefficient and a negative free term except the limiting case of the symmetric potential. This observation allows us to conclude that the given quartic equation has a pair of real roots of opposite sign for any nonzero value of the asymmetry parameter $h_{o;I}$. The infinite sequence of $\Re$S solutions $\mathbf{d}$m of type $\mathbf{d}$ associated with positive roots starts from the basic solution and thereby is referred to as 'primary'. Negative roots describe bound energy states as far as the order of the Routh polynomial in question does not exceed $n_{max}$.

In addition there is another infinite subset $\mathbf{d}'m'$ of $\Re$S solutions of type $\mathbf{d}$ formed by Routh polynomials of order larger than $n_{max}$, where we use prime to distinguish this 'secondary' subset from the primary one. It has been proven in [20] that no Routh polynomial of even order from the primary sequence may have real roots, i.e., the RCSLE associated with the inside branch of the Milson potential has infinitely many nodeless $\Re$S solutions $\mathbf{d}$m. In addition, the latter equation has infinitely many nodeless $\Re$S solutions $\mathbf{d}'m'$ formed by Routh polynomials of sufficiently large order (though the lower bound for the order of Routh polynomial with no real roots in the secondary sequence remains uncertain for now).

## 6. Use of AEH factorization functions for constructing SUSY pairs of rational Liouville potentials

Let us now come back to the general case of RefPFrs defined via (2.13) and (2.15). Obviously the GRef Liouville potentials discussed in previous section represent some particular examples (n = 0) of the PFr beams with $\ell = 0$. Our next step is to construct a ladder of rational SUSY of partner Liouville potential (2.6) starting from RefPFrs ${}^{0}_{\iota}\mathbf{B}$ with $\ell = 0$. A very important observation made by the author, while analyzing Cooper, Ginocchio, and Khare's arguments in [1], is that the CLDT using AEH solution (3.1) as the FF results in a new PFr beam ${}^{1}_{\iota}\mathbf{B}_{\dagger m}$. Its RefPFr is given by (2.13), with ${}^{\ell}_{\iota}\mathbf{B}$, $\ell$, and n changed for ${}^{1}_{\iota}\mathbf{B}_{\dagger m}$, 1, and $n_{\dagger m}$, respectively. In particular, excluding the limiting cases of shape-invariant potentials ($\Im = 0$), any p-step

CLDT acting on the GRef PFr beam $_{\iota}\mathcal{G}$ results in a PFr beam with second-order poles at TP zeros for odd and only odd number of steps, p. The proof is done by applying Suzko's reciprocal formula (A.17) in Appendix A to the FF of the inverse CLDT ${}^{1-\ell}_{\iota}\mathbf{B}_{\dagger m} \to {}^{\ell}_{\iota}\mathbf{B}_{\downarrow\dagger m}$ for each intermediate step. In fact, substituting (3.1) and (2.8) into (A.17) gives

$$ {}^{*}\phi[\xi \mid {}^{\ell}_{\iota}\mathbf{B}^{K\Im 0}_{\downarrow\dagger m}; \dagger m] = \frac{{}_{\iota}\Theta[\xi; -{}_{\iota}\bar{\lambda}_{\dagger m}]\Pi_{n}[\xi; \bar{\xi}^{(n)}_{o}]}{\sqrt{{}_{\iota}a_{K}}\,\Pi^{\iota\rho_{T}(1-\ell)}_{\Im}[\xi; {}_{\iota}\bar{\xi}_{T}]\Pi_{n_{\dagger m}}[\xi; {}^{*}_{\iota}\bar{\xi}^{(n_{\dagger m})}_{\dagger m}]}. \tag{6.1}$$

As pointed to at the beginning of Section 3 we assume that AEH solution (3.1) is irregular at any outer singular point so that the CLDT in question eliminates all the outer second-order poles appearing in the initial RCSLE. In particular this is true for the second-order poles at $\xi = {}_{\iota}\xi_{T;k}$ if the latter exist for the PFr beam ${}^{\ell}_{\iota}\mathbf{B}_{\downarrow\dagger m}$ $(\ell = 1)$. On the contrary, if $\xi = {}_{\iota}\xi_{T;k}$ is a regular point of the initial RCSLE then the partner equation has the second-order pole at this point.

Excluding the anomalous case of the Gendenshtein potential the TP does not have zeros at ${}_{\iota}e_{r}$ ($r$ = 0, |ι|) so that the CLDTs in question keep unchanged the exponent differences for the finite singular points at the factorization energy ${}_{\iota}\varepsilon_{\dagger m}$

$$h_{r}({}_{\iota}\varepsilon_{\dagger m}; {}^{1}_{\iota}h_{o;r}) = h_{r}({}_{\iota}\varepsilon_{\dagger m}; {}_{\iota}h_{o;r}) \text{ for } r = 0, |\iota| \tag{6.2}$$

as well as the asymptotic value of the BI for ι = 0:

$$\lim_{\zeta\to+\infty} I[\zeta; {}_{0}\varepsilon_{\dagger m} \mid {}^{\ell}_{\iota}\mathbf{B}_{\downarrow\dagger m}] = \lim_{\zeta\to+\infty} I[\zeta; {}_{0}\varepsilon_{\dagger m} \mid {}^{1-\ell}_{\iota}\mathbf{B}_{\dagger m}]. \tag{6.3}$$

or, in other words, the preserve the values of the parameters

$$ {}^{1}_{\iota}h_{o;r} = {}_{\iota}h_{o;r} \text{ for } r = 0, |\iota| \tag{6.4}$$

and

$$ {}^{1}\nu_{o} = \nu_{o} \text{ for } \iota = 0. \tag{6.4′}$$

Re-writing (A.19*) as

$$\mathrm{I}^{\mathrm{o}}[\xi\,|\,{}^{1-\ell}_{\iota}\mathbf{B}^{\mathrm{K}\Im 0}_{\dagger\mathrm{m}}] = -ld^{\,2}\left|{}^{\star}\phi[\xi\,|\,{}^{\ell}_{\iota}\mathbf{B}^{\mathrm{K}\Im 0}_{\downarrow\dagger\mathrm{m}};\dagger\mathrm{m}]\right| - \dot{ld}\,\left|{}^{\star}\phi[\xi\,|\,{}^{\ell}_{\iota}\mathbf{B}^{\mathrm{K}\Im 0}_{\downarrow\dagger\mathrm{m}};\dagger\mathrm{m}]\right| - {}_{\iota}\varepsilon_{\dagger\mathrm{m}}\,{}_{\iota}\wp[\xi;\,{}_{\iota}T_2], \tag{6.5}$$

representing the logarithmic derivative of AEH solution (6.4) as the sum

$$\begin{aligned} ld\,\left|{}^{\star}\phi[\xi\,|\,{}^{\ell}_{\iota}\mathbf{B}^{\mathrm{K}\Im 0}_{\downarrow\dagger\mathrm{m}};\dagger\mathrm{m}]\right| &= ld\,{}_{\iota}\Theta[\xi;-{}_{\iota}\bar{\lambda}_{\dagger\mathrm{m}}] + ld\left|\Pi_{\mathrm{n}}[\xi;\xi^{(\mathrm{n})}_{\mathrm{o}}]\right| \\ &\quad - {}_{\iota}\rho_{\mathrm{T}}(1-\ell)ld\left|\Pi_{\Im}[\xi;\,{}_{\iota}\bar{\xi}_{\mathrm{T}}]\right| - ld\left|\Pi_{\mathrm{n}_{\dagger\mathrm{m}}}[\xi;\,{}^{\star}_{\iota}\bar{\xi}^{(\mathrm{n}_{\dagger\mathrm{m}})}_{\dagger\mathrm{m}}]\right|, \end{aligned} \tag{6.6}$$

and keeping in mind that

$$ld^{\,2}\left|\Pi_{\mathrm{n}}[\xi;\xi^{(\mathrm{n})}_{\mathrm{o}}]\right| + \dot{ld}\left|\Pi_{\mathrm{n}}[\xi;\xi^{(\mathrm{n})}_{\mathrm{o}}]\right| = \ddot{\Pi}_{\mathrm{n}}[\xi;\,{}^{\star}_{\iota}\bar{\xi}^{(\mathrm{n})}]/\,\Pi_{\mathrm{n}}[\xi;\xi^{(\mathrm{n})}_{\mathrm{o}}], \tag{6.7}$$

coupled with (3.21) and (3.22), we obtain the following explicit expression for the RefPFr of the partner RCSLE:

$$\begin{aligned} \mathrm{I}^{\mathrm{o}}[\xi\,|\,{}^{1-\ell}_{\iota}\mathbf{B}_{\dagger\mathrm{m}}] &= -ld^{2}\,{}_{\iota}\Theta[\xi;-{}_{\iota}\bar{\lambda}_{\dagger\mathrm{m}}] - \dot{ld}\,{}_{\iota}\Theta[\xi;-{}_{\iota}\bar{\lambda}_{\dagger\mathrm{m}}] + 2\hat{\mathrm{Q}}[\xi;\xi^{(\mathrm{n})}_{\mathrm{o}},(1-\ell)\,{}_{\iota}\bar{\xi}_{\mathrm{T}}] \\ &\quad - 2ld\,{}_{\iota}\Theta[\xi;\,{}_{\iota}\bar{\lambda}_{\dagger\mathrm{m}}]\left\{{}_{\iota}\rho_{\mathrm{T}}(1-\ell)ld\,\Pi_{\Im}[\xi;\,{}_{\iota}\bar{\xi}_{\mathrm{T}}] + ld\left|\Pi_{\mathrm{n}_{\dagger\mathrm{m}}}[\xi;\,{}^{\star}_{\iota}\bar{\xi}^{(\mathrm{n}_{\dagger\mathrm{m}})}_{\dagger\mathrm{m}}]\right|\right\} \\ &\quad + 2\left\{ld\,{}_{\iota}\Theta[\xi;-{}_{\iota}\bar{\lambda}_{\dagger\mathrm{m}}] + ld\left|\Pi_{\mathrm{n}_{\dagger\mathrm{m}}}[\xi;\,{}^{\star}_{\iota}\bar{\xi}^{(\mathrm{n}_{\dagger\mathrm{m}})}_{\dagger\mathrm{m}}]\right| + {}_{\iota}\rho_{\mathrm{T}}(1-\ell)ld\,\Pi_{\Im}[\xi;\,{}_{\iota}\bar{\xi}_{\mathrm{T}}]\right\} \\ &\qquad \times ld\,\Pi_{\mathrm{n}}[\xi;\xi^{(\mathrm{n})}_{\mathrm{o}}] \\ &\quad - \ddot{\Pi}_{\mathrm{n}}[\xi;\,{}^{\star}_{\iota}\bar{\xi}^{(\mathrm{n})}]/\,\Pi_{\mathrm{n}}[\xi;\xi^{(\mathrm{n})}_{\mathrm{o}}] - {}_{\iota}\varepsilon_{\dagger\mathrm{m}}\,{}_{\iota}\wp[\xi;\,{}_{\iota}T_2]. \end{aligned} \tag{6.8}$$

Differentiating the logarithmic derivative

$$ld\,|\,{}_{\iota}\Theta[\xi;-{}_{\iota}\bar{\lambda}_{\dagger\mathrm{m}}]\,| = \sum_{r=0}^{|\iota|}\frac{1-{}_{\iota}\lambda_{r;\dagger\mathrm{m}}}{2(\xi-{}_{\iota}e_r)} + \tfrac{1}{2}\,\delta_{\iota,0}\nu_{\dagger\mathrm{m}} \tag{6.9}$$

($\nu_{\dagger\mathrm{m}} \equiv {}_{0}\lambda_{1;\dagger\mathrm{m}}$) with respect to $\xi$:

$$\dot{l\,d}\,|\,{}_{\iota}\Theta[\xi;-{}_{\iota}\bar{\lambda}_{\dagger\mathrm{m}}]\,| = -\sum_{r=0}^{|\iota|}\frac{1-{}_{\iota}\lambda_{r;\dagger\mathrm{m}}}{2(\xi-{}_{\iota}e_r)^2} \tag{6.10}$$

one finds

$$ld^{2}|\,{}_{\iota}\Theta[\xi;-{}_{\iota}\bar{\lambda}_{\mathbf{t}\mathrm{m}}]|+l\dot{d}\,|\,{}_{\iota}\Theta[\xi;-{}_{\iota}\bar{\lambda}_{\mathbf{t}\mathrm{m}}]|=\sum_{r=0}^{|\iota|}\frac{{}_{\iota}\lambda^{2}_{r;\mathbf{t}\mathrm{m}}-1}{4(\xi-{}_{\iota}e_{r})^{2}}+{}^{1}\!/\!_{4}\,\delta_{\iota,0}\,\nu^{2}_{\mathbf{t}\mathrm{m}}$$

$$+\frac{{}_{\iota}C^{0}_{0}(-{}_{\iota}\lambda_{0;\mathbf{t}\mathrm{m}},-{}_{\iota}\lambda_{1;\mathbf{t}\mathrm{m}})}{\prod_{r=0}^{|\iota|}(\xi-{}_{\iota}e_{r})}, \qquad (6.11)$$

where the coefficient ${}_{\iota}C^{0}_{0}(\lambda_{0},\lambda_{1})$ is defined via (3.34). After substituting the latter relation into the right-hand side of (6.8), together with the conventional formulas

$$h_{r}({}_{\iota}\varepsilon_{\mathbf{t}\mathrm{m}};{}_{\iota}h_{o;r})\equiv{}_{\iota}h_{o;r}-{}_{\iota}\varepsilon_{\mathbf{t}\mathrm{m}} \quad \text{for } r=0,\,|\iota|,\ \nu^{2}_{\mathbf{t}\mathrm{m}}=\nu^{2}_{o}-{}_{0}a_{2}\,{}_{\iota}\varepsilon_{\mathbf{t}\mathrm{m}} \qquad (6.12)$$

it is convenient to explicitly separate second-order poles from the rest of the sum:

$$I^{o}[\xi\,|\,{}^{1-\ell}_{\iota}\mathbf{B}_{\mathbf{t}\mathrm{m}}]=-\frac{{}_{\iota}h_{o;0}}{4(\xi-{}_{\iota}e_{0})^{2}}-|\iota|\frac{{}_{\iota}h_{o;1}}{4(\xi-{}_{\iota}e_{1})^{2}}-{}^{1}\!/\!_{4}\,\delta_{\iota,0}\,\nu^{2}_{o}+2\hat{Q}[\xi;\,{}^{*}_{\iota}\bar{\xi}^{(n_{\mathbf{t}\mathrm{m}})}_{\mathbf{t}\mathrm{m}},(1-\ell)\,{}_{\iota}\bar{\xi}_{T}]$$

$$+\frac{\hat{O}^{\downarrow}_{n_{\mathbf{t}\mathrm{m}}+n+(1-\ell)\Im}[\xi\,|\,{}^{1-\ell}_{\iota}\mathbf{B}_{\mathbf{t}\mathrm{m}}]}{4\prod_{r=0}^{|\iota|}(\xi-{}_{\iota}e_{r})\,\Pi^{1-\ell}_{\Im}[\xi;{}_{\iota}\bar{\xi}_{T}]\,\Pi_{n_{\mathbf{t}\mathrm{m}}}[\xi;\,{}^{*}_{\iota}\bar{\xi}^{(n_{\mathbf{t}\mathrm{m}})}_{\mathbf{t}\mathrm{m}}]\,\Pi_{n}[\xi;\bar{\xi}^{(n)}_{o}]}, \qquad (6.13)$$

except that we keep GPD PFr (3.22) untouched. [As explained in Section 3 the GPD of the RefPFr allows one to slightly simplify free term (3.29) of P-CEQ differential equation (3.15) obtained from the appropriate RCSLE via the appropriate gauge transformation.] The numerator of the fraction in the right-hand side of (6.13) can be handily decomposed as

$$\hat{O}^{\downarrow}_{n_{\mathbf{t}\mathrm{m}}+n+(1-\ell)\Im}[\xi\,|\,{}^{1-\ell}_{\iota}\mathbf{B}_{\mathbf{t}\mathrm{m}}]$$

$$=\Pi_{n}[\xi;\bar{\xi}^{(n)}_{o}]\,{}_{\iota}\hat{O}^{\downarrow}_{n_{\mathbf{t}\mathrm{m}}+(1-\ell)\Im}[\xi;{}_{\iota}\lambda_{0;\mathbf{t}\mathrm{m}},{}_{\iota}\lambda_{1;\mathbf{t}\mathrm{m}};{}_{\iota}\varepsilon_{\mathbf{t}\mathrm{m}};{}^{*}\bar{\xi}^{(n_{\mathbf{t}\mathrm{m}})}_{\mathbf{t}\mathrm{m}},{}_{\iota}\bar{\xi}_{T}]$$

$$+\Delta\,\tilde{O}^{\downarrow}_{n_{\mathbf{t}\mathrm{m}}+n+(1-\ell)\Im+|\iota|}[\xi\,|\,{}^{1-\ell}_{\iota}\mathbf{B}_{\mathbf{t}\mathrm{m}}]. \qquad (6.14)$$

Here

$$
{}_{\iota}\hat{O}^{\downarrow}_{n+(1-\ell)\mathfrak{J}}[\xi;\lambda_0,\lambda_1;\varepsilon;{}^{\star}\bar{\xi}^{(n)},{}_{\iota}\bar{\xi}_T] = -\left[{}_{\iota}d\,{}_{\iota}\varepsilon_{\dagger m} + 4\,{}_{\iota}C^0_0(-{}_{\iota}\lambda_{0;\dagger m},-{}_{\iota}\lambda_{1;\dagger m})\right]
$$

$$
\times\Pi^{1-\ell}_{\mathfrak{J}}[\xi;{}_{\iota}\bar{\xi}_T]\Pi_n[\xi;{}^{\star}_{\iota}\bar{\xi}^{(n)}] \quad (6.15)
$$

$$
+8\hat{B}_1[\xi;\bar{\iota};-\bar{\lambda}]\{\dot{\Pi}_{n_{\dagger,m}}[\xi;{}^{\star}\bar{\xi}^{(n_{\dagger},m)}_{\dagger,m}]\Pi^{1-\ell}_{\mathfrak{J}}[\xi;{}_{\iota}\bar{\xi}_T] + {}_{\iota}\rho_T(1-\ell)\Pi_{n_{\dagger,m}}[\xi;{}^{\star}\bar{\xi}^{(n_{\dagger},m)}_{\dagger,m}]\dot{\Pi}_{\mathfrak{J}}[\xi;{}_{\iota}\bar{\xi}_T]\}
$$

$$
\times\Pi_n[\xi;{}^{\star}_{\iota}\bar{\xi}^{(n)}]
$$

taking into account

$$
ld\,|\,{}_{\iota}\Theta[\xi;-{}_{\iota}\bar{\lambda}_{\dagger m}]\,| = \frac{\hat{B}_1[\xi;\bar{\iota};-{}_{\iota}\bar{\lambda}_{\dagger m}]}{\prod\limits_{r=0}^{|\iota|}(\xi-{}_{\iota}e_r)}, \quad (6.16)
$$

where the polynomial numerator

$$
\begin{aligned}
\hat{B}_1[\xi;\bar{\iota};\bar{\lambda}] = {}^1\!/_2\,|\,\iota\,|\,[(\lambda_0+1)(\xi-{}_{\iota}e_1)+(\lambda_1+1)(\xi-{}_{\iota}e_0)] \\
+ {}^1\!/_2\,\delta_{\iota,0}(\lambda_0+1-\lambda_1\xi)
\end{aligned} \quad (6.17)
$$

is nothing but the first-order polynomial $B_1[\xi;\bar{\iota};\bar{\rho}]$ expressed in $\bar{\lambda}\equiv 2\bar{\rho}-\bar{1}_2$, instead of $\bar{\rho}$. Each of the components forming the second term in the right-hand side of (6.15) contains a derivative of the monomial product $\Pi_n[\xi;\bar{\xi}^{(n)}_o]$:

$$
\Delta\tilde{O}^{\downarrow}_{n_{\dagger m}+n+(1-\ell)\mathfrak{J}+|\iota|-1}[\xi\,|\,{}^{1-\ell}_{\iota}\mathbf{B}_{\dagger m}] \quad (6.18)
$$

$$
=4\{\ddot{\Pi}_n[\xi;\bar{\xi}^{(n)}_o]\Pi_{n_{\dagger,m}}[\xi;{}^{\star}\bar{\xi}^{(n_{\dagger},m)}_{\dagger,m}] - 2\dot{\Pi}_n[\xi;\bar{\xi}^{(n)}_o]\dot{\Pi}_{n_{\dagger,m}}[\xi;{}^{\star}\bar{\xi}^{(n_{\dagger},m)}_{\dagger,m}]\}
$$

$$
\times\prod_{r=0}^{|\iota|}(\xi-{}_{\iota}e_r)\Pi^{1-\ell}_{\mathfrak{J}}[\xi;{}_{\iota}\bar{\xi}_T]
$$

$$
+8B_1[\xi;\bar{\iota};-{}_{\iota}\bar{\lambda}_{\dagger m}]\Pi^{1-\ell}_{\mathfrak{J}}[\xi;{}_{\iota}\bar{\xi}_T]\dot{\Pi}_n[\xi;\bar{\xi}^{(n)}_o]\Pi_{n_{\dagger m}}[\xi;{}^{\star}\bar{\xi}^{(n_{\dagger m})}_{\dagger m}]
$$

$$
-8(1-\ell)\,{}_{\iota}\rho_T\,\dot{\Pi}_{\mathfrak{J}}[\xi;{}_{\iota}\bar{\xi}_T]\dot{\Pi}_n[\xi;\bar{\xi}^{(n)}_o]\prod_{r=0}^{|\iota|}(\xi-{}_{\iota}e_r)\Pi_{n_{\dagger m}}[\xi;{}^{\star}\bar{\xi}^{(n_{\dagger m})}_{\dagger m}].
$$

so that the term disappears as a whole if n = 0.

Substituting (A.20) into the left-hand side of the relation

$$I^{o}[\xi\,|\,{}^{1-\ell}_{\iota}\mathbf{B}^{K\Im 0}_{\dagger m}] - I^{o}[\xi\,|\,{}^{\ell}_{\iota}\mathbf{B}^{K\Im 0}_{\downarrow\dagger m}] \tag{6.19}$$

$$= \sum_{k=1}^{n}\frac{2}{(\xi-\xi^{(n)}_{o;k})^{2}} - \sum_{k=1}^{n_{\dagger m}}\frac{2}{(\xi-{}^{*}_{\iota}\bar{\xi}^{(n_{\dagger m})}_{\dagger m})^{2}} + (2\ell-1)\sum_{k'=1}^{\Im}\frac{{}_{\iota}\rho_{T}({}_{\iota}\rho_{T}-1)}{(\xi-\xi_{T;k'})^{2}}$$

$$+\frac{O^{\downarrow}_{n_{\dagger m}+(1-\ell)\Im}[\xi\,|\,{}^{1-\ell}_{\iota}\mathbf{B}^{K\Im 0}_{\dagger m}]}{4\prod_{r=0}^{|\iota|}(\xi-{}_{\iota}e_{r})\,\Pi^{1-\ell}_{\Im}[\xi;\bar{\xi}_{T}]\,\Pi_{n_{\dagger m}}[\xi;{}^{*}\bar{\xi}^{(n_{\dagger m})}_{\dagger m}]} - \frac{O^{\downarrow}_{n+\ell\Im}[\xi\,|\,{}^{\ell}_{\iota}\mathbf{B}^{K\Im 0}_{\downarrow\dagger m}]}{4\prod_{r=0}^{|\iota|}(\xi-{}_{\iota}e_{r})\,\Pi^{\ell}_{\Im}[\xi;\bar{\xi}_{T}]\,\Pi_{n}[\xi;\bar{\xi}^{(n)}_{o}]}$$

and taking into account that

$$\wp^{-1/2}[\xi;{}_{\iota}T_{K}]\,ld\,\left|\wp^{1/4}[\xi;{}_{\iota}T_{K}]\,\phi_{\dagger m}[\xi\,|\,{}^{\ell}_{\iota}\mathbf{B}^{K\Im 0}_{\downarrow\dagger m}]\right| = \text{const} + O(\xi^{-1}) \tag{6.20}$$

one finds

$$O^{\downarrow}_{n_{\dagger m}+(1-\ell)\Im;n_{\dagger m}+(1-\ell)\Im}\,|\,{}^{1-\ell}_{\iota}\mathbf{B}^{K\Im 0}_{\dagger m}) - O^{\downarrow}_{n+\ell\Im;n+\ell\Im}\,|\,{}^{\ell}_{\iota}\mathbf{B}^{K\Im 0}_{\downarrow\dagger m}) \tag{6.21}$$
$$= 8(n_{\dagger m}-n) - 4(2\ell-1)\Im\,{}_{\iota}\rho_{T}({}_{\iota}\rho_{T}-1),$$

in agreement with (2.34) and (2.35) for $\iota = i$.

It will be proven in Part II that polynomial (6.18) is divisible by the monomial product $\Pi_{n}[\xi;\bar{\xi}^{(n)}_{o}]$ for multi-step rational SUSY partners of GRef potentials so that

$$\Delta\tilde{O}^{\downarrow}_{n_{\dagger},m+n+\ell\Im}[\xi\,|\,{}^{2j+\ell}_{\iota}\boldsymbol{G}^{K\Im 0}_{\{\dagger m\}_{2j+\ell}}] = \Pi_{n}[\xi;\bar{\xi}^{(n)}_{o}]\;\Delta\hat{O}^{\downarrow}_{n_{\dagger},m+\ell\Im}[\xi\,|\,{}^{2j+\ell}_{\iota}\boldsymbol{G}^{K\Im 0}_{\{\dagger m\}_{2j+\ell}}]. \tag{6.22}$$

Keeping in mind that

$${}_{\iota}\rho_{T}\,\dot{\Pi}_{\Im}[\xi;{}_{\iota}\bar{\xi}_{T}] = \xi - {}_{\iota}\xi_{T} \quad \text{for } K=2 \text{ and } \Im = 1 \text{ or } 2 \tag{6.23}$$

one finds

$$\lim_{{}_{\iota}\xi_{T;1},\,{}_{\iota}\xi_{T;2}\to{}_{\iota}\xi_{T}}\hat{Q}^{220}[\xi;{}_{\iota}\bar{\xi}_{T}] = -\frac{1}{(\xi-{}_{\iota}\xi_{T;2})^{2}}, \tag{6.24}$$

$$\lim_{{}_\iota\xi_{T;1},\,{}_\iota\xi_{T;2}\to{}_\iota\xi_T} \hat{O}^{\downarrow}_{n_{\dagger m}+2}[\xi;\,{}_\iota\bar{\rho}_{\dagger m},\nu_{\dagger m};\,{}^{*}\bar{\xi}^{(n_{\dagger m})}_{\dagger m},\,{}_\iota\xi_{T;1},{}_\iota\xi_{T;2}]$$
$$=(\xi-\xi_T)\hat{O}^{\downarrow}_{n_{\dagger m}+1}[\xi;\,{}_\iota\bar{\rho}_{\dagger m},\nu_{\dagger m};\,{}^{*}\bar{\xi}^{(n_{\dagger},m)}_{\dagger m},\,{}_\iota\xi_T] \quad (6.25a)$$

and

$$\lim_{{}_\iota\xi_{T;1},\,{}_\iota\xi_{T;2}\to{}_\iota\xi_T} \Delta\hat{O}^{\downarrow}_{n_{\dagger m}+2+n}[\xi\,|\,{}^{1}_{\iota}\mathbf{B}^{220}_{\dagger m}]$$
$$=(\xi-\xi_T)\Delta\hat{O}^{\downarrow}_{n_{\dagger m}+n+1}[\xi\,|\,{}^{1}_{\iota}\mathbf{B}^{210}_{\dagger m}] \quad (6.25b)$$

so that

$$\lim_{{}_\iota\xi_{T;1},\,{}_\iota\xi_{T;2}\to{}_\iota\xi_T} \hat{O}^{\downarrow}_{n_{\dagger m}+n+2}[\xi\,|\,{}^{1}_{\iota}\mathbf{B}^{220}_{\dagger m}]=(\xi-\xi_T)\,O^{\downarrow}_{n_{\dagger m}+n+1}[\xi\,|\,{}^{1}_{\iota}\mathbf{B}^{210}_{\dagger m}] \quad (6.26)$$

Making ${}_\iota\xi_{T;1}$ and ${}_\iota\xi_{T;2}$ tend to ${}_\iota\xi_T$ in the right-hand side of (6.19) for ${}^{\ell}_{\iota}\mathbf{B}={}^{\ell}_{\iota}\mathbf{B}^{220}$, substituting (6.26), together with (3.26) and (3.26†), into the consequential expression, and comparing the resultant formula with (6.19) for ${}^{\ell}_{\iota}\mathbf{B}={}^{\ell}_{\iota}\mathbf{B}^{210}$, we conclude that

$$\lim_{{}_\iota\xi_{T;1},\,{}_\iota\xi_{T;2}\to{}_\iota\xi_T} I^{o}[\xi\,|\,{}^{\ell}_{\iota}\mathbf{B}^{220}]=I^{o}[\xi\,|\,{}^{\ell}_{\iota}\mathbf{B}^{210}] \quad (6.27)$$

for both odd and even number of steps. We thus confirmed that the DRtTP potentials generated by means of a TP with zero discriminant can be directly obtained from generic GRef potentials $V[\xi\,|\,{}^{\ell}_{\iota}\mathbf{B}^{220}]$ simply by making ${}_\iota\xi_{T;1}$ and ${}_\iota\xi_{T;2}$ tend to ${}_\iota\xi_T$ :

$$V[\xi\,|\,{}^{\ell}_{\iota}\mathbf{B}^{210}]=\lim_{{}_\iota\xi_{T;1},\,{}_\iota\xi_{T;2}\to{}_\iota\xi_T} V[\xi\,|\,{}^{\ell}_{\iota}\mathbf{B}^{220}]. \quad (6.28)$$

Similarly, setting K= $\mathfrak{I}$ = 2 in (3.27) and making ${}_\iota\xi_{T;1}$ and ${}_\iota\xi_{T;2}$ tend to ${}_\iota\xi_T$ shows that

$$\lim_{{}_\iota\xi_{T;1},\,{}_\iota\xi_{T;2}\to{}_\iota\xi_T} C_{n+2}[\xi;\varepsilon\,|\,{}^{\ell}_{\iota}\mathbf{B}^{220};\bar{\sigma}]=(\xi-\xi_T)C_{n+1}[\xi;\varepsilon\,|\,{}^{\ell}_{\iota}\mathbf{B}^{210};\bar{\sigma}]. \quad (6.29)$$

We shall come back to general case of PFr beams $_{\iota}\mathcal{B}$ with n ≥ 0 in next Section to reveal some simple recurrence rules for the α- and β-coefficients [23] in the SUSY ladder of Fuschian equations.

In the particular cases of the PFr beams $_{\iota}\mathcal{B}$ with n = 0, i.e., for the GRef PFr beams $_{\iota}\mathcal{G}^{K\Im 0}$ themselves as well as their basic SUSY partners ${}^{1}_{\iota}\mathcal{G}^{K\Im 0}_{\dagger 0}$ generated by CLDTs with the FFs †0, the RefPFr for their SUSY partners (${}^{1}_{\iota}\mathcal{G}^{K\Im 0}_{\dagger m}$ and ${}^{2}_{\iota}\mathcal{G}^{K\Im 0}_{\dagger_1 0;\dagger_2 m_2}$ accordingly) take the form

$$I^{o}[\xi\,|\,{}^{1}_{\iota}\mathcal{G}^{K\Im 0}_{\dagger m}] = -\frac{{}_{\iota}h_{o;0}}{4(\xi-{}_{\iota}e_0)^2} - |\iota|\frac{{}_{\iota}h_{o;1}}{4(\xi-{}_{\iota}e_1)^2} - \tfrac{1}{4}\delta_{\iota,0}\,\nu_o^2 + 2\hat{Q}[\xi;{}_{\iota}\bar{\xi}_{\dagger m},{}_{\iota}\bar{\xi}_{T}]$$

$$+\frac{\hat{O}^{\downarrow}_{m+\Im}[\xi\,|\,{}^{1}_{\iota}\mathcal{G}^{K\Im 0}_{\dagger 0}]}{4\prod\limits_{r=0}^{|\iota|}(\xi-{}_{\iota}e_r)\,\Pi_{\Im}[\xi;{}_{\iota}\bar{\xi}_{T}]\,\Pi_{m}[\xi;{}_{\iota}\bar{\xi}_{\dagger m}]} \qquad (6.30a)$$

and

$$I^{o}[\xi\,|\,{}^{2}_{\iota}\mathcal{G}^{K\Im 0}_{\dagger_1 0;\dagger_2 m_2}] = -\frac{{}_{\iota}h_{o;0}}{4(\xi-{}_{\iota}e_0)^2} - |\iota|\frac{{}_{\iota}h_{o;1}}{4(\xi-{}_{\iota}e_1)^2} - \tfrac{1}{4}\delta_{\iota,0}\,\nu_o^2 + 2\hat{Q}[\xi;{}_{\iota}\bar{\xi}_{\dagger_1 0;\dagger_2 m_2};1]$$

$$+\frac{\hat{O}^{\downarrow}_{{}_{\iota}\upsilon_{\dagger_1 0;\dagger_2 m_2}}[\xi\,|\,{}^{2}_{\iota}\mathcal{G}^{K\Im 0}_{\dagger_1 0;\dagger_2 m_2}]}{4\prod\limits_{r=0}^{|\iota|}(\xi-{}_{\iota}e_r)\,\Pi_{{}_{\iota}\upsilon_{\dagger_1 0;\dagger_2 m_2}}[\xi;{}_{\iota}\bar{\xi}_{\dagger_1 0;\dagger_2 m_2}]} \qquad (6.30b)$$

whereas polynomials (6.15) can be simplified as follows:

$$\hat{O}^{\downarrow}_{m+\Im}[\xi\,|\,{}^{1}_{\iota}\mathcal{G}^{K\Im 0}_{\dagger m}] = \hat{O}^{\downarrow}_{m+\Im}[\xi;{}_{\iota}\lambda_{0;\dagger m},{}_{\iota}\lambda_{1;\dagger m};{}_{\iota}\varepsilon_{\dagger m};\bar{\xi}_{\dagger m},{}_{\iota}\bar{\xi}_{T}]$$

$$= -\left[{}_{\iota}d\,{}_{\iota}\varepsilon_{\dagger m} + 4\,{}_{\iota}C^{0}_{0}(-{}_{\iota}\lambda_{0;\dagger m},-{}_{\iota}\lambda_{1;\dagger m})\right]\Pi_{\Im}[\xi;{}_{\iota}\bar{\xi}_{T}]\Pi_{m}[\xi;{}_{\iota}\bar{\xi}_{\dagger m}] \qquad (6.30a')$$

$$+8\,{}_{\iota}\hat{B}_1[\xi;\bar{\iota},-{}_{\iota}\bar{\lambda}_{\dagger m}]\left\{\dot{\Pi}_{m}[\xi;{}_{\iota}\bar{\xi}_{\dagger m}]\Pi_{\Im}[\xi;{}_{\iota}\bar{\xi}_{T}] + {}_{\iota}\rho_{T}\Pi_{m}[\xi;{}_{\iota}\bar{\xi}_{\dagger m}]\dot{\Pi}_{\Im}[\xi;{}_{\iota}\bar{\xi}_{T}]\right\}$$

and

$$\hat{O}^{\downarrow}_{\iota\upsilon_{\dagger_1 0;\dagger_2 m_2}}[\xi\,|\,{}^{2}_{\iota}\mathcal{G}^{K\Im 0}_{\dagger_1 0;\dagger_2 m_2}] = -[{}_{\iota}d\,{}_{\iota}\varepsilon_{\dagger_2 m_2} + 4\,{}_{\iota}C^{0}_{0}(-{}_{\iota}\lambda_{0;\dagger_2 m_2}, -\lambda_{1;\dagger_2 m_2})] \times \Pi_{\iota\upsilon_{\dagger_1 0;\dagger_2 m_2}}[\xi;{}_{\iota}\bar{\xi}_{\dagger_1 0;\dagger_2 m_2}]$$
$$+8\hat{B}_1[\xi;\bar{\tau};-{}_{\iota}\bar{\lambda}_{\dagger_2 m_2}]\,\dot{\Pi}_{\iota\upsilon_{\dagger_1 0;\dagger_2 m_2}}[\xi;{}_{\iota}\bar{\xi}_{\dagger_1 0;\dagger_2 m_2}]. \qquad (6.30b')$$

A certain inconvenience of the derived formulas for polynomials (6.30a′) and (6.3ba′) is that the algebraic equations for the factorization energies ${}_{\iota}\varepsilon_{\dagger m}$ derived in next Section vary depending on the type ι of the given PFr beam. For this reason it seems more convenient to compute RefPFrs starting from generic formula (A.20), instead of (A.19*). The appropriate alternative expression for polynomial $\hat{O}^{\downarrow}_{m+\Im}[\xi\,|\,{}^{1}_{\iota}\mathcal{G}^{K\Im 0}_{\dagger m}]$ is derived in Appendix C below whereas a similar analysis for the double-step SUSY partners ${}^{2}_{\iota}\mathcal{G}^{K\Im 0}_{\dagger_1 0;\dagger_2 m_2}$ of the *r*- and *c*-GRef PFr beams is postponed for Part II.

As an illustration let us consider the 'sibling' PFr beams generated via CLDTs using the basic AEH FFs $\dagger 0$. The common remarkable feature of the appropriate RLPs $V[\xi\,|\,{}^{1}_{\iota}\mathcal{G}^{K\Im 0}_{\dagger 0}]$ is that they are exactly quantized via GS Heine or GS *c*-Heine polynomials. For this reason we refer to these RLPs as the 'Hi-EQ sibling potentials.' (Remember that we use the term ' P-EQ' if the position of any outer singular point can be varied at fixed values of the RIs.) In particular, if the TP has has a single zero (K=2, ℑ=1 or K=ℑ=1) then the appropriate RLP $V[\xi\,|\,{}^{1}_{\iota}\mathcal{G}^{K10}_{\dagger 0}]$ exactly quantized via *r*-GS Heun (ι= 1) or *c*-GS (ι= 0) Heun polynomials. Also note that $V[z\,|\,{}^{1}_{1}\mathcal{G}^{220}_{\mathbf{c}0}]$ in our classification scheme stands for the CKG potential [1] which initially inspired this series of publications. If the TP has a positive discriminant then the CKG potential has three siblings $V[\xi\,|\,{}^{1}_{\iota}\mathcal{G}^{220}_{\dagger 0}]$, with $\dagger$ = **a**, **b**, or **d**. (As illuminated in detail in [83] the potential $V[z\,|\,{}^{1}_{1}\mathcal{G}^{210}_{\mathbf{c}0}]$ has only two sibling partners , with $V[z\,|\,{}^{1}_{1}\mathcal{G}^{210}_{\mathbf{d}0}]$ as one of them.)

Assuming that the GRef PFr beam is constructed by means of the generic TP with two distinct roots the RefPFr in question takes the form

$$I^{o}[\xi\,|\,{}^{1}_{\iota}\mathcal{G}^{220}_{\dagger 0}] = I^{o}[\xi\,|\,{}_{\iota}G^{o}_{\downarrow\dagger 0}] - \frac{3}{4(\xi-\xi_{T;1})^2} - \frac{3}{4(\xi-\xi_{T;2})^2} - \frac{1}{2\Pi_2[\xi;\bar{\xi}_T]} \tag{6.31}$$

$$+\frac{\Delta O_2[\xi\,|\,{}^{1}_{\iota}\mathcal{G}^{220}_{\dagger 0}]}{4\prod_{r=0}^{|\iota|}(\xi-{}_{\iota}e_r)\Pi_2[\xi;\bar{\xi}_T]},$$

where, according to (C.8*) with $\mathfrak{I}$=2 and m=0,

$$\Delta O_2[\xi\,|\,{}^{1}_{\iota}\mathcal{G}^{220}_{\dagger 0}] = 4[\,|\,\iota\,|\,{}_{\iota}\lambda_{0;\dagger 0} - (-1)^{|\iota|}\,{}_{\iota}\lambda_{1;\dagger 0}]\Pi_2[\xi;{}_{\iota}\bar{\xi}_T]$$

$$+8\,{}_{\iota}\rho_T\,{}_{\iota}\hat{B}_1[\xi;\bar{\iota},-{}_{\iota}\bar{\lambda}_{\dagger 0}]\dot{\Pi}_2[\xi;{}_{\iota}\bar{\xi}_T] \tag{6.32}$$

The appropriate SUSY partner of the *r*-GRef potential can be thus represented as

$$V[z\,|\,{}^{1}_{1}\mathcal{G}^{220}_{\dagger 0}] = V[z\,|\,{}_{1}\mathcal{G}^{220}_{\downarrow\dagger 0}] \tag{6.33}$$

$$+\left\{\frac{3}{\Pi_2^2[z;\bar{z}_T]} + \frac{2}{\Pi_2[z;\bar{z}_T]}\right\}\frac{z^2(1-z)^2}{{}_{1}a_2\Pi_2[z;\bar{z}_T]} - \frac{8(1-{}_{1}\rho_{0;\dagger 0}-{}_{1}\rho_{1;\dagger 0})z(1-z)}{{}_{1}a_2\Pi_2[z;\bar{z}_T]}$$

$$+\frac{8[(1-{}_{1}\rho_{0;\dagger 0})(z-1)+(1-{}_{1}\rho_{1;\dagger 0})z]z(1-z)(z-z_T)}{{}_{1}a_2\Pi_2^2[z;\bar{z}_T]}$$

To compare (6.33) with (4.3) in [1], first note that the CGK parameters $\alpha_n$, $\beta_n$, and $\delta_n$ are nothing but

$$\alpha_n \equiv \mu_{\mathbf{c}n},\ \beta_n \equiv {}_{1}\lambda_{0;\mathbf{c}n},\ \delta_n \equiv {}_{1}\lambda_{1;\mathbf{c}n} = \sqrt{-{}_{1}\varepsilon_{\mathbf{c}n}} \tag{6.34}$$

in our terms so that

$$\varepsilon_0 + \{[(\beta_0+\delta_0)^2-1]z(z-1) + (\beta_0^2-1)(1-z) + (\delta_0^2-1)z+1]\}/R$$
$$\equiv -I^{o}[z\,|\,{}_{1}\mathcal{G}^{220}_{\downarrow\dagger 0}]\,(z')^2. \tag{6.35}$$

We have however problems with matching other terms in the cited formula which are expected to differ from the additional terms in (6.33) by the Schwartz derivative. In particular the coefficient of the term $z^2(1-z)^2/\Pi_2^3[z;\bar{z}_T]$ in (6.33) is equal to $3/{}_1a_2$, instead of $\Delta_T/(2\,{}_1a_2^3)$ in [1] (subtracting the contribution $5\Delta_T/(4\,{}_1a_2^3)$ coming from the Schwartz derivative). Since this singularity is governed by second-order poles in the RefPFr (6.31) at the TP zeros $z_{T;1}$ and $z_{T;2}$ we doubt that the error is on our side.

For $\iota$ equal to either 1 or 0 the terms dependent of the TP coefficients in the right-hand side of (6.31) vanish in the limit $|\,{}_\iota\xi_{T;1}|, |\,{}_\iota\xi_{T;2}| \to \infty$ and the PFr in question turns into the RefPFr for the RM or Morse potential, respectively, with the same values of both RIs.

Similarly the potentials $V[\eta\,|\,{}^1_i\mathcal{G}^{220}_{\mathbf{c}0}]$ and $V[\eta\,|\,{}^1_i\mathcal{G}^{22s}_{\mathbf{t}0}]$, with $\mathbf{t} = \mathbf{c}$ or $\mathbf{d}$, collapse into the shape-invariant Gendenshtein potential in the limit $\eta_{T;1} \to -i$, $\eta_{T;2} \to +i$ $({}_i c_0 \to 0)$ but the latter has different values of the complex parameter $h_o$. In fact, making use of (5.3.15a) we can represent (6.31) for two basic SUSY partners of the Milson potential as

$$I^o[\eta\,|\,{}^1_i\mathcal{G}^{22s}_{\mathbf{t}0}] = I^o[\eta; h_o] - \frac{3}{4(\eta+i\sqrt{\kappa_+})^2} - \frac{3}{4(\eta-i\sqrt{\kappa_+})^2} - \frac{1}{2(\eta^2+\kappa_+)}$$
$$+\frac{2\,{}_i\lambda_{\mathbf{t}0;R}}{\eta^2+1} - \frac{2[({}_i\lambda_{\mathbf{t}0;R}-1)\eta + {}_i\lambda_{\mathbf{t}0;I}]\eta}{(\eta^2+1)(\eta^2+\kappa_+)}. \qquad (6.36)$$

In particular one can directly verify that the terms of the order of $\eta^{-2}$ compensate each other the correction to $I^o[\eta; h_o]$ as expected from the fact that the potentials $V[\eta(x)\,|\,{}^1_i\mathcal{G}^{22s}_{\mathbf{t}0}]$, with $\mathbf{t} = \mathbf{c}$ and $\mathbf{d}$, must vanish at infinity. In the limit ${}_i c_0 \to 0$ the complex parameter ${}_i\lambda_{\mathbf{t}0}$ turns into square root of the (also complex) RI $h_o+1$:

$$\lim_{{}_i c_0\to 0} {}_i\lambda_{\mathbf{t}0} = \sigma_{\mathbf{t}0}\sqrt{h_o+1} \equiv \sigma_{\mathbf{t}0}\,{}_i\lambda_o \quad (Re\,{}_i\lambda_o > 0) \qquad (6.37)$$

so that Ref PFr (6.36) takes the form

$$\lim_{\kappa_+\to 1} I^o[\eta\,|\,{}^1_i\mathcal{G}^{22s}_{\mathbf{t}0}] = I^o[\eta; h_o] - \frac{\sigma_{\mathbf{t}0}\,{}_i\lambda_o+1}{2(\eta+i)^2} - \frac{\sigma_{\mathbf{t}0}\,{}_i\lambda_o^*+1}{2(\eta-i)^2} + \frac{\sigma_{\mathbf{t}0}\,{}_i\lambda_{o;R}+\frac{1}{2}}{\eta^2+1}. \qquad (6.38)$$

By expressing the right-hand side of (2.35) in terms of the complex RI ${}_i\lambda_o$ one can alternatively represent the parameter ${}_i O^0_0$ as

$$ {}_i O^o_0 = 2({}_i\lambda^2_{o;R} - {}_i\lambda^2_{o;I}) - 1 \tag{6.39} $$

and then verify that the right-hand side of (6.36) is nothing but the RefPFr $I^o[\eta; {}^*h_{o;\dagger 0}]$, where we set

$$ {}^*h_{o;\dagger 0} \equiv ({}_i\lambda_o + \sigma_{\dagger 0})^2 - 1. \tag{6.40} $$

We thus conclude that that sequential DTs erasing the ground energy states in the Gendenshtein potential one by one decrease the real part of ${}_i\lambda_0$ by 1 until the latter becomes smaller than 1 so that the discrete energy spectrum disappears. It is crucial that the CLDTs in question (**†** = **c**, **d**, or **d′**) change the characteristic exponents at both finite singular points. This is a direct consequence of the fact that the exponent differences at both singular points become energy independent in the limit ${}_i c_0 \to 0$. As a result the Heine polynomials describing bound states of multi-step $\Re$p-CEQ SUSY partners of the Gendenshtein potential form finite sets of orthogonal polynomials. In Appendix D we illustrate this anomalous feature of the Gendenshtein potential using single-step rational SUSY partners of its nearly-symmetric reduction as an example.

In the limiting case of the TP with zero discriminant RefPFr (6.7) takes the form:

$$ I^o[\xi\,|\,{}^1_\iota \boldsymbol{G}^{210}_{\dagger 0}] = I^o[\xi\,|\,{}_\iota \boldsymbol{G}^{210}_{\downarrow\dagger 0}] - \frac{2}{(\xi - {}_\iota\xi_T)^2} + \frac{\Delta O_1[\xi\,|\,{}^1_\iota \boldsymbol{G}^{210}_{\dagger 0}]}{4\prod\limits_{r=0}^{\iota}(\xi - {}_\iota e_r)(\xi - {}_\iota\xi_T)}, \tag{6.41} $$

where

$$ \Delta O_1[\xi\,|\,{}^1_\iota \boldsymbol{G}^{210}_{\dagger 0}] = 4[\,|\iota|\,{}_\iota\lambda_{0;\dagger 0} - (-1)^{|\iota|}\,{}_\iota\lambda_{1;\dagger 0}]\}(\xi - \xi_T) + 8\,{}_\iota\hat{B}_1[\xi; \bar{\iota}, -{}_\iota\bar{\lambda}_{\dagger 0}] \tag{6.41*} $$

Comparing (6.41*) with (6.31) shows that

$$\lim_{{}_\iota\xi_{T;1},\,{}_\iota\xi_{T;2}\to{}_\iota\xi_T} \Delta O_2[\xi\,|\,{}^1_\iota\boldsymbol{G}^{220}_{\mathbf{t}0}] = \Delta O_1[\xi\,|\,{}^1_\iota\boldsymbol{G}^{210}_{\mathbf{t}0}](\xi - {}_\iota\xi_T) \tag{6.42}$$

as expected. In particular, substituting (6.32) into the left-hand side of this relation for $\iota=1$ we find

$$\Delta O_1[z\,|\,{}^1_1\boldsymbol{G}^{210}_{\mathbf{t}0}] \equiv O_1[z\,|\,{}^1_1\boldsymbol{G}^{210}_{\mathbf{t}0}] - {}_1O^o_0(z - z_T) \tag{6.43}$$

$$= 8({}_1\rho_{0;\mathbf{t}0} + {}_1\rho_{1;\mathbf{t}0} - 1)(z - z_T) + 8[(1 - {}_1\rho_{0;\mathbf{t}0})(z-1) + (1 - {}_1\rho_{1;\mathbf{t}0})z]. \tag{6.43*}$$

As originally pointed to by us in [98] and then elaborated in more detail in [32] the common remarkable feature of the DRtTP potentials $V[z\,|\,{}^1_1\boldsymbol{G}^{210}_{\mathbf{t}0}]$, with $\mathbf{t}$ = $\mathbf{a}$, $\mathbf{b}$, $\mathbf{c}$, and $\mathbf{d}$, is that they are all exactly quantized by GS Heun [32] polynomials. Similarly the potentials $V[\zeta\,|\,{}^1_0\boldsymbol{G}^{210}_{\mathbf{t}0}]$, again for $\mathbf{t}$ = $\mathbf{a}$, $\mathbf{b}$, $\mathbf{c}$, or $\mathbf{d}$, are exactly quantized by GS *c*-Heun polynomials.

## 7. SUSY Networks of GS Heine and GS *c*-Heine polynomials

Suppose that the RCSLE with BI (2.24) has another AEH solution

$$\phi_{\mathbf{t}'m'}[\xi\,|\,{}^\ell_\iota\mathbf{B}_{\downarrow\mathbf{t}'m'}] = \frac{{}_\iota\Theta[\xi;\,{}_\iota\bar{\lambda}_{\mathbf{t}'m'}]\Pi_{n_{\mathbf{t}'m'}}[\xi;\,{}^*_\iota\bar{\xi}^{(n_{\mathbf{t}'m'})}_{\mathbf{t}'m'}]}{\Pi^{\ell}_{\mathfrak{I}}{}^{{}_\iota\rho_T}[\xi;\,{}_\iota\bar{\xi}_T]\,\Pi_n[\xi;\bar{\xi}^{(n)}_o]}. \tag{7.1}$$

Then, according to (A.22) in Appendix A, the function

$$\phi_{*\mathbf{t}'*m'}[\xi\,|\,{}^{1-\ell}_{\quad\iota}\mathbf{B}_{\mathbf{t}m\downarrow\mathbf{t}'m'}] \equiv \frac{W\{\phi_{\mathbf{t}m}[\xi\,|\,{}^\ell_\iota\mathbf{B}_{\downarrow\mathbf{t}m;\mathbf{t}'m'}],\,\phi_{\mathbf{t}'m'}[\xi\,|\,{}^\ell_\iota\mathbf{B}_{\downarrow\mathbf{t}m;\mathbf{t}'m'}]\}}{{}_\iota\wp^{1/2}[\xi;\,{}_\iota T_K]\;\phi_{\mathbf{t}m}[\xi\,|\,{}^\ell_\iota\mathbf{B}_{\downarrow\mathbf{t}m;\mathbf{t}'m'}]} \tag{7.2}$$

is a solution of the partner RCSLE associated with the PFr beam ${}^{1-\ell}_{\quad\iota}\mathbf{B}_{\mathbf{t}m}$. Our observation that the latter solution also has an AEH form serves as the foundation for this series of studies on rational Liouville potentials quantized by GS Heine polynomials and their confluent counterparts. To prove this assertion, let us first express this solution in terms of the so-called 'polynomial determinant' (PD)

$$P_{n_{\dagger m}+n_{\dagger' m'}+1}[\xi \mid {}^{\ell}_{\iota}\mathbf{B}_{\downarrow \dagger m;\dagger' m'}; \dagger m; \dagger' m'] \tag{7.3}$$

$$\equiv \begin{vmatrix} \Pi_{n_{\dagger m}}[\xi; {}^{*}\bar{\xi}^{(n_{\dagger m})}_{\dagger m}] & \Pi_{n_{\dagger' m'}}[\xi; {}^{*}\bar{\xi}^{(n_{\dagger' m'})}_{\dagger' m'}] \\ {}_{\iota}P_{n_{\dagger m}+1}[\xi; \bar{\iota}; \bar{\lambda}_{\dagger m}; {}^{*}\bar{\xi}^{(n_{\dagger m})}_{\dagger m}] & {}_{\iota}P_{n_{\dagger' m'}+1}[\xi; \bar{\iota}; \bar{\lambda}_{\dagger' m'}; {}^{*}\bar{\xi}^{(n_{\dagger' m'})}_{\dagger' m'}] \end{vmatrix},$$

where

$${}_{\iota}P_{n_{\tilde{\dagger}\tilde{m}}+1}[\xi; \bar{\iota}; \bar{\lambda}_{\tilde{\dagger}\tilde{m}}; {}^{*}\bar{\xi}^{(n_{\tilde{\dagger}\tilde{m}})}_{\tilde{\dagger}\tilde{m}}] \tag{7.4}$$

$$\equiv \prod_{r=0}^{|\iota|} (\xi - {}_{\iota}e_r) \{\dot{\Pi}_{n_{\tilde{\dagger}\tilde{m}}}[\xi; {}^{*}\bar{\xi}^{(n_{\tilde{\dagger}\tilde{m}})}_{\tilde{\dagger}\tilde{m}}] + \Pi_{n_{\tilde{\dagger}\tilde{m}}}[\xi; {}^{*}\bar{\xi}^{(n_{\tilde{\dagger}\tilde{m}})}_{\tilde{\dagger}\tilde{m}}]\, ld \mid {}_{\iota}\Theta[\xi; {}_{\iota}\bar{\lambda}_{\tilde{\dagger}\tilde{m}}\,|\}$$

or alternatively, using (6.16),

$$\begin{aligned} {}_{\iota}P_{n_{\tilde{\dagger}\tilde{m}}+1}[\xi; \bar{\iota}; \bar{\lambda}_{\tilde{\dagger}\tilde{m}}; {}^{*}\bar{\xi}^{(n_{\tilde{\dagger}\tilde{m}})}_{\tilde{\dagger}\tilde{m}}] &= \hat{B}_1[\xi; \bar{\iota}; \bar{\lambda}_{\tilde{\dagger}\tilde{m}}]\Pi_{n_{\tilde{\dagger}\tilde{m}}}[\xi; {}^{*}\bar{\xi}^{(n_{\tilde{\dagger}\tilde{m}})}_{\tilde{\dagger}\tilde{m}}] \\ &+ \prod_{r=0}^{|\iota|} (\xi - {}_{\iota}e_r)\ \dot{\Pi}_{n_{\tilde{\dagger}\tilde{m}}}[\xi; {}^{*}\bar{\xi}^{(n_{\tilde{\dagger}\tilde{m}})}_{\tilde{\dagger}\tilde{m}}]. \end{aligned} \tag{7.4'}$$

The explicit expression of solution (7.2) in terms of PD (7.3):

$$\begin{aligned} \phi_{*\dagger' * m'}[\xi \mid {}^{\ell+1}_{\iota}\mathbf{B}^{K\Im 0}_{\dagger m \downarrow \dagger' m'}] &= \frac{2}{\sqrt{{}_{\iota}a_K}}\, {}_{\iota}\Theta[\xi; {}_{\iota}\bar{\lambda}_{\dagger' m'}] \\ &\times \frac{P_{n_{\dagger m}+n_{\dagger' m'}+1}[\xi \mid {}^{\ell}_{\iota}\mathbf{B}^{K\Im 0}_{\downarrow \dagger m;\dagger' m'}; \dagger m; \dagger' m']}{\Pi^{(1+\ell)\iota\rho_T}_{\Im}[\xi; {}_{\iota}\bar{\xi}_T]\Pi_{n_{\dagger m}}[\xi; {}^{*}\bar{\xi}^{(n_{\dagger m})}_{\dagger m}]\Pi_n[\xi; \bar{\xi}^{(n)}_o]} \end{aligned} \tag{7.5}$$

directly follows from conventional formula for Wroskians [see (A.25) in Appendix A] which gives

$$W\{\phi_{\dagger m}[\xi \mid {}^{\ell}_{\iota}\mathbf{B}_{\downarrow \dagger m;\dagger' m'}], \phi_{\dagger' m'}[\xi \mid {}^{\ell}_{\iota}\mathbf{B}_{\downarrow \dagger m;\dagger' m'}]\} \tag{7.6}$$

$$\begin{aligned} &= \phi_{\dagger m}[\xi \mid {}^{\ell}_{\iota}\mathbf{B}_{\downarrow \dagger m;\dagger' m'}]\ {}_{\iota}\Theta[\xi; {}_{\iota}\bar{\lambda}_{\dagger' m'}] \\ &\times \frac{P_{n_{\dagger m}+n_{\dagger' m'}+1}[\xi \mid {}^{\ell}_{\iota}\mathbf{B}_{\downarrow \dagger m;\dagger' m'}; \dagger m; \dagger' m']}{\Pi_{n_{\dagger m}}[\xi; {}^{*}\bar{\xi}^{(n_{\dagger m})}_{\dagger m}]\Pi^{\ell\,\iota\rho_T}_{\Im}[\xi; {}_{\iota}\bar{\xi}_T]\Pi_n[\xi; \bar{\xi}^{(n)}_o]}. \end{aligned}$$

Though solution (7.5) is reminiscent of (3.1) the numerator of the PFr in its right-hand side is not a Heine (*c*-Heine) polynomial yet for |ι|=1 (ι = 0) because the PFr beam ${}^{p+1}_{\iota}\boldsymbol{G}^{K\Im 0}_{\{\mathbf{t}m\}_p;\mathbf{t}'m}$ is nonsingular at any of the finite singular points of the PFr beam ${}^{p}_{\iota}\boldsymbol{G}^{K\Im 0}_{\{\mathbf{t}m\}_p\downarrow\mathbf{t}m}$ other than the end points of the quantization interval. (As mentioned in previous Section an accurate proof of this assertion will be provided in Part II.) We thus conclude that PD (7.3) must be divisible by $\Pi_n[\xi;\bar{\xi}_o^{(n)}]$. In addition, since the RCSLE associated with the PFr beam ${}^{2j+2}_{\iota}\boldsymbol{G}^{K\Im 0}_{\{\mathbf{t}m\}_{2j+1};\mathbf{t}'m}$ does not have singularities at the TP zeros ${}_{\iota}\xi_{T;k}$ the PD in question is also divisible by the TP. Thereby PD (7.3) can be decomposed as follows

$$P_{n_{\mathbf{t}m}+n_{\mathbf{t}'m'}+1}[\xi\,|\,{}^{\ell}_{\iota}\mathbf{B}^{K\Im 0}_{\downarrow\mathbf{t}m;\mathbf{t}'m'};\mathbf{t}m;\mathbf{t}'m'] \propto \Pi^{2(1-\ell)\,{}_{\iota}\rho_T}_{\Im}[\xi;{}_{\iota}\bar{\xi}_T]\Pi_n[\xi;\bar{\xi}_o^{(n)}]$$
$$\times Hi_{n_{\mathbf{t}m}+n_{\mathbf{t}'m'}-n+1-\ell K}[\xi\,|\,{}^{\ell}_{\iota}\mathbf{B}^{K\Im 0}_{\mathbf{t}m\downarrow\mathbf{t}'m'};\mathbf{t}'m'], \quad (7.7)$$

where the GS Heine polynomial in the right-hand side satisfies differential equation (3.29), with ${}^{\ell}_{\iota}\mathbf{B}_{\downarrow\mathbf{t}m}$ changed for ${}^{1-\ell}_{\iota}\mathbf{B}^{K\Im 0}_{\downarrow\mathbf{t}m;\mathbf{t}'m'}$. When deriving (7.7) we also took into account that

$$2\Im\,{}_{\iota}\rho_T = K \quad (7.8)$$

for any of three combinations ${}_{\iota}\rho_T = \tfrac{1}{2}$, $K = \Im = 1$ or $2$ and ${}_{\iota}\rho_T = 1$, $K = 2\Im = 2$.

Let us now prove that the CLDTs under consideration do not change the type of any AEH solution regular at one end, i.e., $^{*}\mathbf{t}' = \mathbf{t}'$ in the left-hand side of (7.5) if $\mathbf{t}' = \mathbf{a}$ or $\mathbf{b}$. First the solution regular at any finite endpoint ${}_{\iota}e_r$ of the quantization interval must remain regular at this singular point after the transformation because neither TP nor both polynomials $\Pi_n[\xi;\bar{\xi}_o^{(n)}]$ and $\Pi_{n_{\mathbf{t}m}}[\xi;{}^{*}\bar{\xi}^{(n_{\mathbf{t}m})}_{\mathbf{t}m}]$ forming AEH solution (3.1) are allowed to have zeros at ${}_{\iota}e_r$. (Remember that $\bar{\xi}_o^{(n)}$ stands for a set of outer singularities whereas the solution type $\mathbf{t}$ is selected under assumption that the polynomial $\Pi_{n_{\mathbf{t}m}}[\xi;{}^{*}\bar{\xi}^{(n_{\mathbf{t}m})}_{\mathbf{t}m}]$ does not vanish at any finite endpoint.) It has been proven in previous Section that the CLDTs do not change the ExpDiffs at the singular point ${}_{\iota}e_r$ so that ${}_{\iota}\rho_{\mathbf{t}'m';r}$ coincides with one of the ChExps of the latter singular point. Since AEH solution (7.1) is regular at ${}_{\iota}e_r$ the second ChExp must be smaller than ${}_{\iota}\rho_{\dagger'\mathrm{m}';r}$ so that the numerator of the PFr in the right-hand side of (7.5) may not vanish at the given endpoint. This confirms that the ChExp of any AEH solution (7.1) regular at ${}_{\iota}e_r$ is unaffected by the CLDT. On other hand, the AEH solution must remain irregular at the second end – otherwise it becomes an eigenfunction which would contradict to the fact that we deal with a pair of isospectral SUSY partners.

The assertion that any AEH solution regular at the infinite end of the quantization interval $0 < \zeta < +\infty$ for $\iota = 0$ preserves its type under the CLDTs of our interest is nearly trivial and we leave the rest of the proof to the reader.

In Part II we shall extend (7.5) to ratios of Krein discriminants [85]. Namely, it will be proven that the seed monomial products $\Pi_{\tilde{\mathrm{m}}}[\xi;\bar{\xi}_{\dagger\tilde{\mathrm{m}}}]$ and their polynomial supplements ${}_{\iota}P_{\tilde{\mathrm{m}}+1}[\xi;\bar{\iota};\bar{\lambda}_{\dagger\tilde{\mathrm{m}}};\bar{\xi}_{\dagger\tilde{\mathrm{m}}}]$ (scaled by some energy-dependent factors in both cases) form, respectively, odd and even rows of the PDs decomposed into the GS Heine polynomials according to (7.7).

By choosing ${}^{\ell+1}_{\iota}\mathbf{B}^{K\Im 0}_{\dagger\mathrm{m}\downarrow\dagger'\mathrm{m}'} = {}^{1}_{\iota}\boldsymbol{G}^{K\Im 0}_{\dagger\mathrm{m}\downarrow\dagger'\mathrm{m}'}$ we come to the expression for GS Heine polynomials

$$\mathrm{Hi}_{\mathrm{m}+\mathrm{m}'+\iota}[\xi\,|\,{}^{1}_{\iota}\boldsymbol{G}^{K\Im 0}_{\dagger\mathrm{m}\downarrow\dagger'\mathrm{m}'};\dagger'\mathrm{m}'] = \mathrm{Hi}_{\mathrm{m}+\mathrm{m}'+\iota}[\xi\,|\,{}^{1}_{\iota}\boldsymbol{G}^{K\Im 0}_{\dagger'\mathrm{m}'\downarrow\dagger\mathrm{m}};\dagger\mathrm{m}] \tag{7.9*}$$

$$\propto {}_{\iota}\hat{g}({}_{\iota}\lambda_{0;\dagger'\mathrm{m}'} - {}_{\iota}\lambda_{0;\dagger\mathrm{m}}, {}_{\iota}\lambda_{1;\dagger'\mathrm{m}'} - {}_{\iota}\lambda_{1;\dagger\mathrm{m}}; \bar{\xi}_{\dagger\mathrm{m}})\Pi_{\mathrm{m}'}[\xi;\bar{\xi}_{\dagger'\mathrm{m}'}] \tag{7.9}$$

$$= -\,{}_{\iota}\hat{g}({}_{\iota}\lambda_{0;\dagger\mathrm{m}} - {}_{\iota}\lambda_{0;\dagger'\mathrm{m}'}, {}_{\iota}\lambda_{1;\dagger\mathrm{m}} - {}_{\iota}\lambda_{1;\dagger'\mathrm{m}'}; \bar{\xi}_{\dagger'\mathrm{m}'})\Pi_{\mathrm{m}}[\xi;\bar{\xi}_{\dagger\mathrm{m}}] \tag{7.9'}$$

introduced by us in [19] for $\iota = 0$ and 1, where the so-called ‘generic Heine polynomial generator’ ($g$–HiPG) is defined for $\iota = 0$, 1, and $i$ via the generalized relation

$$ {}_{\iota}\hat{g}(\Delta\lambda, \Delta\nu; \bar{\xi}_{\dagger\mathrm{m}}) \equiv -\prod_{r=0}^{|\iota|}(\xi - {}_{\iota}e_r)\left[\Pi_{\mathrm{m}}[\xi;\bar{\xi}_{\dagger\mathrm{m}}]\frac{d}{d\xi} - \dot{\Pi}_{\mathrm{m}}[\xi;\bar{\xi}_{\dagger\mathrm{m}}]\right]$$

$$ - {}_{\iota}\widehat{B}_{1}[\xi;\bar{\iota};\Delta\lambda,\Delta\nu]\Pi_{\mathrm{m}}[\xi;\bar{\xi}_{\dagger,\mathrm{m}}]. \tag{7.10}$$

Note that we can reach Heine polynomial (7.9*) in two different way. This is a typical feature of the network GS Heine and GS $c$-Heine polynomials: the same polynomial can be reached via different paths using ordered sequences of GS solutions which differ from each other only by a single element.

In the particular case of basic FFs HPG (7.10) takes the form

$$_{\iota}\hat{g}(\Delta\lambda,\Delta\nu) \equiv -\prod_{r=0}^{|\iota|}(\xi - {}_{\iota}e_{r})\frac{d}{d\xi} - {}_{\iota}\hat{B}_{1}[\xi;\bar{\tau};\Delta\lambda,\Delta\nu]. \qquad (7.10^*)$$

It can be used as a uniform tool for constructing polynomial solutions of the differential equation

$$\left\{{}^{\ell}_{\iota}\hat{D}\{\bar{\rho}(\varepsilon\,|\,{}^{1}_{\iota}G^{K\Im 0}_{\mathbf{t}0};\bar{\sigma});\,{}_{\iota}\bar{\xi}_{T}\} + C_{\Im}[\xi;\varepsilon\,|\,{}^{1}_{\iota}G^{K\Im 0}_{\mathbf{t}0};\bar{\sigma}]\right\}F[\xi;\varepsilon\,|\,{}^{1}_{\iota}G^{K\Im 0}_{\mathbf{t}0};\bar{\sigma}] = 0, \qquad (7.11)$$

where the second-order differential operator

$$^{\ell}_{\iota}\hat{D}\{{}_{\iota}\bar{\rho}\,{}_{\iota}\bar{\xi}_{T}\} \equiv \prod_{r=0}^{|\iota|}(\xi - {}_{\iota}e_{r})\,\Pi_{\Im}[\,\xi;\,{}_{\iota}\bar{\xi}_{T}]\frac{d^{2}}{d\xi^{2}} + 2\,{}_{\iota}B_{1}[\xi;\bar{\tau};\,{}_{\iota}\bar{\rho}]\frac{d}{d\xi} \qquad (7.12)$$

has $\Im$+|ι| +1 singularities at the fixed points ${}_{\iota}e_{r}$ ($r$ = 0, |ι|), ${}_{\iota}\xi_{T;k}$ (k=1, $\Im$). The most remarkable feature of differential equation (7.11) is that positions of its singular points are independent of the RIs. If $\Im$=1 differential equation (7.11) turns into the Heun or *c*-Heun equation for ι = 1 or 0, accordingly. The appropriate DRtTP (K=2, $\Im$=1) and LTP (K= $\Im$=1) are analyzed in [34] and in Part III of this paper, respectively.

In the particular case of normalizable AEH solutions **t**m = **c**0 and **t**′m′ = **c**n (n > 0) and ι = 1 Heine polynomials (7.9) were originally discovered by Cooper, Ginocchio, and Khare [1] who presented an explicit expression for the Heine polynomial $\mathrm{Hi}_{n+1}[z\,|\,{}^{1}_{1}G^{220}_{\mathbf{c}0\downarrow\mathbf{c}n};\mathbf{c}n]$ (though with some misprints) in terms of hypergeometric polynomials. To relate our expression for the $\mathfrak{J}$S Heine polynomial $\mathrm{Hi}_{n+1}[z\,|\,{}^{1}_{1}G^{220}_{\mathbf{t}0\downarrow\mathbf{c}n};\mathbf{c}n]$ to the polynomial in curly brackets in the right-hand side of (4.4) in [1] let us first represent the monomial product $\Pi_{n}[z;\bar{z}_{\mathbf{c}n}]$ as

$$\Pi_{n}[z;\bar{z}_{\mathbf{c}n}] = \frac{({}_{1}\lambda_{0;\mathbf{c}n}+1)_{n}}{(-1)^{n}(\mu_{\mathbf{c}n}-n)_{n}}F(-n,\mu_{\mathbf{c}n}-n;\,{}_{1}\lambda_{0;\mathbf{c}n}+1;z). \qquad (7.13)$$

Substituting (7.13) into the definition of the $\mathfrak{J}$S Heine polynomial $\mathrm{Hi}_{n+1}[z\,|\,{}^{1}_{1}G^{220}_{\mathbf{t}0\downarrow\mathbf{c}n};\mathbf{c}n]$:

$$(\mu_{\dagger 0} - \mu_{\mathbf{c}n})\mathrm{Hi}_{n+1}[z \mid {}^{1}_{1}\mathcal{G}^{220}_{\dagger 0\downarrow \mathbf{c}n}; \mathbf{c}n]$$
$$= {}_{1}\hat{g}({}_{1}\lambda_{0;\mathbf{c}n} - {}_{1}\lambda_{0;\dagger 0}, {}_{1}\lambda_{1;\mathbf{c}n} - {}_{1}\lambda_{1;\dagger 0})\Pi_n[z; \bar{z}_{\mathbf{c}n}] \quad (7.14)$$

and making use of (15.2.1) in [91] we come to the following explicit expression of this Heine polynomial in terms of two hypergeometric polynomials

$$\frac{(-1)^n(\mu_{\dagger 0} - \mu_{\mathbf{c}n})(\mu_{\mathbf{c}n} - n)_n}{({}_{1}\lambda_{0;\mathbf{c}n} + 2)_{n-1}} \mathrm{Hi}_{n+1}[z \mid {}^{1}_{1}\mathcal{G}^{220}_{\dagger 0\downarrow \mathbf{c}n}; \mathbf{c}n] \quad (7.15)$$

$$= n({}_{1}\lambda_{0;\mathbf{c}n} + {}_{1}\lambda_{1;\mathbf{c}n} + n + 1)z(z-1)$$
$$\times F(1-n, \mu_{\mathbf{c}n} - n + 1; {}_{1}\lambda_{0;\mathbf{c}n} + 2; z)$$
$$+ \tfrac{1}{2}({}_{1}\lambda_{0;\mathbf{c}n} + 1)[({}_{1}\lambda_{0;\mathbf{c}n} - {}_{1}\lambda_{0;\dagger 0})(z-1) + ({}_{1}\lambda_{1;\mathbf{c}n} - {}_{1}\lambda_{1;\dagger 0})z]$$
$$\times F(-n, \mu_{\mathbf{c}n} - n; {}_{1}\lambda_{0;\mathbf{c}n} + 1; z).$$

Comparing the polynomial in the curly brackets in (4.4) in [1] with the right-hand side of (7.15) reveals several discrepancies – some (but not all) of them can be apparently eliminated simply by moving the left square bracket and placing $\beta_n - \beta_0$ in parentheses.

Let us now study more carefully transformation properties of $\mathfrak{J}$S Heine and $\mathfrak{L}$S (Laguerre-seed) *c*-Heine polynomials under the limiting transition to the DRtTP PFr beams for the SUSY partners of *r*-and *c*-GRef potentials. In case of PFr beams ${}^{1}_{1}\mathcal{G}^{220}_{\dagger 0}$ discussed above this would imply the limiting transition from the $\mathfrak{J}$S Heine and $\mathfrak{L}$S *c*-Heine polynomials $\mathrm{Hi}_{n+1}[\xi \mid {}^{1}_{\iota}\mathcal{G}^{220}_{\dagger 0\downarrow \mathbf{c}n}; \mathbf{c}n]$ for $\iota = 1$ or 0, respectively, to the Heun and *c*-Heun polynomials $\mathrm{Hp}_{n+1}[\xi \mid {}^{1}_{\iota}\mathcal{G}^{210}_{\dagger 0\downarrow \mathbf{c}n}; \mathbf{c}n]$.
First, keeping in mind that

$$\lim_{{}_{\iota}\xi_{T;1}, {}_{\iota}\xi_{T;2} \to {}_{\iota}\xi_T} \Pi_2^{2\, {}_{\iota}\rho_T}[\xi; {}_{\iota}\xi_{T;1}, {}_{\iota}\xi_{T;2}] \equiv \lim_{{}_{\iota}\xi_{T;1}, {}_{\iota}\xi_{T;2} \to {}_{\iota}\xi_T} \Pi_2[\xi; {}_{\iota}\xi_{T;1}, {}_{\iota}\xi_{T;2}]$$
$$= \Pi_1^2[\xi; {}_{\iota}\xi_T] \quad (7.16)$$

we conclude that

$$\lim_{{}_{\iota}\xi_{T;1},\,{}_{\iota}\xi_{T;2}\to{}_{\iota}\xi_T} \mathrm{Hi}_{n_{\mathbf{t}m}+n_{\mathbf{t}'m'}+1-n-2\ell}[\xi\,|\,{}^{\ell+1}_{\iota}\mathbf{B}^{220}_{\mathbf{t}m\downarrow\mathbf{t}'m'};\mathbf{t}'m']$$

$$= \mathrm{Hi}_{n_{\mathbf{t}m}+n_{\mathbf{t}'m'}-n+1-2\ell}[\xi\,|\,{}^{\ell+1}_{\iota}\mathbf{B}^{210}_{\mathbf{t}m\downarrow\mathbf{t}'m'};\mathbf{t}'m'].$$

To explicitly extract the Heine polynomials in the right-hand side of (7.7) from the PD on the left in the general case n > 0 it is convenient to represent differential equation (3.29) for the GS Heine polynomial

$$\mathrm{Hi}_{n_{\tilde{\mathbf{t}}\tilde{m}}}[\xi\,|\,{}^{\ell}_{\iota}\mathbf{B}^{K\Im 0}_{\downarrow\tilde{\mathbf{t}}\tilde{m}};\tilde{\mathbf{t}}\tilde{m}] \equiv \Pi_{n_{\tilde{\mathbf{t}}\tilde{m}}}[\xi;\,{}^{*}\bar{\xi}^{(n_{\tilde{\mathbf{t}}\tilde{m}})}_{\tilde{\mathbf{t}}\tilde{m}}] \tag{7.18}$$

as the following polynomial relation:

$$\Pi_n[\xi;\bar{\xi}^{(n)}_o]\Pi^{\ell}_{\Im}[\xi;{}_{\iota}\bar{\xi}_T]Y_{n_{\tilde{\mathbf{t}}\tilde{m}}}[\xi\,|\,{}^{\ell}_{\iota}\mathbf{B}_{\downarrow\tilde{\mathbf{t}}\tilde{m}};\tilde{\mathbf{t}}\tilde{m}] + \tfrac{1}{4}\breve{O}^{\downarrow}_{n+\ell\Im}[\xi\,|\,{}^{\ell}_{\iota}\mathbf{B}_{\downarrow\tilde{\mathbf{t}}\tilde{m}}]\Pi_{n_{\tilde{\mathbf{t}}\tilde{m}}}[\xi;\,{}^{*}\bar{\xi}^{(n_{\tilde{\mathbf{t}}\tilde{m}})}_{\tilde{\mathbf{t}}\tilde{m}}]$$

$$-\Omega_{n+\ell\Im-1}[\xi\,|\,{}^{\ell}_{\iota}\mathbf{B}_{\downarrow\tilde{\mathbf{t}}\tilde{m}}]P_{n_{\tilde{\mathbf{t}}\tilde{m}}+1}[\xi\,|\,{}^{\ell}_{\iota}\mathbf{B}_{\downarrow\tilde{\mathbf{t}}\tilde{m}};\tilde{\mathbf{t}}\tilde{m}] = 0 \tag{7.19}$$

between polynomials (7.4′) and the newly introduced auxiliary polynomials

$$Y_{n_{\tilde{\mathbf{t}}\tilde{m}}}[\xi\,|\,{}^{\ell}_{\iota}\mathbf{B}_{\downarrow\tilde{\mathbf{t}}\tilde{m}};\tilde{\mathbf{t}}\tilde{m}] \equiv \prod_{r=0}^{|\iota|}(\xi-{}_{\iota}e_r)\,\ddot{\Pi}_{n_{\tilde{\mathbf{t}}\tilde{m}}}[\xi;\,{}^{*}\bar{\xi}^{(n_{\tilde{\mathbf{t}}\tilde{m}})}_{\tilde{\mathbf{t}}\tilde{m}}] \tag{7.20}$$

$$+2\,B_1[\xi;\bar{\iota};\,{}_{\iota}\bar{\lambda}_{\tilde{\mathbf{t}}\tilde{m}}]\dot{\Pi}_{n_{\tilde{\mathbf{t}}\tilde{m}}}[\xi;\,{}^{*}\bar{\xi}^{(n_{\tilde{\mathbf{t}}\tilde{m}})}_{\tilde{\mathbf{t}}\tilde{m}}] + {}_{\iota}c_{\tilde{\mathbf{t}}\tilde{m}}\Pi_{n_{\tilde{\mathbf{t}}\tilde{m}}}[\xi;\,{}^{*}\bar{\xi}^{(n_{\tilde{\mathbf{t}}\tilde{m}})}_{\tilde{\mathbf{t}}\tilde{m}}],$$

where the first-order polynomials are defined via (6.17),

$${}_{\iota}c_{\tilde{\mathbf{t}}\tilde{m}} \equiv 2\rho_{0;\tilde{\mathbf{t}}\tilde{m}}[\delta_{|\iota|,1}\;{}_{\iota}\rho_{1;\tilde{\mathbf{t}}\tilde{m}} - \tfrac{1}{2}\delta_{\iota,0}\nu_{\tilde{\mathbf{t}}\tilde{m}}] + \tfrac{1}{4}\,{}_{\iota}d\;{}_{\iota}\varepsilon_{\tilde{\mathbf{t}}\tilde{m}}, \tag{7.21}$$

and

$$\Omega_{n+\ell\Im-1}[\xi\,|\,{}^{\ell}_{\iota}\mathbf{B}^{K\Im 0}] \equiv 2\{\Pi^{\ell}_{\Im}[\xi;{}_{\iota}\bar{\xi}_T]\dot{\Pi}_n[\xi;\bar{\xi}^{(n)}_o] + \ell\,{}_{\iota}\rho_T\Pi_n[\xi;\bar{\xi}^{(n)}_o]\,\dot{\Pi}_{\Im}[\xi;{}_{\iota}\bar{\xi}_T]\}. \tag{7.22}$$

By choosing $\tilde{m} = m$ and m′ in (7.19) and subtracting the resultant expressions we then come to the polynomial relation for PD (7.3)

$$\begin{aligned}\Omega_{n+\ell\Im-1}[\xi\,|\,{}^{\ell}_{\iota}\mathbf{B}^{K\Im 0}_{\downarrow\mathbf{t}m;\mathbf{t}'m'}]P_{n_{\mathbf{t}m}+n_{\mathbf{t}'m'}+1}[\xi\,|\,{}^{\ell}_{\iota}\mathbf{B}^{K\Im 0}_{\downarrow\mathbf{t}m;\mathbf{t}'m'};\mathbf{t}m;\mathbf{t}'m'] \\ = \Upsilon_{m+m'}[\xi\,|\,{}^{\ell}_{\iota}\mathbf{B}^{K\Im 0}_{\downarrow\mathbf{t}m;\mathbf{t}'m'};\mathbf{t}m;\mathbf{t}'m']\Pi_n[\xi;\bar{\xi}^{(n)}_o]\Pi^{\ell}_{\Im}[\xi;{}_{\iota}\bar{\xi}_T],\end{aligned} \tag{7.23}$$

where the polynomial of order $m+m'$ in the right-hand side is defined as follows

$$\Upsilon_{n_{\dagger m}+n_{\dagger' m'}}[\xi\,|\,{}^{\ell}_{\iota}\mathbf{B}_{\downarrow\dagger m;\dagger' m'};\dagger m;\dagger' m'] \tag{7.24}$$
$$\equiv \Pi_{n_{\dagger' m'}}[\xi;\,{}^{*}\bar{\xi}^{(n_{\dagger' m'})}_{\dagger' m'}]\,Y_{n_{\dagger m}}[\xi\,|\,{}^{\ell}_{\iota}\mathbf{B}_{\downarrow\dagger m;\dagger' m'};\dagger m]$$
$$-\,\Pi_{n_{\dagger m}}[\xi;\,{}^{*}\bar{\xi}^{(n_{\dagger m})}_{\dagger m}]\,Y_{n_{\dagger' m'}}[\xi\,|\,{}^{\ell}_{\iota}\mathbf{B}_{\downarrow\dagger m;\dagger' m'};\dagger' m'].$$

Since $\xi^{(n)}_{o;k} \neq {}_{\iota}\xi_{T;k'}$ for k=1,…,n and k′ =1 or ℑ polynomial (7.19) may not have zeros at either $\xi^{(n)}_{o;k}$ or ${}_{\iota}\xi_{T;k'}$ we thus confirm that PD (7.3) must be divisible by $\Pi_n[\xi;\bar{\xi}^{(n)}_{o}]$ as well as by TP (7.6) for $\ell=1$ as specified by decomposition (7.8) For the same reason polynomial (7.17) must be divisible by polynomial (7.22):

$$Hi_{n_{\dagger m}+n_{\dagger' m'}-n-2\ell+1}[\xi\,|\,{}^{1-\ell}_{\iota}\mathbf{B}^{220}_{\dagger m\downarrow\dagger' m'};\dagger m;\dagger' m']$$
$$=\frac{\Upsilon_{n_{\dagger m}+n_{\dagger' m'}}[\xi\,|\,{}^{\ell}_{\iota}\mathbf{B}^{220}_{\downarrow\dagger m;\dagger' m'};\dagger m;\dagger' m']}{\Omega_{n+2\ell-1}[\xi|{}^{\ell}_{\iota}\mathbf{B}^{220}_{\downarrow\dagger m;\dagger' m'}]}. \tag{7.25}$$

On other hand, setting K=2, ℑ=2 and $\ell\,{}_{\iota}\rho_{T}=1$ in (7.7) and substituting the resultant expression into the left-hand side of (7.23) we find

$$Hi_{n_{\dagger m}+n_{\dagger' m'}-n-1}[\xi\,|\,{}^{0}_{\iota}\mathbf{B}^{210}_{\dagger m\downarrow\dagger' m'};\dagger m;\dagger' m']$$
$$=\frac{\Upsilon_{n_{\dagger m}+n_{\dagger' m'}}[\xi\,|\,{}^{1}_{\iota}\mathbf{B}^{210}_{\downarrow\dagger m;\dagger' m'};\dagger m;\dagger' m']}{(\xi-{}_{\iota}\xi_{T})\Omega_{n}[\xi|{}^{1}_{\iota}\mathbf{B}^{210}_{\downarrow\dagger m;\dagger' m'}]}. \tag{7.25*}$$

Taking into account that

$$\lim_{{}_{\iota}\xi_{T;1},\,{}_{\iota}\xi_{T;2}\to{}_{\iota}\xi_{T}} \Omega_{n+1}[\xi|{}^{1}_{\iota}\mathbf{B}^{220}] = (\xi-\xi_{T})\Omega_{n}[\xi|{}^{1}_{\iota}\mathbf{B}^{210}] \tag{7.26}$$

we thus conclude that Heine polynomials associated with the DRtTP potential $V[\xi\,|\,{}_{\iota}\mathbf{B}^{210}]$ can be obtained from the generic ones (K=ℑ=2) via limiting transition (7.10) iff

$$\lim_{\iota\xi_{T;1},\iota\xi_{T;2}\to\iota\xi_T} \mathrm{Hi}_{n_{\mathbf{t}m}+n_{\mathbf{t}'m'}-n-1}[\xi\,|\,{}^{0}_{\iota}\mathbf{B}^{220}_{\mathbf{t}m\downarrow\mathbf{t}'m'};\mathbf{t}m;\mathbf{t}'m'] = \mathrm{Hi}_{n_{\mathbf{t}m}+n_{\mathbf{t}'m'}-n-1}[\xi\,|\,{}^{0}_{\iota}\mathbf{B}^{210}_{\mathbf{t}m\downarrow\mathbf{t}'m'};\mathbf{t}m;\mathbf{t}'m']. \quad (7.27)$$

It should be stressed that the presented arguments do not assure that the limit in question exists. As already shown in [34] one sequence of AEH solutions disappears as TP discriminant becomes equal to 0.

Note that the same Heine polynomial can serve as a component for AEH solutions of different types depending on the path used to construct the appropriate solved-by-polynomials differential equation. For example, if $\mathbf{t}m$ is a regular ($\mathbf{t}$ = $\mathbf{a}$ or $\mathbf{b}$) AEH solution below the ground energy level of the potential $V[\xi|{}^{\ell}_{\iota}\mathbf{B}_{\downarrow\mathbf{t}m}]$, with ι=1 or 0, then the Heine polynomial $\mathrm{Hi}_{v+m-2\ell+1}[\xi\,|\,{}^{1-\ell}_{\iota}\mathbf{B}_{\mathbf{t}m\downarrow\mathbf{c}v};\mathbf{t}m;\mathbf{c}v]$ describes the $v^{th}$ bound energy state in the potential $V[\xi\,|\,{}^{1-\ell}_{\iota}\mathbf{B}_{\mathbf{c}v\downarrow\mathbf{t}m}]$. On other hand, the same polynomial appears in an irregular AEH solution of the RCSLE with the Liouville potential $V[\xi\,|\,{}^{1-\ell}_{\iota}\mathbf{B}_{\mathbf{c}v\downarrow\mathbf{t}m}]$ having a singularity at each zero of the eigenfunction $\mathbf{c}v$ for the potential $V[\xi|{}^{\ell}_{\iota}\mathbf{B}_{\downarrow\mathbf{c}v}]$.

**8. SUSY Pairs of Fuschian equations with polynomial coefficients**

The purpose of this Section is to derive explicit relations between the energy-dependent coefficients $\alpha(\varepsilon\,|\,{}^{\ell}_{\iota}\mathbf{B}^{2\Im 0}_{\downarrow\mathbf{t}m};\bar{\sigma})$ and $\alpha(\varepsilon\,|\,{}^{\ell+1}_{\iota}\mathbf{B}^{2\Im 0}_{\mathbf{t}m}]$ as well as similar relations between the coefficients $\beta(\varepsilon\,|\,{}^{\ell}_{\iota}\mathbf{B}^{2\Im 0}_{\downarrow\mathbf{t}m};\bar{\sigma})$ and $\beta(\varepsilon\,|\,{}^{\ell+1}_{\iota}\mathbf{B}^{2\Im 0}_{\mathbf{t}m}]$ for |ι| = 1. We only discuss PFr beams generated using second-order TPs (K=2). An extension of this analysis to the LP potentials $V[z\,|\,{}_{1}\mathbf{B}^{110}]$ will be presented in Part III. Likewise we postpone the discussion of the shape-invariant limiting case of the latter potential (K=0) represented by the RM potential $V[z\,|\,{}_{1}\boldsymbol{G}^{000}]$.

Making use of generic representations (A.19) and (A.19*) for the RefPFrs of the regular ($|\iota|$=1) PFr beam ${}^{0}_{\iota}\mathbf{B}^{2\Im 0}_{\downarrow\dagger m}$ and its SUSY partner ${}^{1}_{\iota}\mathbf{B}^{2\Im 0}_{\dagger m}$ and comparing the asymptotics of AEH solutions (3.1) and (6.1) at large $\xi$,

$$\phi_{\dagger m}[\xi \mid {}^{0}_{\iota}\mathbf{B}^{2\Im 0}_{\downarrow\dagger m}] \sim \xi^{\frac{1}{2}({}_{\iota}\lambda_{0;\dagger m} + {}_{\iota}\lambda_{1;\dagger m}) + n_{\dagger m} - n + 1} \tag{8.1}$$

and

$$\phi_{*\dagger n}[\xi \mid {}^{1}_{\iota}\mathbf{B}^{2\Im 0}_{\dagger m}] \sim \xi^{-\frac{1}{2}({}_{\iota}\lambda_{0;\dagger m} + {}_{\iota}\lambda_{1;\dagger m}) - m + n} \tag{8.1*}$$

respectively, we conclude that

$$\lim_{\xi\to\infty} \{\xi^2 I^o[\xi \mid {}^{1}_{\iota}\mathbf{B}^{2\Im 0}_{\dagger m}]\} = \lim_{\xi\to\infty} \{\xi^2 I^o[\xi \mid {}^{0}_{\iota}\mathbf{B}^{2\Im 0}_{\downarrow\dagger m}]\} \tag{8.2}$$

and therefore

$$\begin{aligned} \tfrac{1}{4}[\hat{O}^{\downarrow}_{n_{\dagger m}+\Im; n_{\dagger m}+\Im} \mid {}^{1}_{\iota}\mathbf{B}^{2\Im 0}_{\dagger m}) - \hat{O}^{\downarrow}_{n;n} \mid {}^{0}_{\iota}\mathbf{B}^{2\Im 0}_{\downarrow\dagger m})] \\ = 2 \lim_{\xi\to\infty} \{\xi^2 \big(\hat{Q}[\xi; \bar{\xi}^{(n)}_{o}] - \hat{Q}^{2\Im 0}[\xi; {}^{*}_{\iota}\bar{\xi}^{(n_{\dagger m})}_{\dagger m}; {}_{\iota}\bar{\xi}_{T}]\big)\}. \end{aligned} \tag{8.3}$$

Substituting

$$2 \lim_{\xi\to\infty} \{\xi^2 \hat{Q}[\xi; \bar{\xi}^{(n)}_{o}\} = -n(n+1) \tag{8.4}$$

and

$$2 \lim_{\xi\to\infty} \{\xi^2 \hat{Q}^{K\Im 0}[\xi; {}^{*}_{\iota}\bar{\xi}^{(n_{\dagger m})}_{\dagger m}; {}_{\iota}\bar{\xi}_{T})]\} = -(n_{\dagger m} + \Im\, {}_{\iota}\rho_{T})(n_{\dagger m} + \Im\, {}_{\iota}\rho_{T} + 1) \tag{8.4*}$$

thus gives

$$\begin{aligned} \tfrac{1}{4}[\hat{O}^{\downarrow}_{n_{\dagger m}+\Im; n_{\dagger m}+\Im} \mid {}^{1}_{\iota}\mathbf{B}^{2\Im 0}_{\dagger m}) - \hat{O}^{\downarrow}_{n;n} \mid {}^{0}_{\iota}\mathbf{B}^{2\Im 0}_{\downarrow\dagger m})] \\ = (n_{\dagger m} + 1)(n_{\dagger m} + 2) - n(n+1). \end{aligned} \tag{8.5}$$

Keeping in mind that the leading coefficient of first-order polynomial (3.28) is equal to

$$\Xi(\varepsilon \mid {}_{\iota}\boldsymbol{G}^{K\Im 0}; \bar{\sigma}) \equiv \rho_0(\varepsilon \mid {}_{\iota}\boldsymbol{G}^{K\Im 0}; \bar{\sigma}) + \rho_1(\varepsilon \mid {}_{\iota}\boldsymbol{G}^{K\Im 0}; \bar{\sigma}) \tag{8.6}$$

for |ι|=1, it directly follows from (3.27) that

$$C_{n;n}(\varepsilon\,|\,{}^{0}_{\iota}\mathbf{B}^{K\Im 0};\bar{\sigma}) = \tfrac{1}{4}[\hat{O}^{\downarrow}_{n;n}|\,{}^{0}_{\iota}\mathbf{B}^{K\Im 0}) + {}_{\iota}d\,\varepsilon] \quad (8.7)$$
$$+ 2\rho_0(\varepsilon\,|\,{}_{\iota}\boldsymbol{G}^{K\Im 0};\bar{\sigma})\,\rho_1(\varepsilon\,|\,{}_{\iota}\boldsymbol{G}^{K\Im 0};\bar{\sigma}) - 2n\,\Xi(\varepsilon\,|\,{}_{\iota}\boldsymbol{G}^{K\Im 0};\bar{\sigma}).$$

Similarly, using again (3.27), but with n changed for m, one finds

$$C_{n_{\dagger m}+\Im;n_{\dagger m}+\Im}(\varepsilon\,|\,{}^{1}_{\iota}\mathbf{B}^{K\Im 0}_{\dagger m};\bar{\sigma}] = \tfrac{1}{4}[\hat{O}^{\downarrow}_{n_{\dagger m}+\Im;n_{\dagger m}+\Im}|\,{}^{1}_{\iota}\mathbf{B}^{K\Im 0}_{\dagger m}) + \tfrac{1}{4}\,{}_{\iota}d\varepsilon] \quad (8.7^*)$$
$$+2\rho_0(\varepsilon\,|\,{}_{\iota}\boldsymbol{G}^{K\Im 0}_{\downarrow\dagger m};\bar{\sigma})\;\rho_1(\varepsilon\,|\,{}_{\iota}\boldsymbol{G}^{K\Im 0}_{\downarrow\dagger m};\bar{\sigma}) - 2\Xi(\varepsilon\,|\,{}_{\iota}\boldsymbol{G}^{K\Im 0}_{\downarrow\dagger m};\bar{\sigma})(m + \Im\,{}_{\iota}\rho_T),$$

Subtracting (8.7) from (8.7*) and substituting (8.5) into the difference gives

$$C_{n_{\dagger m}+\Im;n_{\dagger m}+\Im}(\varepsilon\,|\,{}^{1}_{\iota}\mathbf{B}^{2\Im 0}_{\dagger m};\bar{\sigma}) = C_{n;n}(\varepsilon\,|\,{}^{0}_{\iota}\mathbf{B}^{2\Im 0}_{\downarrow\dagger m};\bar{\sigma}) \quad (8.8)$$
$$-(n_{\dagger m} - n + 1)[2\Xi(\varepsilon\,|\,{}_{\iota}\boldsymbol{G}^{2\Im 0}_{\downarrow\dagger m};\bar{\sigma}) + n_{\dagger m} + n + 1].$$

Making use of (4.1), with $\ell$ set to 0, we can alternatively represent (8.8) as

$$\alpha(\varepsilon\,|\,{}^{1}_{\iota}\mathbf{B}^{2\Im 0}_{\dagger m};\bar{\sigma})\;\beta(\varepsilon\,|\,{}^{1}_{\iota}\mathbf{B}^{2\Im 0}_{\dagger m};\bar{\sigma}) \quad (8.8^{\dagger})$$
$$= \left[\alpha(\varepsilon\,|\,{}^{0}_{\iota}\mathbf{B}^{2\Im 0}_{\downarrow\dagger m};\bar{\sigma}) + n - n_{\dagger m} - 1\right] \times \left[\beta(\varepsilon\,|\,{}^{0}_{\iota}\mathbf{B}^{2\Im 0}_{\downarrow\dagger m};\bar{\sigma}) + n - n_{\dagger m} - 1\right].$$

Setting $\ell$ =1 in (4.3*) and changing n for m shows that

$$\alpha(\varepsilon\,|\,{}^{1}_{\iota}\mathbf{B}^{2\Im 0}_{\dagger m};\bar{\sigma}) + \beta(\varepsilon\,|\,{}^{1}_{\iota}\mathbf{B}^{2\Im 0}_{\dagger m};\bar{\sigma}) = 2\Xi(\varepsilon\,|\,{}_{\iota}\boldsymbol{G}^{2\Im 0}_{\downarrow\dagger m};\bar{\sigma}) \quad (8.9)$$
$$- 2\Im\,{}_{\iota}\rho_T - 2n_{\dagger m} - 1$$

so that

$$\alpha(\varepsilon\,|\,{}^{1}_{\iota}\mathbf{B}^{2\Im 0}_{\dagger m};\bar{\sigma}) + \beta(\varepsilon\,|\,{}^{1}_{\iota}\mathbf{B}^{2\Im 0}_{\dagger m};\bar{\sigma}) \quad (8.9^{\dagger})$$
$$= \alpha(\varepsilon\,|\,{}^{0}_{\iota}\mathbf{B}^{2\Im 0}_{\downarrow\dagger m};\bar{\sigma}) + \beta(\varepsilon\,|\,{}^{0}_{\iota}\mathbf{B}^{2\Im 0}_{\downarrow\dagger m};\bar{\sigma}) + 2n - 2n_{\dagger m} - 2.$$

Comparing ($8.8^{\dagger}$) and ($8.9^{\dagger}$) we finally conclude that

$$\alpha(\varepsilon\,|\,{}^{1}_{\iota}\mathbf{B}^{2\Im0}_{\dagger m};\bar{\sigma}) = \alpha(\varepsilon\,|\,{}^{0}_{\iota}\mathbf{B}^{2\Im0}_{\downarrow\dagger m};\bar{\sigma}) + n - n_{\dagger m} - 1 \tag{8.10a}$$

and

$$\beta(\varepsilon\,|\,{}^{1}_{\iota}\mathbf{B}^{2\Im0}_{\dagger m};\bar{\sigma}) = \beta(\varepsilon\,|\,{}^{0}_{\iota}\mathbf{B}^{2\Im0}_{\downarrow\dagger m};\bar{\sigma}) + n - n_{\dagger m} - 1. \tag{8.10b}$$

Again selection of the appropriate root of quadratic equation (4.2) is made via the condition

$$\beta(\varepsilon\,|\,{}^{1}_{\iota}\mathbf{B}^{2\Im0}_{\dagger m};\bar{\sigma}) - \alpha(\varepsilon\,|\,{}^{1}_{\iota}\mathbf{B}^{2\Im0}_{\dagger m};\bar{\sigma}) = \beta(\varepsilon\,|\,{}^{0}_{\iota}\mathbf{B}^{2\Im0}_{\downarrow\dagger m};\bar{\sigma}) - \alpha(\varepsilon\,|\,{}^{0}_{\iota}\mathbf{B}^{2\Im0}_{\downarrow\dagger m};\bar{\sigma}) > 0. \tag{8.11}$$

The derived transformation properties of the energy-dependent α- and β-parameters for PFr beams generated using second-order TP allow us to confirm that the CLDT in question converts a GS Heine polynomial of order $n_{\dagger' m'}$ into the partner polynomial of order $n_{\dagger' m'} + n_{\dagger m} - n - 2\ell + 1$, in agreement with (7.18) and (7.18*).

It has been demonstrated in subsection 5.1 that AEH solutions $\dagger m$ of Class β introduced in do exist in the subdomain $A_0$ of the r-GRef PFr beam ${}_{1}\mathcal{G}^{2\Im0}_{\downarrow \mathbf{b} m}$ at least if ${}_{1}c_{0} > 1$ and the TP leading coefficient ${}_{1}a_{2}$ is chosen to be sufficiently small depending on the value of m and therefore the solution $\phi_{\mathbf{b} m}[\xi\,|\,{}^{1}_{\iota}\mathcal{G}^{K\Im0}_{\dagger 0 \downarrow \mathbf{b} m}]$, with RIs inside the subdomain $A_0$, must also belong to Class β under the same constraints imposed on the TP coefficients.

## 9. Conclusions and further developments

The presented study examines the most important features of the theory of rational multi-step SUSY partners of three families of GRef potentials algebraically quantized by *classical* Jacobi (ι=1), *classical* generalized Laguerre (ι=0), and Romanovski-Routh (ι=*i*) polynomials with generally energy-dependent indexes. The preview is done under the assumption that the appropriate RCSLEs has only second-order poles. Each new rational potential in the given ladder is obtained via the CLDT with an AEH FF. Making use of the corresponding gauge transformation we converted the given RCSLE into the solved-by-polynomials equation. As discussed in detail in Section 4 the latter equation has only regular singular points (including infinity) for $\iota = 1$ or $i$ and therefore belongs to the class of second-order Fuschian equations. In Section 8 we derived the recurrence relations for the energy-dependent $\alpha$- and $\beta$-coefficients whose product specifies the free term of each equation.

A serious obstacle to extending the outlined procedure to multistep CLDTs comes from the fact that the second-order poles associated with outer singularities disappear at each step and it remains unclear whether the resultant RCSLE still has first-order poles at these points or these outer singularities simply go away. In Part II we will take advantage of the alternative approach based on the Crum-Krein transformation [99, 85] to confirm that single-step CLDTs indeed eliminate the mentioned singularities so that the new RCSLE has only second-order poles as stated above. Based on the analysis presented in Section 3 this implies that rational potentials constructed in such a way are conditionally exactly quantized by GS Heine or GS *c*-Heine polynomials.

It will be also shown that each AEH solution of our interest can be represented as the product of a positive energy-dependent weight and a ratio of two polynomial determinants formed by Jacobi ($\iota$=1), generalized Laguerre ($\iota$=0), and Routh ($\iota$=*i*) polynomials. In the particular case of multi-step SUSY partners of the RM ($\iota$=1) and Morse ($\iota$=0) potentials rational potentials generated via CLDTs with nodeless FFs form multi-index families of solvable potentials recently sketched by Odake and Sasaki in [100, 89]. Their analysis specifically took advantage of the fact that GS solutions in question can be represented as explicit functions of the variable x used to define the potential in the Schrödinger equation. As a result the terminology used in [100, 89] is strictly controlled by quantum-mechanical applications. For example GS solutions regular at least at one of the end points of the quantization interval are referred to as 'virtual state wavefunctions' whereas GS solutions of type **d** are called 'pseudo- virtual state wavefunctions'. We shall come back to a more detailed discussion of the cited works in Part III specifically dealing with these two limiting cases of the LTP (K=1) potentials $V[\xi \mid {}_{\iota}\boldsymbol{G}^{1\Im 0}]$.

We moved a study on SUSY ladders of the LTP potentials into a separate paper because they have some specific features which do not fit the general pattern for the generic *r*- and

*c*-GRef potentials $V[\xi \mid {}_{\iota}G^{2\Im 0}]$ generated using second-order TPs. For this reason we also postponed any study on its shape-invariant limiting cases (already addressed in the literature [82, 14, 15, 101, 100, 89, 90]).

An analysis of the LTP *r*- and c-GRef potentials is significantly simplified by the fact that energies of AEH solutions are determined by roots of quadratic (instead of quartic) equations so that we can directly formulate constraints selecting regular AEH solutions below the ground-energy level. When the TPs turn into constants the derived constraints become equivalent to the parameter ranges obtained by Quesne [14, 15] for the RM ($\iota$=1) and Morse ($\iota$=0) potentials. This would present a convenient opportunity to more precisely relate Quesne's works to our general approach. Without going into details let us only mention two other astounding attributes of the LTP *r*- and c-GRef potentials:

i) their single-step SUSY partners constructed by means of CLDTs with one of four basic FFs exactly quantized in terms of Heun or *c*-Heun polynomials for $\iota$=1 or 0, respectively;
ii) both *r*- and c-GRef potentials preserve their form under some double-step CLDTs with basic GS solutions.

One of the most important results of this study on Hi-CEQ potentials *on line* is that the CLDTs with AEH FFs preserve ChExps at the common finite singular points ${}_{\iota}e_{r}$, i. e., at finite end points of the quantization interval for *r*- and *c*-GRef potentials and at ${}_{\iota}e_0 = -i$, ${}_{\iota}e_1 = +i$ for the Milson potential [22]. The only exception from this rule is the Gendenshtein (Scarf II) potential [3] which is constructed using the TP with zeros at the singular points $-i$ and $+i$ of the given RCSLE. To a large extent its rational SUSY partners (see Appendix D for examples) are reminiscent of radial Hi-CEQ potentials analyzed in a separate series of publications [19, 33, 51].) The remarkable feature of this exceptional family of rational potentials on the line is that ChExps at the singular points $-i$ and $+i$ of the RCSLE are energy-independent and as result each of the mentioned SUSY partners is quantized by a finite set of orthogonal polynomials.

In subsection 5.2 we showed that the CT0 branch of the *c*-GRef potential (having a Coulomb tail approaching 0 at $+\infty$) exhibits many features similar to the *r*-GRef potential. However, since this branch has an infinite number of bound energy states there are infinitely many AEH solutions of each type. It has been proven that any AEH solution from the primary sequence $\mathbf{a}$m lies below the ground energy level and therefore can be used a seed for multi-step CLDTs.

Let us end this brief survey of the outcome of our study on implicit Hi-CEQ potentials by making some concluding comments on the so-called ‘shape-invariant’ GRef potentials generated by means of TPs which zeros (if any) coincide with singular points of the appropriate RCSLE. As a result the CLDTs with basic FFs $\mathbf{t}$0 preserve the form of these potentials [19, 51] as it initially discovered by Gendenshtein [3] in the particular case of DTs with the FFs $\mathbf{c}$0. Another important consequence of this distinctive attribute of the TPs used to construct ‘shape-invariant’ GRef potentials is that the variable ξ(x) obtained by integration of first-order differential equation (2.7) can be written in an explicit form [2]. As a result the Schwartz derivative $\{{}_\iota\xi,x\}$ has a much simpler form compared with its expression for the generic GRef potentials. This *convenient* feature of the RM (ι=1) and Morse (ι=0) potentials has been explicitly exploited in an extension [103, 13, 81, 101, 90] of the SUSY theory of RLPs to the Riccati-Schrödinger equation expressed in terms of the new variable ξ(x):

$$\begin{aligned} {}_\iota\wp^{-1/2}[\xi;\,{}_\iota T_K]\,\dot{\omega}[\xi;\varepsilon\,|\,{}_\iota\mathbf{B}] - \omega^2[\xi;\varepsilon\,|\,{}_\iota\mathbf{B}] \\ = \varepsilon + {}_\iota\wp^{-1}[\xi;\,{}_\iota T_K]\,I^{o}[\xi\,|\,{}_\iota\mathbf{B}] + \tfrac{1}{2}\{{}_\iota\xi, x\}, \end{aligned} \qquad (9.1)$$

where

$$\omega[\xi;\varepsilon\,|\,{}_\iota\mathbf{B}] \equiv -ld\,\Phi[\xi;\varepsilon\,|\,{}_\iota\mathbf{B}]. \qquad (9.2)$$

As a rule we also tried to avoid references to the rapidly expanding literature on the exceptional orthogonal polynomials other than some historical remarks at the very beginning of this paper. It should be stressed that, contrary to the statement made in Introduction in [90], eigenfunctions for rational multi-step SUSY partners of GRef potentials discussed in this paper (including SUSY partners of the RM and Morse potentials) are not expressible in terms of orthogonal polynomials of any known type (except possibly some very exotic case, yet to be found).

While postponing a more accurate analysis of exceptional orthogonal polynomials for a separate publication let us only mention that it was Ho, Odake, and Sasaki [104] who first noticed that the $X_m$ Jacobi polynomials are global solutions of the Fuschian equation with m+3 regular singular points. This observation brought them to the Heine-Stieltjes theorem [52, 105, 53] for polynomial solutions of a Fuschian equation with rational coefficients (the Heine-type equation in our terms). Since ChExps of finite singular points as well as positions of the singular points are energy-independent in this particular case the $X_m$ Jacobi polynomials can be obtained simply by varying the energy parameter in the free term. This implies that they form a subset of the given set of Heine polynomials. (The $X_m$ Jacobi polynomials are certainly not Heine-Stieltjes polynomials since all the ChExps in the latter case are required to be positive [105, 53].)

**Appendix A**

**SUSY pairs of the Sturm-Liouville Equations Written in Canonical Form**

Let us briefly review some overall features of the SLE written in the 'normal' [23, 24] or 'canonical' [22] form (CSLE)

$$\left\{ \hat{d}_\xi^2 + I^o[\xi; Q^o] + \varepsilon \wp[\xi] \right\} \Phi[\xi; \varepsilon; Q^o] = 0, \tag{A.1}$$

where

$$\hat{d}_\xi \equiv \frac{d}{d\xi} \tag{A.2}$$

and the density function $\wp[\xi]$ is assumed to be positive inside the given quantization interval. The LT [24-26] using the change of variable

$$x = \int d\xi \, \wp^{1/2}[\xi] \;, \tag{A.3}$$

converts CSLE (A.1) into the Schrödinger equation:

$$\left\{\frac{d^2}{dx^2} + V_L[\xi(x);Q^o] + \varepsilon\right\}\Psi[\xi(x);\varepsilon;Q^o] = 0 \tag{A.4}$$

with the Liouville potential

$$V_L(x;Q^o) = V_L[\xi(x);Q^o] = -\wp^{-1}[\xi(x)]\, I^o[\xi(x);Q^o] - \tfrac{1}{2}\{\xi, x\}, \tag{A.5}$$

where the symbol $\{\xi, x\}$ denotes the so-called Schwartz derivative (see, i.g., [60]). General solutions of equations (A.1) and (A.4) are related in a simple manner:

$$\Psi[\xi;\varepsilon;Q^o] = \wp^{1/4}[\xi]\Phi[\xi;\varepsilon;Q^o]. \tag{A.6}$$

In this paper we only consider reflective Liouville potentials with the zero point energy chosen in such a way that

$$V_L(-\infty;Q^o) \geq V_L(+\infty;Q^o) = 0. \tag{A.7}$$

Let

$$\psi_\tau(x;Q^o_{\downarrow\tau}) \equiv \psi_\tau[\xi(x);Q^o_{\downarrow\tau}] = \wp^{1/4}[\xi(x)]\phi_\tau[\xi(x);Q^o_{\downarrow\tau}] \tag{A.8}$$

be a solution of the Schrödinger equation at the energy $\varepsilon = \varepsilon_\tau$, with subscript $\downarrow\tau$ indicating that the solution in question generally exists only in a sub-domain of the space formed by parameters $Q^o$. FF (A.8) is assumed to satisfy the following boundary condition at $+\infty$:

$$\lim_{x\to+\infty} ld\,|\psi_\tau(x;Q^o_{\downarrow\tau})| = \sigma_{1;\tau}\sqrt{-\varepsilon_\tau} \qquad (\sigma_{1;\tau} = +\text{ or }-), \tag{A.9}$$

where the symbol *ld* stands for the logarithmic derivative. Substituting (A.7) and (A.9) into the Ricatti equation

$$\frac{d}{dx} ld\,|\psi_\tau(x;Q^o_{\downarrow\tau})| + ld^2\,|\psi_\tau(x;Q^o_{\downarrow\tau})|] = V_L(x;Q^o_{\downarrow\tau}) - \varepsilon_\tau \tag{A.10}$$

then gives

$$\lim_{x\to+\infty} \frac{d}{dx} ld\,|\psi_\tau(x\,|\,Q^o_{\downarrow\tau})| = 0. \tag{A.11}$$

By expressing the conventional Darboux operator in terms of the variable $\xi$ (x):

$$\hat{D}\{\psi_\tau[\xi;Q^o_{\downarrow\tau}]\} = \wp^{-1/2}[\xi]\ \hat{d}\{\psi_\tau[\xi;Q^o_{\downarrow\tau}]\}, \tag{A.12}$$

where

$$\hat{d}\{\chi[\xi]\} \equiv \hat{d}_\xi - ld\ \chi[\xi]\ , \tag{A.13}$$

one can factorize CSLE (A.1) and the partner equation for the function

$$*\Phi[\xi;\varepsilon;Q^o_{\downarrow\tau}] = \hat{L}_\tau \Phi[\xi;\varepsilon;Q^o_{\downarrow\tau}] \tag{A.14}$$

as

$$(*\hat{L}_\tau\ \hat{L}_\tau + \varepsilon_\tau - \varepsilon)\,\Phi[\xi;\varepsilon;Q^o_{\downarrow\tau}] = 0 \tag{A.15}$$

and

$$(\hat{L}_\tau *\hat{L}_\tau + \varepsilon_\tau - \varepsilon)*\Phi[\xi;\varepsilon;Q^o_{\downarrow\tau}] = 0, \tag{A.15*}$$

with $\hat{L}_\tau$ and $*\hat{L}_\tau$ standing for the 'canonical Liouville-Darboux' operators (CLDO)

$$\hat{L}_\tau \equiv \hat{D}\{\phi_\tau[\xi;Q^o_{\downarrow\tau}]\} \tag{A.16a}$$

$$= \wp^{1/4}[\xi]\hat{D}\{\psi_\tau[\xi;Q^o_{\downarrow\tau}]\}\wp^{-1/4}[\xi] \tag{A.16b}$$

and

$$*\hat{L}_\tau \equiv \hat{D}\{*\phi_\tau[\xi;Q^o_{\downarrow\tau}]\} \tag{A.16a*}$$

$$= \wp^{1/4}[\xi]\hat{D}\{\psi_\tau^{-1}[\xi;Q^o_{\downarrow\tau}]\}\wp^{-1/4}[\xi]. \tag{A.16b*}$$

Here the FFs $\phi_\tau[\xi;Q^o_{\downarrow\tau}]$ and $*\phi_\tau[\xi;Q^o_{\downarrow\tau}]$ are related via Suzko's reciprocal formula [39-41]:

$$*\phi_\tau[\xi;Q^o_{\downarrow\tau}] = \wp^{-1/2}[\xi]/\phi_\tau[\xi;Q^o_{\downarrow\tau}]. \tag{A.17}$$

which plays a pivotal role in our analysis of characteristic exponents associated with regular singular points in the partner CSLE. Taking into account that

$$\hat{D}\{*\phi_\tau[\xi;Q^o_{\downarrow\tau}]\}\wp^{-1/2}[\xi] = \wp^{-1/2}[\xi]\hat{D}\{\phi_\tau^{-1}[\xi;Q^o_{\downarrow\tau}]\} \tag{A.18}$$

one can directly verify that (A.15) is nothing but CSLE (A.1) and therefore (A.15*) must be itself the SLE written in the canonical form.

Operators (A.12a) -- the so-called 'generalized Darboux transformations' (GDTs) –were originally introduced by Rudyak and Zakhariev [35] for some exactly-solvable scattering models and then extended to the generic CSLE by Leib and Schnizer [36-38] and independently by Suzko [39-41]. As already mentioned in Introduction the straightforward analysis of the appropriate intertwining relations was more recently presented by Suzko and Giorgadze [42] who refer to the generic CSLE as the generalized Schrödinger equation and focus solely on physical applications which do not require introducing new variables via LTs.

Since the FFs $\phi_\tau[\xi;Q^o_{\downarrow\tau}]$ and $^*\phi_\tau[\xi;Q^o_{\downarrow\tau}]$ are solutions of CSLEs (A.15) and (A.15*) at the energy $\varepsilon=\varepsilon_\tau$ the free terms of these equations can be represented as

$$I^o[\xi;Q^o_{\downarrow\tau}] = -\frac{\ddot{\phi}_\tau[\xi;Q^o_{\downarrow\tau}]}{\phi_\tau[\xi;Q^o_{\downarrow\tau}]} - \varepsilon_\tau\,\wp[\xi] \tag{A.19}$$

and

$$^1I^o[\xi;Q^o_{\downarrow\tau}\,|\,\tau] = -\frac{{}^*\ddot{\phi}_\tau[\xi;Q^o_{\downarrow\tau}]}{{}^*\phi_\tau[\xi;Q^o_{\downarrow\tau}]} - \varepsilon_\tau\,\wp[\xi]. \tag{A.19*}$$

Substituting (A.17) into the latter formula and making some trivial algebraic manipulations one comes to the following explicit expression between the two

$$^1I^o[\xi;Q^o_{\downarrow\tau}\,|\,\tau] - I^o[\xi;Q^o_{\downarrow\tau}] = 2\,\wp^{1/2}[\xi]\frac{d}{d\xi}\left\{\wp^{-1/2}[\xi]\,ld\left|\wp^{1/4}[\xi]\,\phi_\tau[\xi;Q^o_{\downarrow\tau}]\right|\right\}, \tag{A.20}$$

in agreement with Darboux' conventional formula for the appropriate Liouville potentials

$$V_L[\xi(x);Q^o_{\downarrow\tau}] - {}^1V_L[\xi(x);Q^o_{\downarrow\tau}\,|\,\tau] = 2\frac{d}{dx}\,ld\left|\psi_\tau[\xi(x);Q^o_{\downarrow\tau}]\}\right|. \tag{A.21}$$

Let $\phi_{\tau'}[\xi;Q^o_{\downarrow\tau,\tau'}]$ be another solution of CSLE (A.1) at the energy $\varepsilon_{\tau'}$ . Then the function

$$^1\phi_{\tau'}[\xi;Q^o_{\downarrow\tau,\tau'}] = \frac{W\{\phi_\tau[\xi;Q^o_{\downarrow\tau,\tau'}],\phi_{\tau'}[\xi;Q^o_{\downarrow\tau,\tau'}]\}}{\wp^{1/2}[\xi]\,\phi_\tau[\xi;Q^o_{\downarrow\tau,\tau'}]} \tag{A.22}$$

where $W\{\psi_1[\xi],\psi_2[\xi]\}$ is Wroskian of the functions $\psi_1[\xi]$ and $\psi_2[\xi]$ of $\xi$, is a solution of CSLE (A.14) at the same energy $\varepsilon_{\tau'}$. The easiest way to verify this assertion is to take advantage of the standard Darboux expression

$$ {}^1\psi_{\tau'}[\xi(x);Q^o_{\downarrow\tau,\tau'}\,|\,\tau] = \frac{W\{\psi_\tau[\xi(x);Q^o_{\downarrow\tau,\tau'}],\psi_{\tau'}[\xi(x);Q^o_{\downarrow\tau,\tau'}]\}}{\psi_\tau[\xi(x);Q^o_{\downarrow\tau,\tau'}]} \qquad \text{(A.23)} $$

for the solution of the Schrodinger equation with the Liouville potential ${}^1V_L[\xi(x);Q^o_{\downarrow\tau}\,|\,\tau]$, where the Wroskian is computed by differentiating the appropriated functions with respect to x, so that

$$ W\{\psi_\tau[\xi(x)],\psi_{\tau'}[\xi(x)]\} = \wp^{-1/2}[\xi(x)]\,W\{\psi_\tau[y],\psi_{\tau'}[y]\}\big|_{y=\xi(x)} \qquad \text{(A.24)} $$

Making use of (A.8), with $\tau$ changed for $\tau'$, and taking into account that (A.24)

$$ W\{f[\xi]\phi_1[\xi],\, f[\xi]\phi_2[\xi]\} = f^2[\xi]W\{\phi_1[\xi],\phi_2[\xi]\}, \qquad \text{(A.25)} $$

with $f[\xi]=\wp^{1/4}[\xi]$, then leads us directly to (A.22).

Finally let us briefly compare our results with Gomez-Ullate, Kamran, and Milson's factorizations (13) and (16) in [50] for an arbitrary second-order differential operator

$$ \hat{T} \equiv p[\xi]\hat{d}^2_\xi + q[\xi;Q^o]\hat{d}_\xi + r[\xi;Q^o]\,. \qquad \text{(A.25)} $$

First note that the gauge transformation

$$ \Theta[\xi;\varepsilon;Q^o] = g^{-1}[\xi;Q^o]\,\Phi[\xi;\varepsilon;Q^o]\,, \qquad \text{(A.26)} $$

where

$$ p[\xi] = -\wp^{-1}[\xi]\,, \qquad \text{(A.27a)} $$

$$ q[\xi;Q^o]/\,p[\xi] = -2ld\, g[\xi;Q^o]\,, \qquad \text{(A.27b)} $$

$$ r[\xi;Q^o]/\,p[\xi] = -\wp^{-1}[\xi]\{I^o[\xi;Q^o] + \ddot{g}[\xi;Q^o]\}\,, \qquad \text{(A.27c)} $$

converts the eigenproblem

$$(\hat{T} - \varepsilon)\,\Theta[\xi;\varepsilon;Q^{o}] = 0 \tag{A.28}$$

into CSLE (A.1) so that operators (14) and (15) in [50] are related to the CLDOs $\hat{L}_{\tau}$ and $^{*}\hat{L}_{\tau}$ via the similarity transformation. The direct consequence of this observation is that the SUSY partner of operator (14) must have the same coefficients for both first and second derivatives. If

$$\theta_{\tau}[\xi;Q^{o}_{\downarrow\tau}] = g^{-1}[\xi;Q^{o}_{\downarrow\tau}]\,\phi_{\tau}[\xi;Q^{o}_{\downarrow\tau}] \tag{A.29}$$

is an eigenfunction of operator (A.27) with an eigenvalue $\varepsilon_{\tau}$, then its SUSY partner is thus given by a similar expression

$$^{*}\theta_{\tau}[\xi;Q^{o}_{\downarrow\tau}] = g^{-1}[\xi;Q^{o}_{\downarrow\tau}]\,^{*}\phi_{\tau}[\xi;Q^{o}_{\downarrow\tau}] \tag{A.29*}$$

and

$$\theta_{\tau}[\xi;Q^{o}_{\downarrow\tau}]\,^{*}\theta_{\tau}[\xi;Q^{o}_{\downarrow\tau}] = g^{-2}[\xi;Q^{o}_{\downarrow\tau}]\,\wp^{-1/2}[\xi] \tag{A.30}$$

By applying the similarity transformation to CLDOs (A.12a) and (A.16a*) represented as

$$\hat{L}_{\tau} \equiv \hat{D}\{\theta_{\tau}[\xi;Q^{o}_{\downarrow\tau}]\} - ld\;g[\xi;Q^{o}_{\downarrow\tau}] \tag{A.31}$$

and

$$^{*}\hat{L}_{\tau} = \wp^{-1}[\xi]\{\hat{d}_{\xi} + ld\;\theta_{\tau}[\xi;Q^{o}_{\downarrow\tau}] + ld\;g[\xi;Q^{o}_{\downarrow\tau}]\}\,\wp^{1/2}[\xi]\;, \tag{A.31*}$$

one can directly verify that the resultant operators match definitions (14) and (15) for the operators A and B in [48], respectively, with b standing for $\wp^{-1/2}[\xi]$. The reader can also easily verify that the cited expression for the operator A turns into Schulze-Halberg's GDT (6) in [106], with f and g standing for $-p[\xi]$ and $-q[\xi;Q^{o}]$, respectively. In Part II we will provide some additional details concerning the relationship between higher-order GDTs introduced in [106] and our approach focused on multi-step DTs of RLPs exactly quantized by polynomials.

**Appendix B**

**Disappearance of the upper energy level along one of the c/a′ ZFE separatrix**

Let us study more carefully behavior of AEH solutions near ZFE straight-lines (5.1.31). First note that the signed exponent difference $\lambda_{0;\dagger m}$ has a nonzero absolute value positive for $\lambda_o > 0$ and therefore it must have the same sign on both sides of each straight-line (except the limiting case of the AL potential curve). This implies that the only possible change of solution type from one side of the separatrix to another is either **a** $\leftrightarrow$ **c** or **b** $\leftrightarrow$ **d**.

For each natural number m≥0 there are three ZFE separatrices

$$\mu_o = \lambda_o + 2m + 1, \tag{B.1a}$$

$$\mu_o + \lambda_o = 2m + 1, \tag{B.1b}$$

$$\mu_o + 2m + 1 = \lambda_o. \tag{B.1c}$$

The sectors formed by these separatrices are referred to

Area $A_m$: $\lambda_o < \mu_o - 2m - 1$; (B.2a)

Area $B_m$: $\mu_o + \lambda_o < 2m + 1$; (B.2b)

Area $C_m$: $\lambda_o > \mu_o + 2m + 1$; (B.2c)

Area $D_m$: otherwise, (B.2d)

as illustrated by Figure 1 for m = 2. Area $A_2$ represents the region where the *r*-GRef potential generated using TP with positive discriminant has at least 3 bound energy levels. The number of bound energy levels on the D-side of the **c/a′** ZFE separatrix may not exceed 3. No discrete energy spectrum exists below the solid line representing the **c/a′** ZFE separatrix $\lambda_o + 1 = \mu_o$.

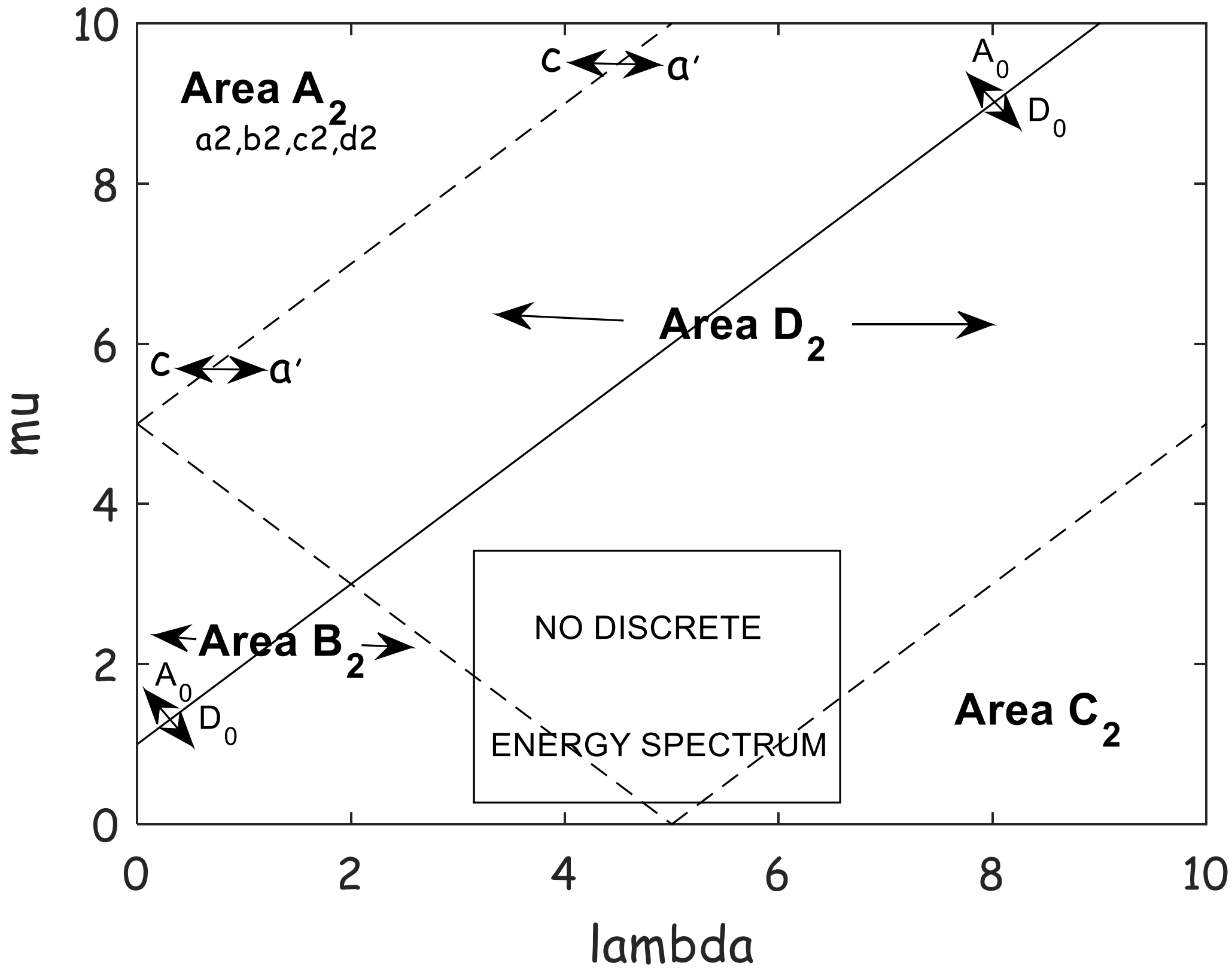

**Figure 1. Four major Areas carved by ZFE separatrices (dashed straight-lines) for the TP with positive discriminant**

It is theoretically possible that normalizable AEH solutions re-appear outside Area $A_0$ after a pair of AEH solutions of the same type merge so quartic polynomial (5.1.20a) has complex roots. However all the closed-formed examples analyzed by us so far support the assumption that the discrete energy spectrum exists only in Area $A_0$.

In this paper we are only interested in **c**/**a′** ZFE separatrices

$$\mu_o = \lambda_o + 2m + 1 \tag{B.4}$$

where the upper bound energy level disappears one by one as $\lambda_o$ monotonically increases at the fixed value of the RI $\mu_o$. By neglecting terms proportional to $\lambda^2$ in the right-hand side of (5.1.20a) one can approximately decompose the quartic polynomial as

$$G_4^{(m)}[\lambda\,|\,{}_1\boldsymbol{G}] = [\mu_o^2 - (\lambda_o + 2\lambda + 2m + 1)(\lambda_o + 2m + 1)]$$
$$\times [\mu_o^2 - (\lambda_o - 2\lambda - 2m - 1)(\lambda_o - 2m - 1)] + O(\lambda^2) \qquad \text{(B.5)}$$

so that the root changing its sign on the m$^{th}$ $\boldsymbol{c}/\boldsymbol{a}'$ ZFE separatrix can be approximated as

$$\lambda_{1;\dagger m} \approx \frac{\mu_o^2 - (2m + 1 + \lambda_o)^2}{2(2m + 1 + \lambda_o)}, \qquad \text{(B.6)}$$

where

$$\dagger = \begin{cases} \boldsymbol{c} & \text{if } \mu_o < \lambda_o + 2m + 1, \\ \boldsymbol{a}' & \text{if } \mu_o > \lambda_o + 2m + 1. \end{cases} \qquad \text{(B.6*)}$$

By representing the quartic polynomial in the vicinity of the m$^{th}$ $\boldsymbol{c}/\boldsymbol{a}'$ ZFE as

$$G_4^{(m)}[\lambda\,|\,{}_1\boldsymbol{G}] = [{}_1d\lambda^2 + 2(2m + 1)(\lambda_o - \lambda)]^2 - 4(\lambda + 2m + 1)^2({}_1c_0\,\lambda^2 + \lambda_o^2)$$
$$+ O(\mu_o - \lambda_o - 2m - 1) \qquad \text{(B.7)}$$

one finds that nonzero roots are defined by the cubic equation

$$\left\{[{}_1d_1\lambda_{1;\dagger m} - 2(\lambda_o + 2m + 1)]^2 - 4\,{}_1c_0({}_1\lambda_{1;\dagger m} + 2m + 1)^2\right\}{}_1\lambda_{1;\dagger m}$$
$$+ 4({}_1\lambda_{1;\dagger m} + 2m + 1)\lambda_o[{}_1d_1\lambda_{1;\dagger m} - 2(\lambda_o + 2m + 1)] = 0 \qquad \text{(B.8)}$$

with a positive leading coefficient and a negative free term, in agreement with the assertion that it has one positive and two negative roots. In the limiting case of the DRtTP potential $V[z|_1\boldsymbol{G}^{210}]$ the leading coefficient vanishes and Eq. (B.8) turns into the quadratic equation.

## Appendix C

## A closer look at reference polynomial fractions generated via single-step CLDTs of GRef potentials

As mentioned in Section 6 use of generic formula (A.20), instead of (A.19*), allows one to exclude the factorization energies ${}_\iota\varepsilon_{\dagger m}$ from the polynomial $\hat{O}^{\downarrow}_{m+\Im}[\xi\,|\,{}^{1}_{\iota}G^{K\Im 0}_{\dagger m}]$ in a straightforward way. Indeed an analysis of the formula

$$I^{o}[\xi\,|\,{}^{1}_{\iota}G^{K\Im 0}_{\dagger m}] - I^{o}[\xi\,|\,{}_{\iota}G^{K\Im 0}_{\downarrow\dagger m}] = \tfrac{1}{2}\,l\dot{d}\;{}_{\iota}\wp[\xi;\,{}_{\iota}T_{2}] - \tfrac{1}{4}\,ld^{2}\;{}_{\iota}\wp[\xi;\,{}_{\iota}T_{2}] \qquad \text{(C.1)}$$

$$+2\,l\dot{d}\left|\phi_{\dagger m}[\xi\,|\,{}_{\iota}G^{K\Im 0}_{\downarrow\dagger m}]\right| - ld\;{}_{\iota}\wp[\xi;\,{}_{\iota}T_{2}]\;ld\left|\phi_{\dagger m}[\xi\,|\,{}_{\iota}G^{K\Im 0}_{\downarrow\dagger m}]\right|,$$

where

$$ld\,|\,\phi_{\dagger m}[\xi\,|\,{}_{\iota}G^{K\Im 0}_{\downarrow\dagger m}]\,| = ld\left|{}_{\iota}\Theta[\xi;\,{}_{\iota}\lambda_{0;\dagger m},\,{}_{\iota}\lambda_{1;\dagger m}]\right| + ld\;|\,\Pi_{m}[\xi;\,\bar{\xi}_{\dagger m}]\,|, \qquad \text{(C.2)}$$

shows that RefPF $I^{o}[\xi\,|\,{}^{1}_{\iota}G^{K\Im 0}_{\dagger m}]$ (6.30a′) and therefore the polynomial $\hat{O}^{\downarrow}_{m+\Im}[\xi\,|\,{}^{1}_{\iota}G^{K\Im 0}_{\dagger m}]$ are explicitly expressible in terms of signed ExpDiffs ${}_{\iota}\lambda_{0;\dagger m}$ and ${}_{\iota}\lambda_{1;\dagger m}$ $({}_{0}\lambda_{1;\dagger m} \equiv \nu_{\dagger m})$.

Differentiating the logarithmic derivative

$$ld\;{}_{\iota}\wp[\xi;\,{}_{\iota}T_{K}] = {}_{\iota}\rho_{T}\,\frac{\dot{\Pi}_{\Im}[\xi;\,{}_{\iota}\bar{\xi}_{T}]}{\Pi_{\Im}[\xi;\,{}_{\iota}\bar{\xi}_{T}]} - 2\sum_{r=0}^{|\iota|}(\xi - {}_{\iota}e_{r})^{-1} \qquad \text{(C.3)}$$

with respect to ξ one can represent the first two terms in the right-hand side of (C.1) as

$$\tfrac{1}{2}\,l\dot{d}\;{}_{\iota}\wp[\xi;\,{}_{\iota}T_{2}] - \tfrac{1}{4}\,ld^{2}\;{}_{\iota}\wp[\xi;\,{}_{\iota}T_{2}] = 2\hat{Q}[\xi;\,{}_{\iota}\bar{\xi}_{T};\,{}_{\iota}\rho_{T}] - 2\,\frac{|\,\iota\,|}{\prod_{r=0}^{|\iota|}(\xi - {}_{\iota}e_{r})} \qquad \text{(C.4)}$$

$$+\,2\,{}_{\iota}\rho_{T}\,\frac{\dot{\Pi}_{\Im}[\xi;\,{}_{\iota}\bar{\xi}_{T}]}{\Pi_{\Im}[\xi;\,{}_{\iota}\bar{\xi}_{T}]}\sum_{r=0}^{|\iota|}(\xi - {}_{\iota}e_{r})^{-1},$$

where the PFr $\hat{Q}[\xi; {}_\iota\bar{\xi}_T; {}_\iota\rho_T]$ is defined via (3.23). Similarly differentiating the logarithmic derivative

$$ld \mid {}_\iota\Theta[\xi; {}_\iota\bar{\lambda}_{\dagger m}] \mid = \sum_{r=0}^{|\iota|} \frac{{}_\iota\lambda_{r;\dagger m} + 1}{2(\xi - {}_\iota e_r)} + \tfrac{1}{2}\delta_{\iota,0}\nu_{\dagger m} \tag{C.5}$$

$(\nu_{\dagger m} \equiv {}_0\lambda_{1;\dagger m})$ with respect to $\xi$:

$$l\dot{d} \mid {}_\iota\Theta[\xi; {}_\iota\bar{\lambda}_{\dagger m}] \mid = -\sum_{r=0}^{|\iota|} \frac{{}_\iota\lambda_{r;\dagger m} + 1}{2(\xi - {}_\iota e_r)^2} \tag{C.6}$$

gives

$$2l\dot{d}\left|{}_\iota\Theta[\xi; {}_\iota\bar{\lambda}_{\dagger m}]\right| + 2\,ld\left|{}_\iota\Theta[\xi; {}_\iota\bar{\lambda}_{\dagger m}]\right| \sum_{r=0}^{|\iota|} \frac{1}{\xi - {}_\iota e_r} = \frac{|\iota| \sum_{r=0,1} ({}_\iota\lambda_{r;\dagger m} + 1) - \delta_{\iota,0}\,\nu_{\dagger m}}{\prod_{r=0}^{|\iota|} (\xi - {}_\iota e_r)} \tag{C.7}$$

so that the PFr in question does not contain second-order poles at $\xi = {}_\iota e_r$. Substituting (C.2)-(C.4) and (C.7) into the right-hand side of (C.1) we can thus represent the RefPF of our interest as

$$\begin{aligned} I^o[\xi \mid {}^1_\iota\boldsymbol{G}^{K\Im 0}_{\dagger m}] - I^o[\xi \mid {}_\iota\boldsymbol{G}^{K\Im 0}_{\downarrow\dagger m}] = {} & 2Q[\xi; \bar{\xi}_{\dagger m}] + 2\hat{Q}[\xi; {}_\iota\bar{\xi}_T; {}_\iota\rho_T] + 2\Delta\hat{Q}^{K\Im 0}[\xi; \bar{\xi}_{\dagger m}; {}_\iota\bar{\xi}_T] \\ & + \frac{|\iota| \sum_{r=0,1} ({}_\iota\lambda_{r;\dagger m} + 1) - \delta_{\iota,0}\,\nu_{\dagger m}}{\prod_{r=0}^{|\iota|} (\xi - {}_\iota e_r)}, \end{aligned} \tag{C.8}$$

where the first three PFrs in the right-hand side are defined via (3.26), (3.21), and (3.23$^\dagger$), respectively. Changing the QPFr for PFr (3.21) via (3.26*) and making use of the identity

$$B_1[\xi; \bar{\iota}; \lambda_0, \lambda_1] + B_1[\xi; \bar{\iota}; -\lambda_0, -\lambda_1] \equiv \sum_{r=0}^{|\iota|} (\xi - {}_\iota e_{1-r})^{|\iota|} \tag{C.9}$$

we can finally represent the RefPFr in question as

$$I^{o}[\xi \,|\, {}_{\iota}^{1}G^{K\Im 0}_{\dagger m}] = I^{o}[\xi \,|\, {}_{\iota}G^{K\Im 0}_{\downarrow\dagger m}] + 2\hat{Q}[\xi; {}_{\iota}\bar{\xi}_{\dagger m}, {}_{\iota}\bar{\xi}_{T}] \tag{C.10}$$

$$+ \frac{\Delta\hat{O}^{\downarrow}_{m+\Im}[\xi \,|\, {}_{\iota}^{1}G^{K\Im 0}_{\dagger 0}]}{4\prod_{r=0}^{|\iota|}(\xi - {}_{\iota}e_{r})\,\Pi_{\Im}[\xi; {}_{\iota}\bar{\xi}_{T}]\,\Pi_{m}[\xi; {}_{\iota}\bar{\xi}_{\dagger m}]},$$

with

$$\Delta\hat{O}^{\downarrow}_{m+\Im}[\xi \,|\, {}_{\iota}^{1}G^{K\Im 0}_{\dagger m}] = 4[\,|\iota|\,({}_{\iota}\lambda_{0;\dagger m} + {}_{\iota}\lambda_{1;\dagger m}) - \delta_{\iota,0}\ \nu_{\dagger m}]\,\Pi_{\Im}[\xi; {}_{\iota}\bar{\xi}_{T}]\Pi_{m}[\xi; {}_{\iota}\bar{\xi}_{\dagger m}]$$

$$+8\,{}_{\iota}\rho_{T}\,\hat{B}_{1}[\xi; \bar{\iota}; -{}_{\iota}\bar{\lambda}_{\dagger m}]\dot{\Pi}_{\Im}[\xi; {}_{\iota}\bar{\xi}_{T}]\Pi_{m}[\xi; {}_{\iota}\bar{\xi}_{\dagger m}]$$

$$+8\sum_{r=0}^{|\iota|}(\xi - {}_{\iota}e_{1-r})^{|\iota|}\,\Pi_{\Im}[\xi; {}_{\iota}\bar{\xi}_{T}]\dot{\Pi}_{m}[\xi; {}_{\iota}\bar{\xi}_{\dagger m}]$$

$$+4\prod_{r=0}^{|\iota|}(\xi - {}_{\iota}e_{r})\Pi_{\Im}[\xi; {}_{\iota}\bar{\xi}_{T}]\ddot{\Pi}_{m}[\xi; {}_{\iota}\bar{\xi}_{\dagger m}]. \tag{C.10′}$$

Alternatively we can exclude the second derivative of the monomial product $\Pi_{m}[\xi; \bar{\xi}_{\dagger m}]$ from the right-hand side of (C.10):

$$\Delta\hat{O}^{\downarrow}_{m+\Im}[\xi \,|\, {}_{\iota}^{1}G^{K\Im 0}_{\dagger m}] = \{4[\,|\iota|\,({}_{\iota}\lambda_{0;\dagger m} + {}_{\iota}\lambda_{1;\dagger m}) - \delta_{\iota,0}\ \nu_{\dagger m}] - C_{0}({}_{\iota}\varepsilon_{\dagger m} \,|\, {}_{\iota}G_{\downarrow\dagger m}; \bar{\sigma}_{\dagger})\}$$

$$\times\Pi_{\Im}[\xi; {}_{\iota}\bar{\xi}_{T}]\Pi_{m}[\xi; {}_{\iota}\bar{\xi}_{\dagger m}]$$

$$+8\,{}_{\iota}\rho_{T}\,\hat{B}_{1}[\xi; \bar{\iota}; -{}_{\iota}\bar{\lambda}_{\dagger m}]\dot{\Pi}_{\Im}[\xi; {}_{\iota}\bar{\xi}_{T}]\Pi_{m}[\xi; {}_{\iota}\bar{\xi}_{\dagger m}] \tag{C.10*}$$

$$+8\hat{B}_{1}[\xi; \bar{\iota}; -{}_{\iota}\bar{\lambda}_{\dagger m}]\Pi_{\Im}[\xi; {}_{\iota}\bar{\xi}_{T}]\dot{\Pi}_{m}[\xi; {}_{\iota}\bar{\xi}_{\dagger m}]$$

taking into account that they satisfy solved-by-polynomials differential equation (3.35). [While deriving (C.10*) we also once again took advantage of symmetry relation (C.9).] Making use of (3.33), coupled with the identity

$${}_{\iota}C^{0}_{0}(\lambda_{0}, \lambda_{1}) - {}_{\iota}C^{0}_{0}(-\lambda_{0}, -\lambda_{1}) \equiv |\iota|\,\lambda_{0} - (-1)^{|\iota|}\lambda_{1} \tag{C.11}$$

one can directly verify that polynomial (6.30a′) satisfies the relation

$$\hat{O}^{\downarrow}_{m+\Im}[\xi\,|\,{}_{\iota}^{1}\mathcal{G}^{K\Im 0}_{\mathbf{t}m}] = {}_{\iota}O^{0}_{0}\Pi_{\Im}[\,\xi;\,{}_{\iota}\bar{\xi}_{T}\,]\,\dot{\Pi}_{m}[\xi;\,{}_{\iota}\bar{\xi}_{\mathbf{t}m}\,] + \Delta\hat{O}^{\downarrow}_{m+\Im}[\xi\,|\,{}_{\iota}^{1}\mathcal{G}^{K\Im 0}_{\mathbf{t}m}\,] \qquad (C.10)$$

as expected.

## Appendix D

## Single-step SUSY partners of the Gendenshtein potential conditionally exactly quantized by orthogonal Routh-seed Heine polynomials

Compared with all other GRef potentials on the line, the very specific feature of the Gendenshtein (Scarf II) potential

$$V_{G}(x) \equiv V[\eta(x)\,|\,{}_{i}\mathcal{G}^{201}] = \frac{b^{2} - a(a+1) + b(2a+1)sh\,x}{2ch^{2}x} \qquad (D.1)$$

is that the TP used for its construction vanishes at both singular points –*i* and +*i*. As a result the ChExps of the ℜS solutions defined via (3.1) and (3.2) become energy-independent and the ℜS Heine polynomials describing bound energy states in the multi-step SUSY partners $V[\eta|\,{}_{i}^{p}\mathcal{G}^{201}_{\{\mathbf{t}m\}_p}\,]$ of the Gendenshtein potential form finite sets of multi-index orthogonal polynomials assuming that the potentials in question do not have singularities on the real axis $-\infty < \eta = sh\,x < +\infty$.

In [20] we have already taken advantage of the mentioned distinguished feature of the PFr beam ${}_{i}\mathcal{G}^{201}$ to prove that any ℜS solution $\mathbf{d}', 2\tilde{\imath}'_{1}$ for $\tilde{\imath}'_{1} > a$, or any ℜS solution $\mathbf{d}, 2\tilde{\imath}_{1}$ is nodeless and therefore (as explicitly demonstrated below) can be used as the FF to construct a new Hi-CEQ potential.

Contrary to all other GRef potentials on the line, the CLDTs in question change the ChExps at both finite singular points –*i* and +*i* . To be more precise, the CLDT with the FF

$$\phi[\eta\,|\,{}_{i}\mathcal{G}^{201}_{\downarrow\mathbf{t}_1 m_1};\mathbf{t}_1 m_1] = (1 - i\,\eta)^{i\rho_{\mathbf{t}_1}}\,(1 + i\,\eta)^{i\rho^{*}_{\mathbf{t}_1}}\,\Pi_{m_1}[\eta;\bar{\eta}_{\mathbf{t}_1 m_1}\,] \qquad (D.2)$$

($\mathbf{t}_1 = \mathbf{c}, \mathbf{d}$, or $\mathbf{d}'$)

changes the ChExps of the appropriate RCSLE exactly by one. This corollary is reminiscent of similar behavior of the RCSLEs with radial Liouville potentials [19, 33, 51] which have energy-independent ChExps at the origin.

The characteristic exponent ${}_{i}\rho_{\mathbf{t}}$ and its complex conjugate ${}_{i}\rho^{*}_{\mathbf{t}}$ common for all the $\Re$S solutions from the given sequence (either $\mathbf{t} = \mathbf{c}, \mathbf{d}'$ or $\mathbf{t} = \mathbf{d}$) are defined via the relation

$$2\,{}_{i}\rho_{\mathbf{t}} - 1 \equiv {}_{i}\lambda_{\mathbf{t}} = {}_{i}\sigma_{\mathbf{t}}\;{}_{i}\lambda_{\mathrm{o}}, \tag{D.3}$$

where

$${}_{i}\sigma_{\mathbf{t}} = \begin{cases} -1 \text{ for } \mathbf{t} = \mathbf{c} \text{ or } \mathbf{d}', \\ +1 \text{ for } \mathbf{t} = \mathbf{d} \end{cases} \tag{D.4}$$

and

$${}_{i}\lambda_{\mathrm{o}} \equiv \sqrt{\mathrm{h}_{\mathrm{o}} + 1} \;\; (Re\,{}_{i}\lambda_{\mathrm{o}} > 0), \tag{D.5}$$

or, using the Gendenshtein parameters *a* and *b*,

$${}_{i}\lambda_{\mathrm{o}} = a + \tfrac{1}{2} + i\,b \;\; (a > -\tfrac{1}{2}). \tag{D.5*}$$

The FF for the inverse CLDT has the following AEH form

$$^{*}\phi[\eta\,|\,{}_{i}G^{201}_{\downarrow\mathbf{t}_1\mathrm{m}_1};\mathbf{t}_1\mathrm{m}_1] \propto (1 - i\,\eta)^{\frac{1}{2} - {}_{i}\rho_{\mathrm{o}}}(1 + i\,\eta)^{\frac{1}{2} - {}_{i}\rho^{*}_{\mathrm{o}}} / \Pi_{\mathrm{m}_1}[\eta;\bar{\eta}_{\mathbf{t}_1\mathrm{m}_1}], \tag{D.6}$$

with the characteristic exponent

$${}^{1}_{i}\rho_{\mathbf{t}_1} = \tfrac{1}{2} - {}_{i}\rho_{\mathbf{t}_1} \tag{D.7}$$

at the singular point $-i$ and its complex conjugate ${}^{1}_{i}\rho^{*}_{\mathbf{t}_1}$ at $+i$. Setting

$${}^{1}_{i}\lambda_{\mathbf{t}_1} \equiv 2\,{}^{1}_{i}\rho_{\mathbf{t}_1} - 1, \tag{D.8a}$$

$$\mathrm{h}_{\mathrm{o}}\,|\,{}^{1}_{i}G^{201}_{\mathbf{t}_1\mathrm{m}_1}) \equiv {}^{1}\mathrm{h}_{\mathrm{o};\mathbf{t}_1} \equiv {}^{1}_{i}\lambda^{2}_{\mathbf{t}_1}, \tag{D.8b}$$

$\mathrm{m}_1 = 2\tilde{\mathrm{i}}_1$, and assuming that the monomial product $\Pi_{2\tilde{\mathrm{i}}_1}[\eta;\bar{\eta}_{\mathbf{t}_1,2\tilde{\mathrm{i}}_1}]$ does not have real roots (i.e., $\eta^{*}_{\mathbf{t}_1,2\mathrm{i}'} = \eta_{\mathbf{t}_1,2\mathrm{i}'-1}$ for k=1,…, $\mathrm{j}_1$) we can represent the RefPFr $\mathrm{I}^{\mathrm{o}}[\eta\,|\,{}^{1}_{i}G^{201}_{\mathbf{t}_1,2\tilde{\mathrm{i}}_1}]$ as

$$I^{o}[\eta\,|\,{}^{1}_{i}\boldsymbol{G}^{201}_{\mathbf{t}_1,2\tilde{i}_1}] = -\frac{{}^{1}h_{o;\mathbf{t}_1}(\eta^2-1)}{2(\eta^2+1)^2} - 4\sum_{i'=1}^{\tilde{i}_1}\frac{\eta^2+Re\,\eta^2_{\mathbf{t}_1,2\tilde{i}_1;2i'}}{(\eta^2+|\,\eta_{\mathbf{t}_1,2\tilde{i}_1;2i'}|^2)^2} \tag{D.9}$$

$$+\frac{O^{o}_{2\tilde{i}_1}[\eta\,|\,{}^{1}_{i}\boldsymbol{G}^{201}_{\mathbf{t}_1,2\tilde{i}_1}]}{4(\eta^2+1)^2\,\Pi_{2\tilde{i}_1}[\eta;\bar{\eta}_{\mathbf{t}_1,2\tilde{i}_1}]}.$$

The square root of (D.8) with a positive real part is thus equal to the following complex number

$$Re\,{}^{1}_{i}\lambda_{o;\mathbf{t}_1} = |2Re\,{}_{i}\rho_{\mathbf{t}_1}| = |\,Re\,{}_{i}\lambda_{o} + {}_{i}\sigma_{\mathbf{t}_1}|, \tag{D.10}$$

i.e.,

$${}^{1}_{i}\lambda_{o;\mathbf{t}_1} = \begin{cases} {}_{i}\lambda_{o} + {}_{i}\sigma_{\mathbf{t}_1} \text{ if } \mathbf{t}_1 = \mathbf{d} \text{ or } {}_{i}\lambda_{o} > 1 \\ 1 - {}_{i}\lambda_{o} \text{ if } \mathbf{t}_1 = \mathbf{c} \text{ or } \mathbf{d}' \text{ and } {}_{i}\lambda_{o} < 1. \end{cases} \tag{D.10*}$$

(The case ${}_{i}\lambda_{o} = 1$ should be considered separately.)

If the FF of type **d** is nodeless then AEH solution (D.6) turns into the ground-energy eigenfunction

$$\phi_{\mathbf{c}0}[\eta\,|\,{}^{1}_{i}\boldsymbol{G}^{201}_{\mathbf{t}_1,2\tilde{i}_1\downarrow\mathbf{c}0}] \propto (1-i\,\eta)^{1/2-{}_{i}\rho_{o}}(1+i\,\eta)^{1/2-{}_{i}\rho^{*}_{o}}/\,\Pi_{2\tilde{i}_1}[\eta;\bar{\eta}_{\mathbf{t}_1,2\tilde{i}_1}]$$

$$(\mathbf{t}_1 = \mathbf{d} \text{ or } \tilde{i}_1 > a,\ \mathbf{t}_1 = \mathbf{d}') \tag{D.11}$$

for the potential $V[\eta\,|\,{}^{1}_{i}\boldsymbol{G}^{201}_{\mathbf{t}_1,2\tilde{i}_1}]$. The eigenfunctions describing excited bound energy states are given by the generic formulas

$$\phi_{\mathbf{c},v+1}[\eta\,|\,{}^{1}_{i}\boldsymbol{G}^{201}_{\mathbf{t}_1,2\tilde{i}_1\downarrow\mathbf{c}v}] \propto (1-i\,\eta)^{{}_{i}\rho_{\mathbf{t}_1}-1/2}(1+i\,\eta)^{{}_{i}\rho^{*}_{\mathbf{t}_1}-1/2}$$

$$\times\frac{{}_{i}P_{2\tilde{i}_1+v+1}[\eta\,|\,\overline{\mathbf{t}_1,2\tilde{i}_1;\mathbf{c}v}]}{\Pi_{2\tilde{i}_1}[\eta;\bar{\eta}_{\mathbf{t}_1,2\tilde{i}_1}]}, \tag{D.11'}$$

where the PD in the right-hand side is defined via (7.3) and (7.4) in Section 7. In general

$${}_{i}P_{2\tilde{i}_1+v+1}[\eta\,|\,\overline{\mathbf{t}_1,2\tilde{i}_1;\mathbf{c}v}]$$

$$\propto (\eta^2+1)^{\nu_{\mathbf{t}_1,\mathbf{c}}}\,Hi_{2\tilde{i}_1+v+1-2\nu_{\mathbf{t}_1,\mathbf{c}}}[\eta;{}_{i}\lambda_{o}\,|\,\mathbf{t}_1,2\tilde{i}_1;\mathbf{c}v], \tag{D.12}$$

where $\nu_{\mathbf{t}_1,\mathbf{c}}$ stands for a nonnegative integer, i. e., the PD itself is proportional to a Heine polynomial only if it is indivisible by $\eta^2$+1. Since both GS solutions $\mathbf{c}$v and $\mathbf{d}', 2\tilde{\mathrm{i}}_1$ share the same pair of the ChExps, $1-{}_i\rho_0$ and $1-{}_i\rho_0^*$, PD (D.12) for $\mathbf{t}_1 = \mathbf{d}'$ can be analytically decomposed as follows

$$ {}_i\mathrm{P}_{2\tilde{\mathrm{i}}_1+\mathrm{v}+1}[\eta\,|\,\overline{\mathbf{d}',2\tilde{\mathrm{i}}_1;\mathbf{c}\mathrm{v}}] \propto (\eta^2+1)\,\mathrm{W}\{\Pi_{2\tilde{\mathrm{i}}_1}[\eta;\bar{\eta}_{\mathbf{d}',2\tilde{\mathrm{i}}_1}], \Pi_{\mathrm{v}}[\eta;\bar{\eta}_{\mathbf{c}\mathrm{v}}]\} \quad \text{(D.13)} $$

so that eigenfunctions (D.11′) describing excited bound energy states in the potential $\mathrm{V}[\eta\,|\,{}^1_i\mathcal{G}^{201}_{\mathbf{d}',2\tilde{\mathrm{i}}_1}]$ take the form

$$ \phi_{\mathbf{c},\mathrm{v}+1}[\eta\,|\,{}^1_i\mathcal{G}^{201}_{\mathbf{d}',2\tilde{\mathrm{i}}_1\downarrow\mathbf{c}\mathrm{v}}] \propto (1-i\,\eta)^{\,{}_i\rho_{\mathbf{d}'}+1/2}(1+i\,\eta)^{\,{}_i\rho^*_{\mathbf{d}'}+1/2} \quad \text{(D.14)} $$

$$ \times\,\frac{\mathrm{W}\{\Pi_{2\tilde{\mathrm{i}}_1}[\eta;\bar{\eta}_{\mathbf{d}',2\tilde{\mathrm{i}}_1}], \Pi_{\mathrm{v}}[\eta;\bar{\eta}_{\mathbf{c}\mathrm{v}}]\}}{\Pi_{2\tilde{\mathrm{i}}_1}[\eta;\bar{\eta}_{\mathbf{t}_1,2\tilde{\mathrm{i}}_1}]}, $$

On other hand, substituting (D.10*) into the definition

$$ Re\,{}^1_i\lambda_{\mathrm{o};\mathbf{t}_1} = |\,2(Re\,{}_i\rho_{\mathbf{c}} - 1/2 + \nu_{\mathbf{t}_1,\mathbf{c}}) - 1| \quad \text{(D.15)} $$

of the selected root of parameter (D.8b) one finds

$$ 1-2\nu_{\mathbf{t}_1,\mathbf{c}} = {}_i\sigma_{\mathbf{t}_1}\,. \quad \text{(D.16)} $$

We thus conclude that the Wroskian in the right-hand side of (D.14) is not divisible by $\eta^2$+1 and therefore proportional to a ℜS Heine polynomial:

$$ \mathrm{W}\{\Pi_{2\tilde{\mathrm{i}}_1}[\eta;\bar{\eta}_{\mathbf{d}',2\tilde{\mathrm{i}}_1}], \Pi_{\mathrm{v}}[\eta;\bar{\eta}_{\mathbf{c}\mathrm{v}}]\} \propto \mathrm{Hi}_{2\tilde{\mathrm{i}}_1+\mathrm{v}-1}[\eta;\,{}_i\lambda_{\mathrm{o}}\,|\,\mathbf{d}', 2\tilde{\mathrm{i}}_1; \mathbf{c}\mathrm{v}]. \quad \text{(D.17′)} $$

On the contrary, $\nu_{\mathbf{d},\mathbf{c}} = 0$ so that the PD itself is nothing but a scaled ℜS Heine polynomial:

$$ {}_i\mathrm{P}_{2\tilde{\mathrm{i}}_1+\mathrm{v}+1}[\eta\,|\,\overline{\mathbf{d},2\tilde{\mathrm{i}}_1;\mathbf{c}\mathrm{v}}] \propto \mathrm{Hi}_{2\tilde{\mathrm{i}}_1+\mathrm{v}-1}[\eta;\,{}_i\lambda_{\mathrm{o}}\,|\,\mathbf{d}, 2\tilde{\mathrm{i}}_1; \mathbf{c}\mathrm{v}]. \quad \text{(D.17)} $$